\documentclass[a4paper,12pt]{article}

\usepackage{color}
\usepackage{latexsym}
\usepackage{amssymb,euscript, mathrsfs}
\usepackage{amsmath}
\usepackage{bbm} 
\usepackage{verbatim}

\usepackage{jheppub}                   
\makeatletter
\gdef\@fpheader{\ }                    
\makeatother

\newcommand{\bpm}{\begin{pmatrix}}
\newcommand{\epm}{\end{pmatrix}}

\newsavebox{\ns}
\newsavebox{\dbrane}
\newsavebox{\dbshort}

\def\be{\begin{equation}}
\def\ee{\end{equation}}
\def\bea{\begin{eqnarray}}
\def\eea{\end{eqnarray}}

\newcommand{\nn}{\nonumber}

\newcommand\R{\mathbb{R}}
\newcommand\Z{\mathbb{Z}}

\newcommand\ml{\mathcal{L}}

\newcommand{\cO}{\mathcal{O}}

\newcommand\C{\mathbb{C}}

\newcommand\diff{\mathrm{d}}

\newcommand{\dd}{\mathrm{d}}
\newcommand{\me}{\mathrm{e}}
\newcommand{\ii}{\mathrm{i}}

\newcommand{\ex}{\mathrm{e}}
\newcommand{\vol}{\mathrm{vol}}

\newcommand{\zbar}{\overline{z}}

\newlength{\sswidth}

\newcommand{\gr}{\mathrm{grav}}

\newcommand{\ct}{\mathrm{ct}}

\newcommand{\hook}{\mathbin{\rule[.2ex]{.4em}{.03em}\rule[.2ex]{.03em}{.9ex}}}

\newcommand{\Esusy}{E_{\mathrm{susy}}}
\newcommand{\calF}{\mathcal{F}}
\newcommand{\calA}{\mathcal{A}}
\newcommand{\zetacharge}{\chi}
\newcommand{\lambdaNEW}{\mu}
\newcommand{\PsiTwist}{\Xi}

\def\a{\mathtt{a}}
\def\c{\mathtt{c}}
\def\wzero{w}
\def\wone{u}
\def\phizero{a}
\def\cpsi{\gamma}
\def\ct{\gamma'}

\newcommand{\tr}{\rho}
\newcommand{\tx}{x}

\numberwithin{equation}{section}       

\title{Holographic~\hbox{renormalization and supersymmetry}}

\author[a]{Pietro Benetti Genolini,}
\emailAdd{pietro.benettigenolini@maths.ox.ac.uk}
\author[b]{Davide Cassani,}
\emailAdd{davide.cassani@lpthe.jussieu.fr}
\author[c]{Dario Martelli,}
\emailAdd{dario.martelli@kcl.ac.uk}
\author[a]{and James Sparks}
\emailAdd{James.Sparks@maths.ox.ac.uk}

\preprint{KCL-MTH-16-08}

\affiliation[a]{Mathematical Institute, University of Oxford, Woodstock Road, Oxford OX2 6GG, U.K.}

\affiliation[b]{LPTHE, Sorbonne Universit\'es UPMC Paris 6 and CNRS, UMR 7589, F-75005, Paris,~France}

\affiliation[c]{Department of Mathematics, King's College London, The Strand, London, WC2R 2LS,~U.K.}

\abstract{Holographic renormalization is a systematic procedure for 
regulating divergences in observables in asymptotically locally AdS spacetimes. For dual boundary field theories 
which are supersymmetric it is natural to ask whether this defines a supersymmetric renormalization scheme. 
Recent results in localization have brought this question into sharp focus: 
rigid supersymmetry on a curved boundary requires specific geometric structures, and 
general arguments imply that BPS observables, such as the partition function, 
are invariant under certain deformations of these structures. One can then ask if the dual holographic observables are similarly invariant. We study this question in minimal $\mathcal{N}=2$ gauged supergravity in four and five dimensions. In four dimensions we show that holographic renormalization precisely reproduces the expected field theory results. In five dimensions we find that no choice of standard holographic counterterms is compatible with  supersymmetry, which leads us to introduce novel finite boundary terms. For a class of solutions satisfying certain topological assumptions we provide some independent tests of these new boundary terms, in particular showing that they reproduce the expected VEVs of conserved charges.}

\begin{document}
\maketitle

\section{Introduction and summary}

Holographic observables in the AdS/CFT correspondence typically need regularizing. In particular 
divergences often arise near the conformal boundary, which are interpreted as UV divergences in the 
 dual field theory  \cite{Witten:1998qj,Gubser:1998bc}.  The method of holographic renormalization, which 
removes these infinities in gravitational observables via the addition of local boundary counterterms, 
was systematically  developed from the very beginnings of the subject.
This mirrors the corresponding procedure in field theory, and forms 
part of the foundations of the AdS/CFT correpondence.
Early references, incorporating a variety of approaches, include
\cite{Witten:1998qj,Gubser:1998bc,Henningson:1998gx, Balasubramanian:1999re,deBoer:1999tgo,deHaro:2000vlm, Bianchi:2001de, Bianchi:2001kw,Martelli:2002sp, Skenderis:2002wp}. 
However, the existence of \emph{finite} counterterms implies non-uniqueness 
of the renormalization scheme, and in such situations it is generally unclear 
how to match schemes on the two sides. 
Given that the classical gravitational description is typically valid only 
in a strong coupling limit of the field theory, generically it is difficult to directly 
compute observables on both sides, and hence make precise quantitative comparisons.

Precision tests of the AdS/CFT correspondence usually rely on the presence of additional symmetries, 
the notable examples being integrability and supersymmetry. In particular when the 
field theory is supersymmetric, it is natural to ask whether holographic renormalization 
of its dual description is a supersymmetric regularization scheme. Recent exact results 
in supersymmetric quantum field theories defined in curved space, relying on localization techniques \cite{Witten:1988ze,Nekrasov:2002qd,Pestun:2007rz}, have brought this question into sharp focus: many BPS observables 
may be computed exactly and unambiguously in field theory, and these
may then be compared with holographic dual supergravity computations. 
Any ambiguities in defining finite renormalized quantities in gravity 
are then expected to be resolved in making such comparisons.
As well as trying to match precise quantities on both sides, 
there are more general predictions that may also be compared, such as the dependence of BPS observables on given sets of boundary data. 
These latter tests of the correspondence are inherently more  robust than comparing observables in particular 
theories/backgrounds, and will hence be a main focus of this paper.  

We will concentrate on the correspondence between field theory partition function and gravity on-shell action.
In the appropriate large $N$ field theory limit in which semi-classical gravity describes the conformal field theory (CFT), the AdS/CFT correspondence states that
\bea\label{AdSCFTZ}
Z_{\rm CFT}[M_{d}] \,& = &\, \sum\, \ex^{-S[M_{d+1}]}~.
\eea
Here $Z_{\rm CFT}$ is the partition function of the CFT defined on a background $M_d$, 
while $S[M_{d+1}]$ is the holographically renormalized gravity action, evaluated on an asymptotically locally Euclidean AdS (AlEAdS) solution $M_{d+1}$ 
that has conformal boundary $M_d$, with the boundary conditions for the gravity fields corresponding to the CFT background fields.
The sum is over all AlEAdS gravity solutions with these boundary conditions that can be embedded into string theory~\cite{Witten:1998qj, Gubser:1998bc}.
We will study minimal $\mathcal{N}=2$ gauged supergravity in four and five dimensions, 
whose bosonic sectors are simply Einstein-Maxwell theory with a negative cosmological constant 
(and Chern-Simons coupling in dimension five). Solutions to these theories uplift either to M-theory or to type II string theory, for which there are large classes of known field theory duals.

Asymptotically locally AdS (AlAdS) supersymmetric solutions induce a rigid supersymmetric structure on the conformal boundary, which has been studied in both Lorentzian and Euclidean  signature~\cite{Klare:2012gn,Cassani:2012ri}. 
The boundaries $M_3$ of AlEAdS supersymmetric solutions to four-dimensional supergravity have metric of the form
\bea\label{3dmetricintro}
\diff s^2_3 &=& (\diff \psi+a)^2 + 4\ex^{\wzero}\diff z\diff \bar{z}~.
\eea
Here $\partial_\psi$ is a nowhere zero Killing vector on $M_3$, and we have used the 
freedom to make conformal transformations to take this to be a unit norm vector. 
This generates a transversely holomorphic foliation of $M_3$, 
allowing one to introduce a canonical local transverse complex coordinate $z$. 
The function
$\wzero=\wzero(z,\bar{z})$ is in general a local transverse function, while $a=a_z(z,\bar{z})\diff z + \overline{a_z(z,\bar{z})}\diff\bar{z}$ 
is a local one-form. We may also write $\diff a= \ii \wone\,   \ex^{\wzero} \diff z\wedge \diff\bar{z}$, where 
$u=u(z,\bar{z})$. In addition to the background metric (\ref{3dmetricintro}) 
there is also a non-dynamical Abelian  R-symmetry gauge field, which arises 
as the restriction of the bulk Maxwell field to the conformal boundary and whose form is specified by supersymmetry.

It is a general result of \cite{Closset:2013vra,Closset:2014uda} that 
the partition function of any $\mathcal{N}=2$ field theory in three dimensions, with a choice 
of Abelian R-symmetry coupling to the background R-symmetry gauge field, depends on the above background geometry 
only through the choice of transversely holomorphic foliation. 
Concretely, this means that the field theory partition function 
is invariant under deformations $\wzero\rightarrow \wzero + 
\delta \wzero$, $\wone\rightarrow \wone+ 
\delta \wone$, where $\delta \wzero(z,\bar{z})$, $\delta \wone(z,\bar{z})$ 
are \emph{arbitrary} smooth global functions on $M_3$, invariant under 
$\partial_\psi$. This is proven by showing that these 
deformations of the background geometry lead to $\mathcal{Q}$-exact 
deformations of the Lagrangian, where $\mathcal{Q}$ is a supercharge, and a standard argument then 
shows that the partition function is invariant. 
This general result has also been borne out by explicit computations 
of localized partition functions (such as \cite{Alday:2013lba}, where $M_3$ has the topology of $S^3$). 

 The field theory results in the previous paragraph then lead to a very concrete prediction: 
 the holographically renormalized on-shell action of a supersymmetric AlEAdS solution to 
 four-dimensional supergravity, with 
 conformal boundary $M_3$ and metric (\ref{3dmetricintro}),  should be invariant under the arbitrary deformations 
 $\wzero\rightarrow \wzero + \delta \wzero$, $\wone\rightarrow \wone+ \delta \wone$ defined above. 
As we shall review, in four dimensions holographic renormalization leads to a unique set of standard
counterterms for minimal $\mathcal{N}=2$ gauged supergravity -- there are no finite ambiguities\footnote{
More precisely there are no finite diffeomorphism-invariant and gauge-invariant local counterterms constructed 
using the bosonic supergravity fields.} -- 
and we prove that the renormalized on-shell action has indeed the expected invariance properties. 
Since we do this for an \emph{arbitrary} solution, and \emph{arbitrary} deformation, 
this constitutes a robust check of the AdS/CFT correspondence, in 
particular that holographic renormalization corresponds to the (unique) 
supersymmetric renormalization scheme employed implicitly in the  localization computations. 
We also go further, and show that the on-shell action itself 
correctly evaluates to the large $N$ field theory partition function obtained from localization, in the cases where this is known. 

The corresponding situation for five-dimensional supergravity turns out to be more involved. 
We will consider Euclidean conformal boundaries $M_4$ given by the direct product of a circle $S^1$ with $M_3$ equipped with the metric (\ref{3dmetricintro}), although we shall later generalize this slightly to a simple class of twisted backgrounds 
in which $S^1$ is fibred over $M_3$; the boundary value of the Abelian gauge field in the supergravity multiplet is again determined by supersymmetry. 
The general dependence of the four-dimensional field theory partition function on the background is similar to the one in three dimensions: for $\mathcal{N}=1$ theories with an R-symmetry (and thus for any \hbox{$\mathcal{N}=1$} superconformal field theory), the supersymmetric partition function is invariant under deformations $\wzero\rightarrow \wzero + 
\delta \wzero$, $\wone\rightarrow \wone+ 
\delta \wone$ \cite{Closset:2013vra,Closset:2014uda,Assel:2014paa}.  Although contrastingly with the three-dimensional case these ``supersymmetric Ward identities''  {\it a priori} only hold up to anomalies and local finite counterterms, it was shown in~\cite{Assel:2015nca} that the supersymmetric renormalization scheme used in field theory is unique, {\it i.e.}\ free of ambiguities. Moreover the background $M_4$ we consider is such that there are no Weyl and R-symmetry anomalies \cite{Cassani:2013dba}. Therefore the statement on invariance of the partition function holds exactly in our set-up.

In five-dimensional supergravity, holographic renormalization contains a set of diffeomorphism-invariant and gauge-invariant local boundary terms corresponding {\it a priori} to the same ambiguities and anomalies as in field theory~\cite{Witten:1998qj,Henningson:1998gx, Balasubramanian:1999re}. 
One might thus have expected that there is a unique linear combination of the finite 
holographic counterterms that matches the supersymmetric field theory scheme, {\it i.e.}\ such that the renormalized action is invariant under deformations $\wzero\rightarrow \wzero + 
\delta \wzero$, $\wone\rightarrow \wone+ 
\delta \wone$ of $M_4$. 
 Surprisingly, we find that \emph{no choice} of these counterterms 
has this property. If the AdS/CFT correspondence is to hold, 
we must conclude that holographic renormalization \emph{breaks supersymmetry} 
in this case (or, perhaps more precisely, is not compatible with the four-dimensional supersymmetry determining the Ward identities above).  
However, remarkably we are able to write down a set of non-standard, finite boundary terms that do not correspond to the usual diffeomorphism and gauge invariant terms and that give the on-shell action the expected invariance properties. 

The approach we follow in our supergravity analysis starts in Lorentzian signature. In particular we will rely on the existing classification of Lorentzian supersymmetric solutions to minimal gauged supergravity~\cite{Gauntlett:2003fk} to construct a very general AlAdS solution in a perturbative expansion near the boundary. Then we perform a Wick rotation; this generally leads to complex bulk solutions, however we focus on a class with real Euclidean conformal boundary $M_4 \cong S^1\times M_3$. 

The fact that supersymmetric holographic renormalization is more 
subtle in five dimensions was already anticipated, and in fact the issue can be illustrated by considering the simple case of AdS$_5$. In global coordinates, and after compactifying the Euclidean time, the conformal boundary of AdS$_5$ can be taken to be $M_4 \cong S^1\times S^3$, with a round metric on $S^3$. This space is expected to be dual to the vacuum of a superconformal field theory (SCFT) on $M_4$.
In this background, such theories develop a non-ambiguous non-zero vacuum expectation value (VEV) for both the energy and the R-charge operators \cite{Assel:2014paa,Assel:2015nca}. 
On the other hand, standard holographic renormalization unambiguously yields a vanishing electric charge for AdS$_5$, which leads to an immediate contradiction with the field theory result.  In fact this mismatch holds much more generally than just for AdS$_5$ space.
 For instance, in \cite{Cassani:2014zwa} a family of five-dimensional supergravity solutions was constructed, where the conformal boundary comprises a squashed $S^3$, and it was found that no choice of  standard holographic counterterms correctly reproduced the supersymmetric partition function and the corresponding VEV of the energy 
(the supersymmetric Casimir energy). Our general results summarized above 
explain all these discrepancies, and moreover the new counterterms we have introduced 
solve \emph{all} of these issues. In fact we go further, and show that for a general
class of solutions satisfying certain topological assumptions (which may be argued to be required for the solution to correspond to the vacuum state of the dual SCFT), our holographically renormalized 
VEVs of conserved charges quantitatively reproduce the expected field theory results.
Part of these results, with an emphasis on the holographic supersymmetric Casimir energy, were presented in the short communication~\cite{Genolini:2016sxe}.

The rest of the paper is organized as follows. In section~\ref{sec:fieldtheory} we review the relevant field theory backgrounds and the properties of supersymmetric partition functions. In section~\ref{sec:4dgravity} we present our four-dimensional supergravity analysis, showing in particular that standard holographic renormalization does satisfy the supersymmetric Ward identities, and evaluating the on-shell action for a large class of self-dual solutions. In section~\ref{sec:5dsugra} we turn to five-dimensional supergravity. We prove that standard holographic renormalization fails to satisfy the supersymmetric Ward identities and we introduce the new boundary terms curing this issue. Then under some global assumptions we evaluate the renormalized on-shell action and compute the conserved charges, showing that they satisfy a BPS condition. 
Section~\ref{sec:5dexamples} discusses a number of examples in five dimensions, illustrating further the role of our new boundary terms and making contact with the existing literature.
We conclude in section~\ref{sec:outlook}.
Finally, appendix~\ref{app_curvature} contains information on relevant curvature tensors, appendix~\ref{app:constr_5d_sol} illustrates our construction of the five-dimensional perturbative solution, and appendix~\ref{app:Killingspinors} discusses the Killing spinors at the boundary.

\section{Field theory}\label{sec:fieldtheory}

In this paper we are interested in the holographic duals to both three-dimensional and four-dimensional 
supersymmetric field theories, defined on general classes of rigid supersymmetric backgrounds. 
More precisely, these are three-dimensional $\mathcal{N}=2$ theories and 
four-dimensional $\mathcal{N}=1$ theories, in both cases with a choice of Abelian R-symmetry. 
For superconformal field theories, relevant for AdS/CFT, this R-symmetry will be the superconformal R-symmetry. 
Putting such theories on curved backgrounds, in a way that preserves supersymmetry, requires particular geometric structures. 
There are two general approaches: one can either couple the field theory 
to supergravity, and take a rigid limit in which the supergravity multiplet becomes a set of
non-dynamical background fields; or take a holographic approach, realizing 
the background geometry as the conformal boundary of a holographic 
dual supergravity theory \cite{Festuccia:2011ws, Klare:2012gn, Dumitrescu:2012ha, Closset:2012ru}. 
Both lead to the same results, although the holographic approach will be particularly 
relevant for this paper.

We will focus on backgrounds admitting two supercharges  of opposite R-charge. 
The resulting geometric structures in three and four dimensions are very closely 
related, and this will allow us to treat some aspects in parallel. In particular 
certain objects  will appear in both dimensions, and we will use a common notation 
-- the dimension should always be clear from the context.

\subsection{Three-dimensional backgrounds}\label{sec:3dbackgrounds}

The three-dimensional geometries of interest belong to a general class of 
real supersymmetric backgrounds, admitting two supercharges 
related to one another by charge conjugation \cite{Closset:2012ru}. 
If $\zeta$ denotes the Killing spinor then there is an associated Killing vector
\bea\label{xi3d}
\xi &=& \zeta^\dagger \sigma^i\zeta\, \partial_i \ = \ \partial_\psi~.
\eea
In an orthonormal frame 
here the Clifford algebra generators $\sigma^a$ may be taken to be the Pauli matrices, where 
$a=1,2,3$ is an orthonormal frame index. The Killing vector (\ref{xi3d}) is nowhere zero, and thus defines 
a foliation of  the three-manifold $M_3$. This foliation is transversely holomorphic, with 
transverse local complex coordinate~$z$. In terms of these coordinates the background metric is
\bea\label{3dmetric}
\diff s^2_3 &=& \Omega^2\left[(\diff\psi+a)^2 + 4\ex^{w}\diff z\diff\bar{z}\right]~.
\eea
Here $\Omega=\Omega(z,\bar{z})$ is a conformal factor, which is a global nowhere zero function on $M_3$,
$\wzero=\wzero(z,\bar{z})$ is in general a local transverse function, while $a=a_z(z,\bar{z})\diff z + \overline{a_z(z,\bar{z})}\diff\bar{z}$ 
is a local one-form. 
The metric and Riemannian volume form on the two-dimensional leaf space are
\bea
\label{2dMetric}
\diff s^2_{2} &= & 4\ex^{w}\diff z \diff \zbar~, \qquad \vol_{2} = 2\ii\, \ex^{w}\diff z \wedge \diff \zbar~.
\eea
Notice that $a$ is not gauge invariant under local diffeomorphisms 
of $\psi$. On the other hand the one-form
\bea\label{def_eta}
\eta \,&\equiv &\, \diff\psi+a
\eea
is a global almost contact form on $M_3$, where the Killing vector $\xi=\partial_\psi$ is the associated \emph{Reeb vector field}. 
It will be convenient to write 
\bea
\label{eq:3dBoundaryDa}
\diff\eta \, = \, \diff a \,&=&\, \ii\, \wone\, \ex^{\wzero}\diff z\wedge\diff{\bar{z}}~,
\eea
where $\wone=\wone(z,\bar{z})$ is a global function that parametrizes the gauge-invariant 
data in~$a$.  

Since we are mainly interested in conformal theories with gravity duals, 
we will (without loss of generality) henceforth set the conformal factor $\Omega\equiv 1$.
With this choice, the 
 non-dynamical R-symmetry gauge field that couples to the R-symmetry current is
\bea\label{A3d}
A \,&=&\, \frac{\wone}{4} (\diff \psi + a) + \frac{\ii}{4}(\partial_{\bar{z}}\wzero \diff\bar{z} - \partial_z \wzero \diff z)
 + \cpsi\,\diff\psi + \diff\lambda~.
\eea
Notice this is determined entirely by the metric data in (\ref{3dmetric}), apart from the last two terms which are locally pure gauge. Here $\lambda=\lambda(z,\bar{z})$, and the constant $\cpsi$ will play a particularly important role in this paper.
\footnote{Compared to the conventions of \cite{Farquet:2014kma, Farquet:2014bda},
 we have reversed the overall sign of $A$. However, as noted in the first of these references, 
for real $A$ sending $A\rightarrow - A$ is a symmetry of the Killing spinor equation, provided one also 
charge conjugates the spinor $\zeta\rightarrow\zeta^c$. This $\Z_2$ symmetry also reverses the sign of the Killing vector 
(\ref{xi3d}).}

\subsection{Four-dimensional backgrounds}\label{sec:4dbackgrounds}

There is a related class of rigid four-dimensional supersymmetric backgrounds, 
first discussed in~\cite{Klare:2012gn,Dumitrescu:2012ha}. These again have two supercharges
of opposite R-charge, with corresponding Killing spinors $\zeta_\pm$. 
We use the spinor conventions of~\cite{Dumitrescu:2012ha,Assel:2014paa}, in which  the positive/negative chirality $\zeta_\pm$ are two-component spinors with corresponding Clifford algebra generated by $({\sigma_\pm})^a=(\pm\vec{\sigma},-\ii \mathbbm{1}_2)$, 
where $a=1,\ldots,4$ is an orthonormal frame index and $\vec{\sigma}=(\sigma^1,\sigma^2,\sigma^3)$ are 
the Pauli matrices. In particular the generators of $SU(2)_\pm\subset \mathrm{Spin}(4)=SU(2)_+\times SU(2)_-$ are 
$(\sigma_\pm)^{ab} = \frac{1}{4}\left(\sigma_\pm^a\sigma_\mp^b - \sigma_\pm^b\sigma_\mp^a\right)~.
$ 
As in 
(\ref{xi3d}) we may define the vector field
\bea\label{K4d}
K &=& \zeta_+ \sigma_+^i \zeta_- \, \partial_i~.
\eea 
This is a \emph{complex} Killing vector, satisfying $K^i K_i=0$. 
Following \cite{Assel:2014paa, Martelli:2015kuk}, and to parallel the three-dimensional 
discussion in section \ref{sec:3dbackgrounds}, we consider a restricted class 
of these backgrounds in which the metric on $M_4$ takes the product form
\bea\label{bckgnd_metric}
\diff s^2_4  \,&=&\,  \diff \tau^2 + (\diff \psi + a)^2 + 4\me^{\wzero}\diff z \diff\bar z\ .
\eea
Thus $M_4\cong S^1\times M_3$, where $\tau\in [0,\beta)$ parametrizes the circle $S^1=S^1_\beta$. 
More generally one can also introduce an overall conformal factor $\Omega=\Omega(z,\bar{z})$, 
as in (\ref{3dmetric}), and the $\tau$ direction may be fibred over $M_3$, as we will discuss later in section~\ref{sec:twisted_bckgnd}.
The complex Killing vector (\ref{K4d}) takes the form
\bea\label{K4d_explicit}
K &=& \frac{1}{2}(\xi - \ii \partial_\tau)~,
\eea
where again $\xi=\partial_\psi$. The induced geometry on 
$M_3$, on a constant Euclidean time slice $\tau=$ constant, is identical to that 
for rigid supersymmetry in three dimensions. Moreover, the non-dynamical R-symmetry gauge field 
is
\be\label{A4d}
A \,=\, \frac{\wone}{4} (\diff \psi + a) + \frac{\ii}{4}(\partial_{\bar z} \wzero\diff \bar z - \partial_z \wzero \diff z) + \cpsi\,\diff \psi + \diff \lambda +\frac{\ii}{8}\wone\,\diff \tau  - \ii \ct \diff \tau \ .
\ee
We stress that this is the gauge field of  background
\emph{conformal} supergravity, 
 rather than  
the gauge field of new minimal supergravity~\cite{Sohnius:1981tp} used in~\cite{Dumitrescu:2012ha}. The former arises as the restriction
of the bulk graviphoton to the conformal boundary in the holographic approach to 
rigid supersymmetry \cite{Klare:2012gn,Cassani:2012ri}. 
Notice that setting $\tau=$ constant, 
(\ref{A4d}) reduces to the three-dimensional 
gauge field (\ref{A3d}). The last term in (\ref{A4d}), proportional to the (real) constant $\ct$, is 
again locally pure gauge, although via a complex gauge transformation. In contrast 
to three dimensions here $A$ is generically complex, although 
after a Wick rotation $\tau = \ii t$ to Lorentzian signature it becomes real.

The geometry we have described above is \emph{ambi-Hermitian}: the two Killing spinors 
$\zeta_\pm$ equip $M_4$ with two commuting integrable complex structures
\bea
(I_\pm)^{i}{}_{j} \,&=&\, -\frac{2\ii}{|\zeta_\pm|^2}\,\zeta_\pm^\dagger (\sigma_\pm)^{i}{}_{j}\,\zeta_\pm~.
\eea
The metric (\ref{bckgnd_metric}) is Hermitian with respect to both of these, but where the induced orientations 
are opposite. The complex Killing vector (\ref{K4d}) has Hodge  type $(0,1)$ with respect to both complex structures. 
On the other hand, the local one-form $\diff z$ has Hodge type $(1,0)$ with respect to $I_+$, but 
Hodge type $(0,1)$ with respect to $I_-$.

\subsection{Examples}\label{sec:FTexamples}

In both cases the geometry involves a three-manifold $M_3$, equipped with a transversely holomorphic 
foliation generated by the real Killing vector $\xi=\partial_\psi$. 
Any such three-manifold, with any compatible metric of the form (\ref{3dmetric}), defines 
a rigid supersymmetric background in both three and four dimensions. 
If all its orbits close $\xi$ generates a $U(1)$ isometry, and the quotient space 
$\Sigma_2=M_3/U(1)$ is  an orbifold Riemann surface, with induced metric 
\eqref{2dMetric}. 
Such three-manifolds are classified, and are known as \emph{Seifert fibred three-manifolds}. 
If $\xi$ has a non-closed orbit then $M_3$ admits at least 
a $U(1)^2$ isometry, meaning that the transverse metric $\diff s^2_2$ 
also admits a Killing vector. 

The simplest example has $M_3\cong S^3$, with 
$\xi$ generating the Hopf fibration of the round metric on $S^3$.\footnote{Throughout the paper, the symbol $\cong$ means ``diffeomorphic to''. In general, $M_d \cong S^d$ does not imply that the metric is the round metric on $S^d$; we will always specify when this is the case.} In this case $\Sigma_2\cong S^2$, equipped with 
its round metric. More generally one can think of $S^3\subset \C\oplus\C$, and take 
\bea\label{Reeb_S3}
\xi \,&=&\, b_1\partial_{\varphi_1}+b_2\partial_{\varphi_2}~,
\eea
where $\varphi_1$, $\varphi_2$ are standard $2\pi$ periodic azimuthal angles on each copy of $\C$. 
For $b_1=\pm b_2$ this is again the Hopf action on $S^3$, but for $b_1/b_2$ irrational 
the flow of $\xi$ is irregular, with generically non-closed orbits. In this case $\psi$ and $\arg z$ are not good global coordinates on the three-sphere.
It is straightforward to 
write down the general form of a compatible smooth metric in this case, of the form (\ref{3dmetric}) -- 
see \cite{Assel:2014paa}. From the perspective of complex geometry, these manifolds with $S^1\times S^3$ topology (and largely arbitrary Hermitian metric) are primary {\it Hopf surfaces}.

A large and interesting class of examples are given by links of weighted homogeneous hypersurface singularities. 
Here one begins with $\C^3$ with a weighted $\C^*$ action 
$(Z_1,Z_2,Z_3)\rightarrow (q^{w_1}Z_1, q^{w_2} Z_2, q^{w_3}Z_3)$, where 
$w_i\in\mathbb{N}$ are the weights, $i=1,2,3$, and $q\in\C^*$. The hypersurface 
is the zero set
\bea
X &=& \{f \ = \ 0\}\ \subset\ \C^3~,
\eea
where $f=f(Z_1,Z_2,Z_3)$ is a polynomial satisfying
\bea
f(q^{w_1}Z_1,q^{w_2}Z_2,q^{w_3}Z_3) &=& q^d f(Z_1,Z_2,Z_3)~,
\eea
where $d\in\mathbb{N}$ is the degree. For appropriate choices of 
$f$ the \emph{link}
\bea
M_3 &=& X\cap \{|Z_1|^2+|Z_2|^2+|Z_3|^2 \ = \ 1\}
\eea
is a smooth three-manifold. Moreover, the weighted $\C^*$ action 
induces a $U(1)$ isometry of the metric (induced from the flat metric 
on $\C^3$), and the associated Killing vector $\xi$ naturally 
defines a transversely holomorphic foliation of $M_3$. 
Here $\Sigma_2=M_3/U(1)$ is  the orbifold Riemann 
surface given by $\{f=0\}$ in the corresponding 
weighted projective space $\mathbb{WCP}^2_{[w_1,w_2,w_3]}$. 
This construction covers all spherical three-manifolds 
$S^3/\Gamma_{ADE}$, but also many three-manifolds with 
infinite fundamental group. One can further generalize this construction by 
considering links of complete intersections, {\it i.e.} realizing 
$X$ as the zero set of $m$ weighted homogeneous polynomials in $\C^{2+m}$.

\subsection{A global restriction}\label{sec:global}

If we take the product $X_0\equiv \R_{>0}\times M_3$, then 
we may pair the Reeb vector $\xi$ with a radial vector 
$r\partial_r$, where $r$ is the standard coordinate on $\R_{>0}$.
Notice this is particularly natural in four dimensions, 
where we may identify $\tau = \log r$, with $X_0=\R_{>0}\times M_3$ 
being a covering space for $M_4=S^1\times M_3$.   
Then $X_0$ is naturally a complex manifold, with 
the complex vector field $\xi-\ii r\partial_r$ being of Hodge type $(0,1)$. 
In fact $X_0$ may be equipped with either the $I_+$ 
or the $I_-$ complex structure, with the former more natural 
in the sense that $z$ is a local holomorphic coordinate with 
respect to $I_+$. In the following we hence take the $I_+$ complex structure.

The examples in section \ref{sec:FTexamples} all share a common feature: 
in these cases the complex surface $X_0$ admits a global holomorphic $(2,0)$-form. 
That is, its canonical bundle $\mathcal{K}$ is (holomorphically) trivial. 
This is obvious for $S^3$, where $X_0\cong \C^2\setminus\{0\}$, 
while for links of homogeneous hypersurface singularities $X$ 
we may identify $X_0=X\setminus\{o\}$, where 
the isolated singular point $o$ is at the origin $\{Z_1=Z_2=Z_3=0\}$ of $\C^3$.
In this case the holomorphic $(2,0)$-form is 
$\Psi=\diff Z_1\wedge \diff Z_2/ (\partial f/\partial Z_3)$ in a patch where 
$\partial f/\partial Z_3$ is nowhere zero. One can easily check that 
$\Psi$ patches together to give a smooth holomorphic volume form on $X_0$. 
Such singularities $X$ are called \emph{Gorenstein}.

As shown in \cite{Martelli:2015kuk}, the one-form $A$ in (\ref{A3d}) is (in our sign conventions) a connection on $\mathcal{K}^{1/2}$. It follows that when 
the canonical bundle of $X_0$ is trivial $A$ may be taken to be a 
\emph{global} one-form (this is true on $M_3$ or on $M_4\cong S^1\times M_3$). 
This global restriction on $A$ will play an important role in certain computations 
later. For example, the computation of the supersymmetric Casimir energy in 
 \cite{Martelli:2015kuk} requires this additional restriction on $M_4\cong S^1\times M_3$, 
and the same condition will also be needed in our evaluations of the renormalized gravitational actions 
 in four and five dimensions. That said, other computations will not require this restriction, 
 and we shall always make clear when we need the global restriction of this section, and when not.
 
As explained in \cite{Martelli:2015kuk}, when the canonical bundle of $X_0$ is trivial the constant $\cpsi$ in (\ref{A3d}), (\ref{A4d}) may be identified 
with $\frac{1}{2}$ the charge of the holomorphic $(2,0)$-form $\Psi$ under the Reeb vector $\xi$. 
Thus for example we have
\begin{align}
\label{eq:GammaValues}
\gamma &= \begin{cases}\ \frac{1}{2}(b_1+b_2)~, \quad & \quad \mbox{$S^3$ with Reeb vector $\xi=b_1\partial_{\varphi_1}+b_2\partial_{\varphi_2}$}\\[1mm] \ \frac{1}{2}b(-d + \sum_{i=1}^3 w_i)~,\quad & \quad \mbox{$M_3=$ link of weighted 
homogeneous} \\[-1mm]
\ & \qquad \quad  \ \ 
\mbox{hypersurface singularity, $\xi=b\zetacharge\ $.}\end{cases}
\end{align}
Here in the second example  the normalized generator of the $U(1)\subset \C^*$ action 
for the link has been denoted by $\zetacharge$, and $b$ is an 
arbitrary scale factor.
The local function $\lambda(z,\bar{z})$ in  (\ref{A3d}), (\ref{A4d}) is chosen 
so that $A$ is a global one-form on $M_3$. The form of this depends 
on the choice of transverse coordinate $z$, and then $\lambda$ is fixed uniquely 
up to a shift by a global function on $M_3$ that is invariant under $\xi$: 
this is just a small gauge transformation of $A$. Finally, 
on $M_4\cong S^1\times M_3$ the constant $\ct$ introduced in \eqref{A4d}
is fixed by requiring the Killing spinors $\zeta_\pm$ 
to be invariant under $\partial_\tau$. This is necessary 
in order that the Killing spinors survive the compactification 
of $\R\times M_3$ to $S^1\times M_3$. In fact as we show in appendix~\ref{app:Killingspinors} this sets $\ct=0$, 
but it will be convenient to keep this constant since the more general background with $S^1$ fibred over $M_3$ we will discuss in section~\ref{sec:twisted_bckgnd} will require $\ct\neq0$.

In order to compute the four- and five-dimensional on-shell supergravity actions later in the paper, we will 
also need some further expressions for the constant $\gamma$. 
Since we may always approximate an irregular Reeb vector field (with 
generically non-closed orbits) by a quasi-regular Reeb vector field (where 
all orbits close), there is no essential loss of generality in assuming that 
$\xi$ generates a $U(1)$ isometry of $M_3$. 
Equivalently, $M_3$ is the total space of a $U(1)$ principal orbibundle over an orbifold Riemann surface $\Sigma_2$ with metric \eqref{2dMetric} (which is smooth where $U(1)$ acts freely on $M_3$). Since the orbits of $\xi=\partial_\psi$ close, for a generic orbit we may write
 $\psi \sim \psi + 2\pi/b$, with $b\in \R_{>0}$ a constant. This allows us to write the following relation between the almost contact volume and characteristic class
\bea\label{c1L}
\frac{b^2}{(2\pi)^2}\int_{M_3}\eta\wedge\diff\eta &=& \int_{\Sigma_2}c_1\left(\ml\right)~,
\eea
where $c_1\left(\ml\right)\in H^2\left(\Sigma_2,\mathbb{Q}\right)$ is the first Chern class of $\ml$, the orbifold line bundle associated to $S^1\hookrightarrow M_3\rightarrow \Sigma_2$. If the $U(1)$ action generated by $\xi$ is free, then $\Sigma_2$ is a smooth Riemann 
surface and the right hand side of (\ref{c1L}) is an integer; more generally it is a rational number.
Analogously, by definition the first Chern class of $\Sigma_2$ is the first Chern class of its anti-canonical bundle, which integrates to
\bea
\int_{\Sigma_2}c_1(\Sigma_2) \ \equiv \ \int_{\Sigma_2}c_1\left(\mathcal{K}^{-1}_{\Sigma_2}\right) &=& \frac{1}{4\pi}\int_{\Sigma_2}R_{2d}\, \vol_{2}~.
\eea
Here $R_{2d}=-\square \wzero$ is the  scalar curvature of the metric~\eqref{2dMetric} on $\Sigma_2$, expressed in terms of the two-dimensional Laplace operator $\square \equiv \ex^{-\wzero}\partial^2_{z\bar{z}}$ (we are using the notation $\partial^2_{z\zbar}\equiv\partial_z\partial_{\zbar}$). Equivalenty we may write this as an integral over $M_3$:
\bea\label{R2dc1}
\int_{\Sigma_2}c_1(\Sigma_2) \,=\, \frac{b}{8\pi^2}\int_{M_3}R_{2d}\, \eta\wedge\vol_{2}~.
\eea

Given these preliminary formulas, we next
claim that the expression \eqref{A3d} for $A$ describes a globally defined one-form on $M_3$ if and only if $\gamma$ is given by 
\bea
\label{eq:4dGammaFromGauge}
\gamma \,&=&\,  -\frac{b}{2}\frac{\int_{\Sigma_2}c_1(\Sigma_2)}{\int_{\Sigma_2}c_1\left(\ml\right)} \, = \, -\frac{1}{4}\frac{\int_{M_3}R_{2d}\, \eta\wedge\vol_{2}}{\int_{M_3}\eta\wedge\diff\eta}~.
\eea
To see this, recall from our discussion above that 
 $2A$ is a connection on the canonical bundle $\mathcal{K}$ of $X_0$. The latter is (by assumption)  holomorphically trivial, with global holomorphic section a $(2,0)$-form $\Psi$. 
 It follows that $2\gamma$ may be identified with the charge of $\Psi$ under 
the Reeb vector $\xi=\partial_\psi$ \cite{Martelli:2015kuk}. On the other hand, $\Psi$ in turn may be constructed as
a section of  the canonical bundle $\mathcal{K}_{\Sigma_2}$ of $\Sigma_2$, tensored with a section of some power of $\mathcal{L}^*$, say $(\mathcal{L}^*)^p$, where $\mathcal{L}^*$ is the bundle dual to $\mathcal{L}$.  The former must be dual line bundles in order that $\Psi$ is globally defined as a form, meaning that
\bea
p\, c_1(\mathcal{L}^*) &=& - c_1(\mathcal{K}_{\Sigma_2}) \ = \ c_1(\Sigma_2)~.
\eea
Since $\exp(b\, \ii\psi)$ is a section of $\mathcal{L}$, which has charge $b$ under $\xi=\partial_\psi$, and $c_1(\mathcal{L}^*)= - c_1(\mathcal{L})$, this means 
that the charge of $\Psi$ is fixed to be 
\bea
2\gamma &=& b\, p \ = \ - b \, \frac{\int_{\Sigma_2}c_1(\Sigma_2)}{\int_{\Sigma_2}c_1(\mathcal{L})}~.
\eea
Rearranging gives (\ref{eq:4dGammaFromGauge}).
We stress again that although we have derived 
(\ref{eq:4dGammaFromGauge}) for quasi-regular 
Reeb vector fields, by continuity the expression 
for $\gamma$ given by the first equality 
holds also in the irregular case.

These Seifert invariants are readily computed for particular examples. 
For example, in section \ref{sec:FTexamples} we considered $M_3\cong S^3$ with 
Reeb vector $\xi=b_1\partial_{\varphi_1}+b_2\partial_{\varphi_2}$, where $\varphi_1$, $\varphi_2$ 
are standard $2\pi$ periodic coordinates. The foliation is quasi-regular when 
$b_1/b_2=p/q\in \mathbb{Q}$ is rational. Taking $p,q\in\mathbb{N}$ 
with no common factor, we have $\Sigma_2=S^3/U(1)_{p,q}\cong \mathbb{WCP}^1_{[p,q]}$. 
This weighted projective space is topologically a two-sphere, but with orbifold 
singularities with cone angles $2\pi/p$ and $2\pi/q$ at the north and south poles, respectively. 
Recalling that $\mathcal{L}$ is the line bundle associated to $S^1\hookrightarrow S^3\rightarrow \Sigma_2$, 
it is straightforward to compute that
\bea\label{c1S3}
\int_{\Sigma_2}c_1(\mathcal{L}) &=& - \frac{1}{pq}~, \qquad \int_{\Sigma_2}c_1(\Sigma_2) \ = \  \frac{p+q}{pq}~.
\eea
Similarly, for $M_3$ a link of a weighted homogeneous hypersurface singularity, described in section  \ref{sec:FTexamples}, 
one finds
\bea\label{c1whhs}
\int_{\Sigma_2}c_1(\mathcal{L}) &=& - \frac{d}{w_1w_2w_3}~, \qquad \int_{\Sigma_2}c_1(\Sigma_2) \ = \  \frac{d(-d+\sum_{i=1}^3 w_i)}{w_1w_2w_3}~.
\eea
These invariants are also often referred to as the \emph{virtual degree} and \emph{virtual Euler characteristic} of 
the weighted homogeneous hypersurface singularity, respectively. Notice 
that (\ref{c1S3}) may be derived from (\ref{c1whhs}) as a special case: 
we may take weights $(w_1,w_2,w_3)=(p,q,1)$, together with the polynomial 
$f(Z_1,Z_2,Z_3)=Z_3$, which has degree $d=1$. The zero set of 
$f$ is then $\C^2$, with coordinates $Z_1,Z_2$, with weighted Reeb vector $\xi=p\partial_{\varphi_1}+q\partial_{\varphi_2}$. 

Finally, it is worth pointing out there are interesting examples that 
are not covered by the restriction we make in this section. In particular 
setting the connection one-form $a=0$ gives a direct product $M_3\cong S^1\times \Sigma_2$, 
but unless $\Sigma_2\cong T^2$ the canonical bundle of $X_0$ 
is non-trivial (being the pull back of the canonical bundle of $\Sigma_2$). 
This rules out $M_3\cong S^1\times S^2$, where 
the Reeb vector rotates the $S^1$. In this case 
$A$ is a unit charge Dirac monopole on $S^2$. Localized gauge theory partition functions on such backgrounds have been computed in~\cite{Benini:2015noa,Benini:2016hjo,Closset:2016arn}.

\subsection{The partition function and supersymmetric Casimir energy}\label{sec:FTCasimir}

The general results of \cite{Closset:2013vra,Closset:2014uda} imply that 
the supersymmetric partition function 
of an $\mathcal{N}=2$ theory on $M_3$, 
or an $\mathcal{N}=1$ theory on $M_4 \cong S^1\times M_3$, 
depends on the choice of background only via the 
transversely holomorphic foliation of $M_3$. 
Concretely, this means that the partition function 
is invariant under  deformations 
$\wzero\rightarrow \wzero + 
\delta \wzero$, $\wone\rightarrow \wone+ 
\delta \wone$, where $\delta \wzero(z,\bar{z})$, $\delta \wone(z,\bar{z})$ 
are \emph{arbitrary} smooth global functions on $M_3$, invariant under 
$\xi=\partial_\psi$. Rigid supersymmetric backgrounds 
$M_4$ with a single supercharge $\zeta$ are in general Hermitian, 
and more generally the partition function is insensitive to Hermitian metric deformations and depends on the 
background only via the complex structure (up to 
local counterterms and anomalies)~\cite{Closset:2013vra}.  It is important to note that these statements are valid when the new minimal formulation of four-dimensional supergravity~\cite{Sohnius:1981tp} (or its three-dimensional analogue) is used to couple the field theory to the curved background.  
We will refer to these results as {\it supersymmetric Ward identities}. 

The Lagrangians for general vector and chiral multiplets 
on these backgrounds may be found in the original references 
cited above. In \cite{Closset:2013vra,Closset:2014uda} the strategy is 
to show that deformations of the background 
geometry that leave the transversely holomorphic foliation 
 (or more generally in four dimensions the  complex structure) fixed 
 are $\mathcal{Q}$-exact. A standard argument then shows that 
 the partition function is invariant under such deformations 
 (up to invariance of the measure). 

These general statements are supported by explicit computations 
of localized partition functions. In three dimensions 
the simplest case is $M_3\cong S^3$, 
with general Reeb vector~\eqref{Reeb_S3}. 
This was studied in \cite{Alday:2013lba}. The partition function of a general 
$\mathcal{N}=2$ gauge theory coupled to arbitrary matter
localizes to a matrix model for the scalar in the vector multipet, where 
this matrix model depends on the background geometry only via $b_1$, $b_2$.  
The large $N$ limit was computed for a broad class of Chern-Simons-matter theories in 
\cite{Martelli:2011fu} using
saddle point methods. The final result for the free energy 
$\mathscr{F}=-\log Z$ in the large $N$ limit is
\bea\label{FS3}
\mathscr{F} &=& \frac{(b_1+b_2)^2}{4b_1b_2}\cdot \frac{4\pi^2}{\kappa_4^2}~.
\eea
Here 
\bea
\mathscr{F}_{S^3_{\mathrm{round}}} &=& \frac{4\pi^2}{\kappa_4^2}
\eea
is the free energy
on the round $S^3$, which scales as $N^{3/2}$ \cite{Drukker:2010nc}, where 
$\kappa_4^2$ is the four-dimensional effective coupling constant of the gravity dual.
The partition function has also been computed 
on (round) Lens spaces $S^3/\Z_p$  in \cite{Benini:2011nc, Alday:2012au}. 
Here the partition function localizes onto flat 
gauge connections, and thus splits into a sum over 
topological sectors. However, 
in the large $N$ limit of the ABJM theory 
studied in \cite{Alday:2012au} it was shown that 
only certain flat connections contribute, 
all giving the same contribution as the trivial flat connection. 
The upshot is that the large $N$ free energy is simply 
$\frac{1}{p}$ times the free energy on $S^3$. 
As far as the authors are aware, there are no explicit 
results for the partition function, or its large $N$ limit, 
on more general links of homogeneous hypersurface singularities.
However, it is tempting to conjecture that for appropriate classes 
of theories with large $N$ gravity duals, the large $N$ free 
energy may be computed from the sector with trivial gauge connection.
The one-loop determinants here should be relatively straightforward to compute, 
in contrast to the full partition function which localizes onto solutions 
of the Bogomol'nyi equation, {\it i.e.} flat connections (on a closed three-manifold). 

The partition function for general $\mathcal{N}=1$ theories with an R-symmetry, defined on 
Hopf surfaces $M_4\cong S^1\times S^3$, was computed using localization in 
\cite{Assel:2014paa} (the chiral multiplet was also studied in~\cite{Closset:2013sxa}). With two supercharges of opposite 
R-charge one localizes onto flat gauge connections, which on 
$S^1\times S^3$ amount to a constant component of 
the dynamical gauge field along~$S^1$. The resulting matrix model 
is similar to that in three dimensions, albeit with additional 
modes along $S^1$, and indeed in \cite{Assel:2014paa} the results of 
\cite{Alday:2013lba} were used.  
Besides checking explicitly that the supersymmetric partition function depends on the transversely holomorphic foliation defined by the Reeb vector \eqref{Reeb_S3} on $M_3\cong S^3$ and not on the choice of Hermitian metric on the Hopf surface, the main result of~\cite{Assel:2014paa} was that the  partition function factorizes as
\bea\label{ZS1M3}
Z_{S^1_\beta\times S^3}\,&=&\, \ex^{-\beta \Esusy} \cdot\, \mathcal{I}~,
\eea
where $\mathcal{I}$ is the supersymmetric index originally defined in~\cite{Romelsberger:2005eg,Kinney:2005ej} and
\bea\label{EsusyS3}
\Esusy &=& \frac{2}{27}\frac{(b_1+b_2)^3}{b_1b_2}(3\mathtt{c}-2\mathtt{a}) + \frac{2}{3}(b_1+b_2)(\mathtt{a}-\mathtt{c})~
\eea
was dubbed the \emph{supersymmetric Casimir energy}. Here, $\mathtt{a}$ and $\mathtt{c}$ are the usual trace anomaly coefficients for 
a four-dimensional SCFT; more generally, for a supersymmetric theory with a choice of R-symmetry one should replace $\mathtt{a}$ and $\mathtt{c}$  in (\ref{EsusyS3}) by the corresponding 't Hooft anomaly formulae, involving traces over the R-charges of fermions. 
This result has been argued to be scheme-independent, provided one 
uses a supersymmetric regularization scheme, hence $E_{\rm susy}$ is an intrinsic observable~\cite{Assel:2014tba, Assel:2015nca}. One can see that $\Esusy$ corresponds to a Casimir energy by showing that it is the vacuum expectation value of the Hamiltonian generating translations along the Euclidean time, in the limit $\beta\to\infty$~\cite{Lorenzen:2014pna,Assel:2015nca}.
 
For field theories admitting a large $N$ gravity dual in type IIB 
supergravity, to leading order in the large $N$ limit one has 
$\mathtt{a}=\mathtt{c}=\pi^2/\kappa_5^2$, where $\kappa_5^2$ is the 
five-dimensional gravitational coupling constant and we have set the AdS radius to 1. Moreover, one can see that the index $\mathcal{I}$ does not contribute at leading order~\cite{Kinney:2005ej}. Then at large $N$ the field theory partition function reduces to
\be\label{Esusy_Hopf_gravity}
-\frac{1}{\beta}\log Z_{S^1_\beta\times S^3} \ = \ \Esusy \ =\ \frac{2(b_1+b_2)^3}{27b_1b_2} \frac{\pi^2}{\kappa_5^2}~.
\ee
The right hand side is expressed in terms of the five-dimensional gravitational coupling constant, 
and one of our aims will be to reproduce this formula from a dual supergravity computation.
For the locally conformally flat $S^1_\beta \times S^3_{r_3}$, where 
$M_3\cong S^3_{r_3}$ is equipped with the standard round metric of radius $r_3$, 
we have $b_1=b_2=1/r_3$, leading to 
\bea\label{Esusy_round}
-\frac{1}{\beta}\log Z_{S^1_\beta\times S^3_{r_3}} \ = \ {\Esusy}_{, \, S^1_\beta\times S^3_{r_3}} &= & \frac{16}{27r_3}\frac{\pi^2}{\kappa_5^2}~.
\eea

Following \cite{Assel:2015nca,Lorenzen:2014pna}, in \cite{Martelli:2015kuk} the supersymmetric Casimir 
energy was studied on the more general class of $M_4\cong S^1_\beta\times M_3$ 
backgrounds, by reducing to a supersymmetric quantum mechanics.\footnote{Other methods to extract the supersymmetric Casimir energy on Hopf surfaces use equivariant integration of anomaly polynomials \cite{Bobev:2015kza} or exploit properties of the supersymmetric index \cite{Ardehali:2015hya,Brunner:2016nyk}. See also~\cite{Nishioka:2014zpa} for localization on backgrounds with more general topologies.}
The short multiplets that contribute to $\Esusy$ were shown to be 
in 1-1 correspondence with holomorphic functions 
on $X_0\cong \R_{>0}\times M_3$, with their contribution being determined by the 
charge under the Reeb vector $\xi$. This makes it manifest 
that $\Esusy$ depends on the background only via 
the choice of transversely holomorphic foliation on $M_3$.
From this it follows 
that $\Esusy$ may be computed from an index-character 
that counts holomorphic functions on $X_0$ according to their Reeb charge. 
Again, more precisely this is true in the sector with trivial 
flat gauge connection, while more generally one should 
look at holomorphic sections of the corresponding flat 
holomorphic vector bundles. In any case, in the 
sector with trivial flat connection on $M_3$ 
one can use this result to show that
for links of homogeneous hypersurface singularities
\bea\label{Esusylink}
\Esusy &=& \frac{2b}{27}\frac{ d\,c_1^3}{w_1w_2w_3}(3\mathtt{c}-2\mathtt{a}) 
+ \frac{b}{3}\frac{d\,c_1}{w_1w_2w_3}(c_1^2-c_2)(\mathtt{a}-\mathtt{c})~.
\eea
Here we have defined 
\bea
c_1 &=& -d+\sum_{i=1}^3w_i~, \qquad c_2 \ = \ -d^2 + \sum_{i=1}^3 w_i^2~.
\eea
In particular, $c_1$ is precisely the charge of the holomorphic $(2,0)$-form 
under the generator $\zetacharge$  of the $U(1)$ action. Equivalently, this is 
the orbifold first Chern number of the orbifold anti-canonical bundle 
of the orbifold Riemann surface $\Sigma_2=M_3/U(1)$, which is an integer
version of the second invariant in
(\ref{c1whhs}). 
Again, for theories with a large $N$ gravity dual, in the large $N$ limit this becomes
\bea\label{EsusylargeNlink}
\Esusy &=&  \frac{2b}{27}\frac{ d\,c_1^3}{w_1w_2w_3} \frac{\pi^2}{\kappa_5^2}~.
\eea
Assuming that the dominant contribution comes from this 
sector with trivial flat connection, (\ref{EsusylargeNlink}) 
is hence the prediction for the gravity dual.

An aim of this paper will be to reproduce these field theory results holographically from supergravity.

\section{Four-dimensional supergravity}\label{sec:4dgravity}

In this section we are interested in the gravity duals to three-dimensional $\mathcal{N}=2$ field theories on the 
backgrounds $M_3$ described in section \ref{sec:3dbackgrounds}. The gravity solutions are constructed 
in $\mathcal{N}=2$ gauged supergravity in four dimensions. The general form 
of (real) Euclidean supersymmetric solutions to this theory was studied in \cite{Dunajski:2010uv}. In particular 
they admit a Killing vector, which for asymptotically locally Euclidean AdS solutions restricts 
on the conformal boundary $M_3$ to the Killing vector $\xi$ defined in  (\ref{xi3d}).
Indeed, we will see that the conformal boundary of a general supersymmetric supergravity solution is equipped with the same 
geometric structure described in section~\ref{sec:3dbackgrounds}.  We show that the renormalized on-shell supergravity action, regularized according to standard holographic renormalization, depends on the boundary geometric data only via the transversely holomorphic foliation, 
thus agreeing with the general field theory result summarized in section~\ref{sec:FTCasimir}. Moreover, 
for \emph{self-dual} supergravity solutions we show that the holographic free energy correctly reproduces the large $N$ field theory partition function (in the cases where 
this is available) described in  section 
\ref{sec:FTCasimir}. We thus find very general agreement between large $N$ localized field theory calculations, 
on general supersymmetric backgrounds $M_3$, and dual supergravity computations. 

\subsection{Supersymmetry equations}\label{sec:SUSYeqns}

The Euclidean action for the bosonic sector of four-dimensional $\mathcal{N}=2$ gauged supergravity \cite{Freedman:1976aw} is
\bea
\label{eq:4dSUGRAaction}
S_{\rm bulk} \,&=&\, - \frac{1}{2\kappa_4^2}\int \diff^4 x\sqrt{G} \left( R_G + 6 - \calF_{\mu\nu}\calF^{\mu\nu}\right)~.
\eea
Here $R_G$ is the Ricci scalar of the four-dimensional metric $G_{\mu\nu}$, $\calF=\diff\calA$ is the field strength of the Abelian graviphoton $\calA$, and the cosmological constant has been normalized to $\Lambda = -3$.\footnote{Our curvature conventions are summarized in appendix~\ref{app_curvature}.} The equations of motion are
\bea
R_{\mu\nu} + 3G_{\mu\nu} &= & 2 \left(\calF_{\mu}{}^{\rho}\calF_{\nu\rho} - \frac{1}{4}\calF_{\rho\sigma}\calF^{\rho\sigma} G_{\mu\nu}\right)~,\nn \\ \label{eq:4dSUGRA}
\diff *_4 \calF &=& 0~.
\eea
A supergravity solution is supersymmetric if it admits a non-trivial Dirac spinor $\epsilon$ satisfying the Killing spinor equation
\be
\label{eq:4dKillingSpinor}
\left( \nabla_{\mu} + \frac{\ii}{4}\calF_{\nu\rho}\Gamma^{\nu\rho}\Gamma_{\mu} + \frac{1}{2} \Gamma_{\mu} + \ii \calA_{\mu} \right) \epsilon = 0~,
\ee
where $\Gamma_{\mu}$ generate Cliff$(4)$ in an orthonormal frame, so $\{\Gamma_{\mu},\Gamma_{\nu}\} = 2G_{\mu\nu}$. Locally, any such solution can be uplifted to a supersymmetric solution of eleven-dimensional supergravity in a number of ways, as explained in \cite{Gauntlett:2007ma}. Strictly speaking the latter reference discusses the Lorentzian signature case, while the 
corresponding Euclidean signature result was studied in~\cite{Farquet:2014bda}. We also note that 
there may be global issues in uplifting some solutions, as discussed in detail in~\cite{Martelli:2012sz}. However, 
these considerations will not affect any of the statements and results in the present paper.

Following the analysis of~\cite{Dunajski:2010uv}, real Euclidean supersymmetric solutions to this theory admit a canonically defined local coordinate system in which the metric takes the form
\bea
\label{eq:Generic4dMetric}
\diff s^2_4 \,& = &\, \frac{1}{y^2 UV}(\diff\psi + \phi)^2 + \frac{UV}{y^2}(\diff y^2 + 4\ex^W\diff z\diff \zbar)~.
\eea
Here $\xi=\partial_\psi$ is a Killing vector,  arising canonically as a bilinear from supersymmetry, and 
 $W = W(y,z,\zbar)$, $U = U(y,z,\zbar)$, $V = V(y,z,\zbar)$, while $\phi$ is a local one-form  satisfying $\xi\hook \phi = 0$ and $\mathcal{L}_\xi \phi=0$. In addition, the following equations 
 should be imposed:
\begin{align}
\label{eq:GenericU}
U & \, = \, 1-\frac{y}{4}\partial_yW + \frac{f}{2}\ , \\
\label{eq:GenericDzbarzw}
\partial^2_{z\zbar}W &+ \ex^W\left[\partial_{yy}^2W+\frac{1}{4}(\partial_yW)^2 + 3y^{-2}f^2\right] \, = \, 0\ , \\
\begin{split}
\label{eq:GenericDzbarzf}
\partial^2_{z\zbar}f &+ \frac{\ex^W}{y^2}\bigg[f\left(f^2+2\right) - y\left(2\partial_yf+\frac{3}{2}f\partial_y W\right)\ + \\
& + y^2\left(\partial_{yy}^2f + \frac{3}{2}\partial_yW\partial_yf + \frac{3}{2}f\partial_{yy}^2W + \frac{3}{4}f(\partial_yW)^2\right) \bigg]\, = \, 0 \ ,
\end{split}\\
\begin{split}
\label{eq:Genericdphi}
\diff \phi & \, = \, \ii \, UV \bigg[ \partial_z\log\frac{V}{U} \ \diff y\wedge \diff z - \partial_{\zbar}\log\frac{V}{U} \ \diff y\wedge \diff \zbar \\
&\ \ \ + 2\,\ex^W\left(\partial_y\log\frac{V}{U} + \frac{2}{y}(U-V)\right)\diff z\wedge\diff \zbar\bigg]~,
\end{split}
\end{align}
where we have introduced $f\equiv U-V$. The first equation \eqref{eq:GenericU} defines $U$ in terms of $W$ and $f$, and we could therefore use it to substitute in \eqref{eq:Genericdphi} and conclude that the entire geometry is fixed by a choice of $W$ and $f$ (apart from a possible gauge transformation/diffeomorphism on $\phi$). In deriving this form of the solutions, \eqref{eq:GenericU},  \eqref{eq:GenericDzbarzw} and \eqref{eq:Genericdphi} follow from imposing the Killing spinor equation (\ref{eq:4dKillingSpinor}),  while \eqref{eq:GenericDzbarzf} is required for the equation of motion for $\calF$ (the Maxwell equation) to be satisfied. 

The graviphoton is determined by the above geometry, and is  given by
\bea
\calA & = & \frac{1}{2y}\frac{f}{U(U-f)}(\diff\psi + \phi) + \frac{\ii}{4}(\partial_{\zbar}W\diff\zbar - \partial_z W \diff z)~.
\eea
In general 
this expression is only valid locally, and we will see later that we need to perform a local gauge transformation 
in order that $\calA$ is  regular.

A rich subclass of solutions are the \emph{self-dual} solutions, studied in \cite{Dunajski:2010zp, Farquet:2014kma}. 
Here one imposes  $\calF$ to be anti-self-dual, which together with supersymmetry implies 
that the metric has anti-self-dual Weyl tensor \cite{Dunajski:2010zp}. We adopt the same abuse of terminology as 
\cite{Farquet:2014kma}, and refer to these as ``self-dual'' solutions. 
This amounts to setting
\bea
\label{eq:SelfDualCondition}
f &= &\frac{y}{2}\partial_yW \qquad \mbox{(self-dual case)}.
\eea
This in turn fixes $U\equiv 1$, and therefore self-dual solutions to $\mathcal{N}=2$ gauged supergravity in four dimensions are completely specified by a single function $W=W(y,z,\bar{z})$, which solves  \eqref{eq:GenericDzbarzw}. This turns out to be the $SU(\infty)$ Toda equation.\footnote{Of course for self-dual solutions the Maxwell equation is automatic, and indeed one can check that, with (\ref{eq:SelfDualCondition}) imposed, equation \eqref{eq:GenericDzbarzf} is implied by the other equations.}

\subsection{Conformal boundary}\label{sec:conformalboundary}

In order to apply the gauge/gravity correspondence we  require the solutions described in the previous subsection
to be 
AlEAdS (also known as \emph{asymptotically locally hyperbolic}). This is naturally imposed, with the coordinate $1/y$ playing the 
role of the radial coordinate. Indeed, there is then a conformal boundary at $y=0$, and the 
metric has the leading asymptotic form $\frac{\diff y^2}{y^2} + \frac{1}{y^2}\diff s^2_{M_3}$. 
More precisely, this all follows if we assume that 
$W(y,z,\zbar)$, $f(y,z,\zbar)$ are analytic functions in $y$ around $y=0$:\footnote{Note that this is not true in general. For more details see section 3 of \cite{Farquet:2014kma}.}
\begin{align}
\label{eq:4dExpansionwf}
W(y,z,\zbar)\, &=\, w_{(0)}(z,\zbar) + y w_{(1)}(z,\zbar) + \frac{y^2}{2}w_{(2)}(z,\zbar) + \cO(y^3)~, \nn\\
f (y,z,\zbar) \,&= \,f_{(0)}(z,\zbar) + y f_{(1)}(z,\zbar) + \frac{y^2}{2}f_{(2)}(z,\zbar) + \frac{y^3}{6}f_{(3)}(z,\zbar) + \cO(y^4)~,
\end{align}
and the one-form $\phi$ can be expanded as
\bea
\label{eq:4dExpansiona}
\phi(y,z,\zbar) &= & a_{(0)}(z,\zbar) + y a_{(1)}(z,\zbar) + \frac{y^2}{2}a_{(2)}(z,\zbar) + \cO(y^4)~.
\eea
This implies that to leading order
\bea
\diff s^2_4 \,&=&\, [1+\cO(y)]\frac{\diff y^2}{y^2} + y^{-2}[(\diff\psi + a_{(0)})^2 + 4\ex^{w_{(0)}}\diff z \diff\zbar + \cO(y)]~,
\eea
confirming that the metric is indeed AlEAdS around the boundary $\{y =0\}$. A natural choice of metric (rather than 
conformal class of metrics) on the boundary $M_3$ is therefore
\bea
\label{eq:3dExpansionMetric}
\diff s^2_3 &= &(\diff\psi + a_{(0)})^2 + 4\ex^{w_{(0)}}\diff z \diff\zbar.
\eea
The boundary one-form $\eta \equiv \diff\psi + a_{(0)}$ has exterior derivative
\be
\label{eq:3dExpansionDa}
\diff\eta \,=\, 2\ii\, \ex^{w_{(0)}}f_{(1)} \ \diff z\wedge\diff\zbar, 
\ee
as can be seen by expanding \eqref{eq:Genericdphi} to leading order and using $f_{(0)}=0$, the latter coming from the leading order term in \eqref{eq:GenericDzbarzw}. More specifically, $\eta$ is a global almost-contact one-form and $\xi$ is its Reeb vector field, as
\be
\xi \hook \eta \ =\  1, \qquad \xi \hook \diff\eta \ =\  0~.
\ee
On the conformal boundary $\xi$ is nowhere vanishing, which implies that it foliates $M_3$. This Reeb foliation is transversely holomorphic, with locally defined complex coordinate $z$.
The leading term of the expansion of the bulk Abelian graviphoton is
\bea
\label{eq:3dExpansionGaugeField}
A_{(0)} &\equiv & \calA\mid_{\{y=0\}} \ = \  \frac{f_{(1)}}{2}\left( \diff\psi + a_{(0)} \right) + \frac{\ii}{4}\left( \partial_{\zbar}w_{(0)} \diff\zbar - \partial_z w_{(0)} \diff z\right),
\eea
where as usual this expression is only valid locally, and we are free to perform (local)  gauge transformations.

Of course, we see immediately that we recover the rigid supersymmetric geometry of $M_3$ described in section \ref{sec:3dbackgrounds}. 
More precisely, comparing (\ref{eq:3dExpansionMetric}) and (\ref{3dmetric}) we identify $a_{(0)}=a$, 
$w_{(0)}=w$, with the choice of conformal factor $\Omega = 1$ so that the Killing vector $\xi$ has length 1 (as usual in AdS/CFT, the conformal factor $\Omega$ on the boundary appears as a Weyl rescaling of the radial coordinate $y \rightarrow \Omega^{-1}y$). 
Moreover,  comparing \eqref{eq:3dExpansionDa} and \eqref{eq:3dBoundaryDa} 
we see that
\bea
f_{(1)} &=& \frac{1}{2}\wone~.
\eea
Finally, the background R-symmetry gauge field arises as the restriction to the conformal boundary of the bulk Abelian graviphoton, as shown by comparing  \eqref{eq:3dExpansionGaugeField} and \eqref{A3d}. Thus we identify $A_{(0)}=A$ (up to local gauge transformations). 

By expanding \eqref{eq:GenericDzbarzw}, \eqref{eq:GenericDzbarzf} and \eqref{eq:Genericdphi} to higher order we obtain the relations
\bea
\label{eq:Ricci2dw}
w_{(2)} & = &  -\ex^{-w_{(0)}}\partial^2_{z\zbar}w_{(0)} - 3f_{(1)}^2 - \frac{1}{4}w_{(1)}^2~, \\
f_{(3)} &= &  -3\ex^{-w_{(0)}}\partial^2_{z\zbar}f_{(1)} - \frac{9}{4}f_{(1)}\left(w_{(1)}^2+2w_{(2)}\right) - 3f_{(1)}^3 -\frac{9}{4}f_{(2)}w_{(1)}~, \\[2mm]
\phi_{(2)} & = & \ii \left( \partial_{\zbar}f_{(1)}\diff \zbar - \partial_z f_{(1)}\diff z\right)~.
\eea
This (and expansions to higher orders) allows us to see an interesting difference between the self-dual and non-self-dual case. In general a representative of the boundary conformal class is fixed by the choice of two basic functions $w_{(0)}=w$ and $f_{(1)}=u/2$. 
However, in the general case there are in addition two free functions in the expansion into the bulk, namely $w_{(1)}$ and $f_{(2)}$, 
that appear in the Taylor expansions of $W$ and $f$ in the inverse radial coordinate $y$. 
In general these functions are  not determined by the conformal boundary data, but only by 
regularity of the solution in the deep interior of the bulk solution.  However, 
given $w_{(0)}$, $w_{(1)}$, $f_{(1)}$ and $f_{(2)}$, the series solutions of $W$ and $f$ 
are then uniquely fixed by the supersymmetry equations/equations of motion.
On the other hand, in the self-dual case, 
 instead $f$ and $W$ are related by \eqref{eq:SelfDualCondition}, so that the coefficients of the power series expansion $f_{(n)}$ and $w_{(n)}$ are related by 
\bea
\label{eq:4dSDEcondition}
f_{(n)} & = & \frac{n}{2}\, w_{(n)} \qquad \mbox{(self-dual case)}~.
\eea
Thus the gravitational filling of a given conformal boundary has a unique power series solution with self-dual metric, while there is no such 
uniqueness in the general case (as one would expect).

\subsection{Holographic renormalization}

The Euclidean supergravity action \eqref{eq:4dSUGRAaction}, with the Gibbons-Hawking-York term added to obtain the equations of motion \eqref{eq:4dSUGRA} on a manifold with boundary, diverges for AlEAdS solutions. However, we can use (the by now standard) holographic renormalization to remove these divergences.

In order to obtain a finite value for the on-shell action we need to consider a cut-off space $M_{\epsilon}$, where the coordinate $y>0$ extends to $y=\epsilon$, and add to the regularized action the appropriate local counterterms on the hypersurface $\partial M_{\epsilon} = \{ y=\epsilon\}$.
One then sends  $\epsilon\rightarrow 0$. Explicitly, we write the bulk action~\eqref{eq:4dSUGRAaction} as
\bea
S_{\rm bulk} &= & S_{\rm grav} + S_{\rm gauge}~,
\eea
where
\bea
S_{\rm grav} \, =\,  - \frac{1}{2\kappa_4^2}\int_{M_{\epsilon}}\!\diff^4x\, \sqrt{G}\left( R_G + 6\right)\,, \qquad S_{\rm gauge}  = \, \frac{1}{2\kappa_4^2}\int_{M_{\epsilon}}\!\diff^4x\, \sqrt{G}\, \calF_{\mu\nu}\calF^{\mu\nu}\,. \ \ \  \ \ 
\eea
As we are considering a manifold with boundary we must  add the Gibbons-Hawking-York term to make the Dirichlet variational problem for the metric well-defined,
\bea\label{eq:4dGHterm}
S_{\rm GH} \,&= &\, -\frac{1}{\kappa_4^2}\int_{\partial M_{\epsilon}}\diff^3x \, \sqrt{h}\, K~.
\eea
Here $h$ is the induced metric on $\partial M_{\epsilon}$, and $K$ is the trace of the second fundamental form of $\partial M_{\epsilon}$ with the induced metric. Finally, we add the counterterms
\bea
\label{eq:4dGravCounterterm}
S_{\rm ct} \,& =&\, \frac{1}{\kappa_4^2}\int_{\partial M_{\epsilon}}\diff^3x \, \sqrt{h}\, \Big(2+\frac{1}{2}R\Big),
\eea
where here $R$ is the  scalar curvature of $h$. These counterterms cancel the power-law divergences in the action. Note the absence of logarithmic terms, which are known to be related to the holographic Weyl anomaly, as the boundary is three-dimensional and therefore there is no conformal anomaly.
The on-shell action is the limit of the sum of the four terms above
\bea
\label{eq:4donshellS}
S \,& =&\,  \lim_{\epsilon\to 0}\left( S_{\rm bulk} + S_{\rm GH} + S_{\rm ct} \right)~.
\eea

The holographic energy-momentum tensor is defined as the quasi-local energy-momentum tensor of the gravity solution; that is, the variation of the on-shell gravitational action with respect to the \textit{boundary metric} $g_{ij}$, $i,j =1,2,3$, on $M_3$:
\be
\label{eq:EqnInconsistent}
 T_{ij}  \,=\, -\frac{2}{\sqrt{g}}\frac{\delta S}{\delta g^{ij}}~.
\ee
The holographic energy-momentum tensor can be expressed as a limit of a tensor defined on any surface of constant $y=\epsilon$. In 
our case this is
\be
\label{eq:3dHolographicTensor}
 T_{ij}  \ = \ \frac{1}{\kappa_4^2}\,\lim_{\epsilon\to 0}\,\frac{1}{\epsilon}\left(K_{ij} - K\, h_{ij} + 2h_{ij} - R_{ij} + \frac{1}{2}R\, h_{ij}\right)~,
\ee
where the tensors in the bracket are computed on $\partial M_{\epsilon}$ using $h_{ij}$, the induced metric. One can define a holographic $U(1)_R$ current in a similar way as
\be
 j^i  \ = \ \frac{1}{\sqrt{g}} \frac{\delta S}{\delta A_i}~,
\ee
where $A=A_{(0)}$ is the boundary R-symmetry gauge field.
In three boundary dimensions, this current can be extracted from the expansion of the bulk Abelian graviphoton as
\be
\label{eq:3dHolographicRcurrent}
\calA \ = \ A_{(0)} - \frac{1}{2}\kappa_4^2\,  j  \, y + \mathcal{O}\left(y^2\right).
\ee
The holographic energy-momentum tensor and R-current are identified with the expectation values of the respective field theory operators in the state dual to the supergravity solution under study.

From these definitions, a variation of the renormalized on-shell action can be expressed  as
\bea
\label{eq:4dActionVariation}
\delta S \,& = &\, \int_{M_3}\diff^3x\, \sqrt{g}\left( -\frac{1}{2} T_{ij} \delta g^{ij} +  j^i  \delta A_{(0)i} \right)~.
\eea
This formula can be used to check several holographic Ward identities. Invariance of the action under a boundary gauge transformation gives the conservation equation of the holographic R-current
\be
\label{eq:3dWardDivj}
\nabla_i j^i \  = \ 0~.
\ee
Invariance under boundary diffeomorphisms generated by arbitrary vectors on $M_3$ leads to the conservation equation for the holographic energy-momentum tensor,\footnote{This is easily seen by recalling that if $v^i$ is the boundary vector generating the diffeomorphism, then $\delta g^{ij} = -2 \nabla^{(i}v^{j)}$ and $\delta A_i = v^j \nabla_j A_i + \nabla_i v^j A_j$.}
\be
\label{eq:3dWardConservationT}
\nabla^i T_{ij}  \ = \ F_{(0)ji} j^i ~,
\ee
where $F_{(0)}=\diff A_{(0)}$. Performing a Weyl transformation at the boundary $\delta g_{ij} = 2g_{ij}\delta\sigma$, $\delta A_{(0)}=0$, for infinitesimal parameter function $\sigma$, we obtain for the trace of the holographic energy-momentum tensor,
\be
\label{eq:3dWardTraceT}
 T_i{}^i  \ = \ 0~,
\ee
consistently with the fact that there is no conformal anomaly in three-dimensional SCFTs.

As reviewed in section \ref{sec:fieldtheory}, the field theory supersymmetric Ward identities of \cite{Closset:2013vra,Closset:2014uda} imply that the supersymmetric partition function of $\mathcal{N}=2$ theories on $M_3$ depends on the background only via the transversely holomorphic foliation of $M_3$. AdS/CFT thus implies that the holographically renormalized on-shell supergravity action evaluated on a solution with boundary $M_3$  should also depend on the geometric data of $M_3$ only through its transversely holomorphic foliation. Concretely, this means that the on-shell action should be invariant under arbitrary deformations $w_{(0)} \rightarrow w_{(0)} + \delta w_{(0)}, a_{(0)} \rightarrow a_{(0)} + \delta a_{(0)}$, where $\delta w_{(0)}(z,\zbar)$ is an arbitrary smooth basic global function on $M_3$, and $\delta a_{(0)}(z,\zbar)$ is an arbitrary smooth basic global one-form on $M_3$. Recall that the Reeb foliation induces a basic cohomology on $M_3$: a $p$-form $\alpha$ on $M_3$ is called \textit{basic} if
$\xi \hook \alpha  =  0$, $\mathcal{L}_{\xi}\alpha  = 0$,
and the set of basic forms $\Omega_B^{\bullet}$ together with the exterior derivative $\diff_B = \diff|_{\Omega^{\bullet}_B}$ constitute the basic de Rham complex.

We may now check this directly by evaluating \eqref{eq:4dActionVariation} for the general class of supersymmetric solutions described in sections \ref{sec:SUSYeqns}, \ref{sec:conformalboundary}.
The holographic R-current is obtained from the subleading term in the expansion \eqref{eq:3dHolographicRcurrent}, and 
a computation reveals that this
 is
given by
\bea
\label{eq:3djExplicit}
 j  \,&= & \,-\frac{1}{2\kappa_4^2}\left[ \left(f_{(2)} + f_{(1)}w_{(1)}\right)\eta + \diff_B^c w_{(1)} \right]~.
\eea
We find that the holographic energy-momentum tensor \eqref{eq:3dHolographicTensor} evaluates to
\begin{align}
\label{eq:3dTExplicit}
4\kappa_4^2 \, T \, & =\, \left[ 2 f_{(1)}\left(f_{(2)} + f_{(1)}w_{(1)}\right) + \square w_{(1)} \right] \eta^2  \nn \\[1mm]
\,& \,\quad - 2\left( w_{(1)}\diff_B^c f_{(1)} + \diff_B^cf_{(2)}\right)\odot \eta - \partial_B w_{(0)}\odot \partial_B w_{(1)} - \overline{\partial}_B w_{(0)}\odot \overline{\partial}_B w_{(1)}  \nn \\[1mm]
\,&\, \quad - 2\ex^{w_{(0)}}\left[2 f_{(1)}\left(f_{(2)} + f_{(1)}w_{(1)}\right) + \square w_{(1)}\right]\diff z \diff \zbar~,
\end{align}
where $\odot$ denotes the symmetrized tensor product with weight 1/2.
In writing these expressions we have used the almost contact form on $M_3$, $\eta$, the differential operators of the basic cohomology, $\diff_B = \partial_B+\overline{\partial}_B, \diff_B^c = \ii \left( \overline{\partial}_B - \partial_B \right)$, and the transverse Laplacian $\square = \ex^{-w_{(0)}}\partial^2_{z\zbar}\,$.

We next plug these expressions for the holographic energy-momentum tensor and R-current in~\eqref{eq:4dActionVariation}. We assume that the boundary $M_3$ is compact, which allows us to use Stokes' theorem to simplify  expressions.
Moreover the resulting integrand can be simplified by recalling that all functions are basic, as is the deformation $\delta a_{(0)}$. We find that the general variation of the on-shell action is
\be
\delta S \ = \ \frac{\ii}{2\kappa_4^2}\int_{M_3} \eta \wedge \diff_B\left[\left(f_{(2)}+w_{(1)}f_{(1)}\right) \delta a_{(0)} + \frac{1}{2} *_2 \left(\delta w_{(0)}\, \diff_B w_{(1)}\right) \right]~.
\ee
Notice this {\it a priori} depends on the non-boundary functions $w_{(1)}$, $f_{(2)}$, which (with the exception of self-dual 
solutions) are not determined by the boundary data, but only via regularity of the supergravity solution in the deep interior.

However, this expression vanishes because of an analogue of Stokes' theorem, valid for almost contact structures (for instance, it can be found as Lemma 9.1 of \cite{Futaki:2006cc}). Let $X$ be a $(2m+1)$-dimensional manifold with almost contact one-form $\eta$: if $\alpha$ is a basic $(2m-1)$-form, then
\be
\label{eq:AlmostContactStokes}
\int_X\eta\wedge\diff_B\alpha \ = \ 0~.
\ee

The vanishing of the variation of the action $\delta S=0$ under arbitrary deformations of the background that leave the transversely holomorphic foliation fixed is a very general check of the AdS/CFT relation \eqref{AdSCFTZ}: it shows that both sides 
depend on the same data, which {\it a priori} is far from obvious. Anticipating the (contrasting) results in AdS$_5$/CFT$_4$ 
we shall obtain later in the paper, we might also stress that this means that standard holographic renormalization 
agrees with the supersymmetric renormalization scheme used in the boundary three-dimensional field theory to obtain the results of \cite{Closset:2013vra}.

In the next section we go further, and show that for a suitable class of solutions the holographically renormalized 
action reproduces the known field theory results, the latter obtained by supersymmetric localization methods.

\subsection{Evaluation of the on-shell action}
\label{subsec:Action4d}

In this section we evaluate the  regularized on-shell action \eqref{eq:4donshellS} for  a class of \emph{self-dual} 
supersymmetric AlEAdS solutions. The supergravity equations are simpler in the 
self-dual case,  and moreover the geometry is better understood; there are also more known examples 
\cite{Farquet:2014kma}. However, explicit families of non-self-dual supersymmetric solutions 
are known \cite{Martelli:2012sz}, and it would be interesting to generalize the computations in this section to cover the general case.

 As already mentioned the self-dual condition fixes $U\equiv 1$, so that the metric locally takes the form
\bea
\label{eq:4dSUGRAsolSD}
\diff s^2 & = & \frac{1}{y^2V}\left(\diff \psi + \phi \right)^2 + \frac{V}{y^2} \left( \diff y^2 + 4\ex^W\diff z\diff\zbar\right)~.
\eea
The graviphoton is
\bea\label{calASD}
\calA &=& \frac{1}{2y}\frac{1-V}{V}\left(\diff\psi + \phi\right) + \frac{\ii}{4}\left( \partial_{\zbar}W \diff\zbar - \partial_z W\diff z\right) + \gamma\,\diff \psi + \diff\lambda~,
\eea
where $\lambda=\lambda(y,z,\bar{z})$ is a local basic function. Moreover, the following equations should be imposed
\bea
V &=& 1-\frac{1}{2}y\partial_y W~, \nn\\[1mm]
\diff\phi &=& \ii\,\partial_zV \diff y\wedge\diff z - \ii\, \partial_{\zbar}V \diff y\wedge\diff \zbar + 2\ii \,\partial_y\left(V\ex^W\right) \diff z\wedge\diff \zbar~,\nn \\[1mm]
\label{eq:4dSUGRASDToda}
0 &=& \partial^2_{z\zbar}W + \partial_y^2\ex^W~.
\eea
Here the first equation may be used to eliminate $V$ in terms of $W=W(y,z,\bar{z})$, the second equation simply fixes $\diff\phi$, 
while the final equation is the $SU(\infty)$ Toda equation.
We begin by following part of the global analysis in \cite{Farquet:2014kma} -- the latter reference focused 
on solutions with $U(1)^2$ isometry and $M_4$ diffeomorphic to a ball, with conformal boundary $M_3\cong S^3$, 
but in fact a number of key arguments go through more generally.

First we recall that the coordinate $y$ may be  more invariantly defined as 
\bea
y^2 &=& \frac{2}{\|\PsiTwist\|^2}~, \qquad \mbox{where} \qquad \PsiTwist \ \equiv \ \frac{1}{2}\left(\diff \xi^\flat + *_4\diff\xi^\flat\right)_+~.
\eea
Here the self-dual two-form $\PsiTwist$ is called a \emph{twistor}, and is constructed from the Killing one-form $\xi^\flat=(1/y^2V)(\diff\psi+\phi)$ dual
to the Killing vector $\xi=\partial_\psi$. The conformal boundary is at $y=0$. Assuming the metric is regular in the interior, 
the twistor form is then also regular, and thus $y$ is non-zero in the interior. There can potentially be points at
which $\|\PsiTwist\|=0$, where $y$ then diverges, and indeed there are smooth solutions for which this happens. However, this can \emph{only} happen at fixed points of the Killing vector $\xi$ 
-- see the discussion in section 3.4 of \cite{Farquet:2014kma}. It follows that 
$y$ is a globally well-defined non-zero function on the interior of $M_4\setminus\{\xi=0\}$.  These self-dual solutions are also (locally)
\emph{conformally K\"ahler}, with K\"ahler two-form
\bea
\omega &=& -y^3\PsiTwist\ \ = \ \diff y \wedge (\diff\psi+\phi) +  V \ex^W\, 2\ii \, \diff z\wedge\diff\bar{z}~.
\eea
It follows from the first equality that $\omega$ is also well-defined on the interior of $M_4\setminus\{\xi=0\}$. 
Since $\diff y=-\xi\lrcorner \omega$, we see that $y$ is also a Hamiltonian function for $\xi$, 
and in particular is a Morse-Bott function. This implies that 
$y$ has no critical points on $M_4\setminus\{\xi=0\}$. We may hence extend the $y$ coordinate from 
the conformal boundary $y=0$ up to some $y=y_0>0$ in the interior, 
where on the locus $y=y_0$ the Killing vector $\xi$ has a fixed point 
(this may include $y_0=\infty$). Moreover, 
 the preimage of $(0,y_0)$ in $M_4$
 is topologically simply a product, $(0,y_0)\times M_3$, 
where the Killing vector is tangent to $M_3$ and has 
no fixed points.

With these global properties in hand, we can now proceed to compute the regularized on-shell action. 
We deal with each term in turn. Consider first the gravitational part of the action. Using the equation of motion we may write $R_G=-12$, so that on-shell
\bea
\label{eq:4dGravBulk1}
S_{\gr} \,&= &\, \frac{3}{\kappa_4^2}\int_{M_{\epsilon}}\vol_4~,
\eea
where the Riemannian volume form is
\bea
\vol_4 &=& \frac{1}{y^4}\diff y \wedge (\diff\psi + \phi)\wedge V\ex ^W2\ii \, \diff z\wedge \diff\zbar.
\eea
We can write this as an exact form
\bea
-3\vol_4 &=& \diff\Upsilon,
\eea
with
\bea\label{Gamma}
\Upsilon &=& \frac{1}{2y^2}(\diff\psi + \phi)\wedge\diff\phi + \frac{1}{y^3}(\diff\psi + \phi)\wedge V \ex^W2\ii\, \diff z \wedge \diff \zbar~.
\eea
The global arguments above imply that 
 $\Upsilon$ is well-defined everywhere on $M_4\setminus\{\xi=0\}$: in the first term $y$ is a global regular function and $\xi$ does not vanish, guaranteeing that $\diff\psi + \phi$ is a global one-form. The second term is simply $1/y^3(\diff\psi+\phi)\wedge\omega$, which is also 
globally well-defined and regular on $M_4\setminus\{\xi=0\}$. 
Having written the volume form as a globally exact form on $M_4\setminus\{\xi=0\}$, 
 we can then use Stokes' theorem to write \eqref{eq:4dGravBulk1} in terms of integrals over the 
conformal boundary $M_3\cong \{y=\epsilon\}$, and over the boundary  $\mathcal{T}$ of a small tubular neighbourhood
around the fixed point set of $\xi$. 
Using the expansion of the Toda equation \eqref{eq:4dSUGRASDToda} and \eqref{eq:AlmostContactStokes} near the conformal boundary, we can simplify the resulting expression to 
\bea\label{gravy}
S_{\rm grav} \ = \   \frac{1}{\kappa_4^2} \frac{1}{\epsilon^3}\int_{M_3}\eta\wedge\vol_{2}  + \frac{3}{4\kappa_4^2} \frac{1}{\epsilon^2}\int_{M_3}w_{(1)}\, \eta\wedge\vol_{2}- \frac{ 1 }{\kappa_4^2}\int_{\mathcal{T}}\Upsilon~. \ \ \ 
\eea
Here  $\vol_{2}$ is the two-dimensional volume form \eqref{2dMetric} (with $w_{(0)}=w$).
In general the fixed point set of $\xi$ may have a number of connected components, consisting 
either of fixed points (NUTs) or fixed two-dimensional surfaces (bolts). More precisely the 
last term in (\ref{gravy}) is then a sum over connected components, and 
the integral should be understood as a limit $\lim_{\delta\rightarrow 0}\int_{\mathcal{T}_\delta}$, 
where $\mathcal{T}_\delta$ is the boundary of a tubular neighbourhood, of radius $\delta$, around the fixed point set.

The first two divergent terms in (\ref{gravy}) are
cancelled by the Gibbons-Hawking-York term~\eqref{eq:4dGHterm} and the local counterterms \eqref{eq:4dGravCounterterm}, which in a neighbourhood of infinity become
\bea
S_{\rm GH} + S_{\rm ct} &=& -\frac{1}{32\kappa_4^2}\int_{M_3}\left( w_{(1)}^3+4w_{(1)}\square w_{(0)}\right)\eta\wedge\vol_{2} -\frac{1}{\kappa_4^2} \frac{1}{\epsilon^3}\int_{M_3}\eta\wedge\vol_{2} \nn \\
& & - \frac{3}{4\kappa_4^2} \frac{1}{\epsilon^2}\int_{M_3}w_{(1)}\, \eta\wedge\vol_{2}~,
\eea
where again $\square=\ex^{-w_{(0)}}\partial^2_{z\zbar}$. Overall, the contribution from gravity is hence
\be
\label{eq:4dGravAction}
S_{\gr} + S_{\rm GH} + S_{\rm ct}
\,=\, -\frac{1}{32\kappa_4^2}\int_{M_3}\left(w_{(1)}^3 + 4w_{(1)}\square w_{(0)}\right) \eta\wedge\vol_{2} - \frac{ 1 }{\kappa_4^2}\int_{\mathcal{T}}\Upsilon~.
\ee

Next we turn to the contribution of the gauge field to the on-shell action. 
Here for the first time in this section we impose the additional global assumption in 
section~\ref{sec:global}: that is, we take $A=A_{(0)}=\mathcal{A}\mid_{y=0}$
 to be a global one-form on the conformal boundary $M_3$. 
 Equivalently, $M_4\mid_{(0,y_0)}\cong (0,y_0)\times M_3$ is 
conformally K\"ahler, and we are imposing that the associated canonical bundle 
is trivial. If this is true throughout $M_4\setminus\{\xi=0\}$ then 
$\mathcal{F}=\diff\mathcal{A}$ is globally exact on the latter\footnote{If the canonical bundle is non-trivial in the interior 
of $M_4\setminus\{\xi=0\}$ there would also be contributions from Dirac strings, but we shall not consider that 
further here.}, and 
we may again use Stokes' theorem to deduce
\bea\label{eq:4dGaugeAction}
S_{\rm gauge} \, = \, -\frac{1}{\kappa_4^2}\int_{M_4}\! \calF \wedge \calF \, = \, \frac{1}{\kappa_4^2}\int_{M_3}\!A_{(0)}\wedge F_{(0)}
- \frac{1}{\kappa_4^2}\int_{\mathcal{T}}\calA\wedge\calF~. \ \ \ 
\eea
In order to further evaluate the first term on the right hand side of (\ref{eq:4dGaugeAction}), 
recall that in the self-dual case the boundary gauge field is
\bea\label{A3dagain}
A_{(0)} &=& \frac{1}{4}w_{(1)}\eta + \frac{\ii}{4}(\partial_{\bar{z}}w_{(0)}\diff\bar{z}-\partial_z w_{(0)}\diff z) + \gamma\, \diff \psi + 
\diff\lambda~.
\eea
Carefully integrating by parts then leads to 
\bea\label{eq:4dGaugeActionbetter}
\frac{1}{\kappa_4^2}\int_{M_3}A_{(0)}\wedge F_{(0)}&=& - \frac{\gamma}{4\kappa_4^2}\int_{M_3}R_{2d}\, \eta\wedge \vol_{2} \nn\\ &&+ \frac{1}{32\kappa_4^2}\int_{M_3}\left( w_{(1)}^3 + 4w_{(1)}\square w_{(0)}\right) \eta\wedge\vol_{2}~.
\eea
Here the first term arises by noting that $R_{2d}=-\square w_{(0)}$ is the  scalar curvature for $\Sigma_2$. 
Notice that the second term perfectly cancels the same term in \eqref{eq:4dGravAction}. 
In general the total action,  obtained by summing (\ref{eq:4dGravAction}) and (\ref{eq:4dGaugeAction}),
is thus
\bea\label{Stotnice}
S \, &=&\,  - \frac{\gamma}{4\kappa_4^2}\int_{M_3}R_{2d}\, \eta\wedge \vol_{2} -\frac{1}{\kappa_4^2}\int_{\mathcal{T}}\left(\Upsilon + \mathcal{A}\wedge\mathcal{F}\right)~.
\eea
This hence splits into a term evaluated at the conformal boundary $M_3$, and an integral
around the fixed points of $\xi$.

We may next further evaluate the  first term on the right hand side of (\ref{Stotnice}) using 
some of the results of section \ref{sec:global}. As argued there, since we may approximate 
an irregular Reeb vector field by quasi-regular Reeb vectors, there is no essential loss of generality
(for the formulas that follow) in assuming that 
$M_3$ is quasi-regular. This means that $M_3$ is the total space 
of a circle orbibundle over an orbifold Riemann surface $\Sigma_2$, with associated 
line orbibundle $\mathcal{L}$.
Combining equations (\ref{R2dc1}) and (\ref{eq:4dGammaFromGauge}) then  allows us to write the action (\ref{Stotnice}) as 
\bea\label{Stotnicer}
S \,&=&\, \frac{\pi^2}{\kappa_4^2}\frac{\left(\int_{\Sigma_2}c_1(\Sigma_2)\right)^2}{\int_{\Sigma_2}c_1(\mathcal{L})} - \frac{1}{\kappa_4^2}\int_{\mathcal{T}}\left(\Upsilon + \mathcal{A}\wedge\mathcal{F}\right)~.
\eea
The contribution of the conformal boundary is now written purely in terms of 
topological invariants of the Seifert fibration structure of $M_3$. 
We will not attempt to evaluate the contributions around the fixed points in (\ref{Stotnicer}) in general -- this 
would take us too far from our main focus. Instead we will follow the computation in \cite{Farquet:2014kma}, 
where $M_4$ has the topology of a ball, with a single fixed point at the origin (a NUT). In this case 
$\mathcal{A}$ is a global one-form on $M_4$, and correspondingly $\int_{\mathcal{T}}\mathcal{A}\wedge\mathcal{F}=0$. 
Similarly, since the K\"ahler form $\omega$ is smooth near the NUT, one can argue that the second term in 
$\Upsilon$ in (\ref{Gamma}) does not contribute to the (limit of the) integral in (\ref{Stotnicer}). However, 
the first term in $\Upsilon$ \emph{does} contribute. Using Stokes' theorem we may write this as
\bea\label{NUTcontr}
- \frac{1}{\kappa_4^2}\int_{\mathcal{T}}\Upsilon \,&=&\, - \frac{1}{\kappa_4^2}\cdot\frac{1}{2y_{\mathrm{NUT}}^2}\int_{M_3}\eta\wedge\diff\eta~,
\eea
where $y_{\mathrm{NUT}}$ is the function $y$ evaluated at the NUT. 
Since the Reeb vector $\xi$ has norm $\|\xi\|\sim r$ near the NUT, where 
$r$ denotes geodesic distance from the NUT, one concludes from the form of the metric 
(\ref{eq:4dSUGRAsolSD}) that $V\sim r^{-2}$. Since $\xi\hook\mathcal{A}$ is 
necessarily zero at the NUT in order that $\mathcal{A}$ is smooth there, from 
(\ref{calASD}) we hence deduce that
\bea
\label{eq:yNUTgamma}
0 &=& -\frac{1}{2y_{\mathrm{NUT}}} + \gamma~,
\eea
which allows us to relate $y_{\mathrm{NUT}}$ to $\gamma$.\footnote{The same formula 
was derived in \cite{Farquet:2014kma} using a different, much longer, route. In the latter reference it was concluded that for $M_3\cong S^3$ all cases where $b_1/b_2  > 0$, and $b_1/b_2 = -1$, are regular. The case $b_1/b_2=-1$ is qualitatively different from the former: the NUT is a point at infinity in the conformal K\"ahler metric, and the K\"ahler metric is asymptotically locally Euclidean. The instanton is regular at the NUT because it vanishes there, and $V\sim r^2$, so \eqref{eq:yNUTgamma} does not hold. Nevertheless, a careful analysis shows that the action evaluates to \eqref{4dSfinal}.}
Thus we may also express the contribution to the action from the NUT (\ref{NUTcontr}) purely 
in terms of topological invariants of $M_3$:
\bea
- \frac{1}{\kappa_4^2}\int_{\mathcal{T}}\Upsilon \,&=&\, - \frac{1}{\kappa_4^2} \cdot 2\gamma^2 \cdot \frac{(2\pi)^2}{b^2}\int_{\Sigma_2}c_1(\mathcal{L})
\ = \  - \frac{2\pi^2}{\kappa_4^2}\frac{\left(\int_{\Sigma_2}c_1(\Sigma_2)\right)^2}{\int_{\Sigma_2}c_1(\mathcal{L})}~.
\eea
Thus in this case the total action (\ref{Stotnicer}) becomes simply
\bea\label{4dSfinal}
S &=&  - \frac{\pi^2}{\kappa_4^2}\frac{\left(\int_{\Sigma_2}c_1(\Sigma_2)\right)^2}{\int_{\Sigma_2}c_1(\mathcal{L})}~.
\eea
Using (\ref{c1S3}) we reproduce the result of \cite{Farquet:2014kma}, where recall that $b_1/b_2=p/q$. However, we can now generalize this further: 
in the above computation all that we needed was the existence of a supergravity solution 
with topology $X=C(M_3)$, a real cone over $M_3$, where the tip of the cone is the only fixed 
point of $\xi$, hence a NUT. If $M_3$ is not diffeomorphic to $S^3$ this will not be smooth
at the NUT, but we can formally consider such singular solutions. The assumptions 
we made about the behaviour of the metric near to this point are then satisfied if the metric is conical near to the NUT. 
In this situation all of the above steps are still valid, and we obtain the same formula (\ref{4dSfinal}) for the action. 

In general
\bea
\int_{\Sigma_2} c_1(\Sigma_2) &=& 2-2g - n + \sum_{I=1}^n\frac{1}{k_I}~,
\eea
where the smooth Riemann surface associated to $\Sigma_2$ has genus $g$, and there 
are $n$ orbifold points with cone angles $2\pi/k_I$, $k_I\in\mathbb{N}$, $I=1,\ldots,n$. 
When the first Chern class above is positive, $\Sigma_2$ hence necessarily has genus $g=0$
and so is topologically $S^2$. It then follows that $M_3\cong S^3/\Lambda$, where $\Lambda$ 
is a finite group. This shows that the class of weighted homogeneous hypersurface singularities 
with $-d+\sum_{i=1}^3w_i>0$ have links $M_3$ which are all quotients of $S^3$ by finite groups. 
Corresponding supergravity solutions can hence be constructed very simply as quotients by $\Lambda$ of 
smooth solutions $M_4$ with ball topology. The supergravity action should then be $1/|\Lambda|$
times the action for the ball solution. It is simple to check this is indeed the case from the formula (\ref{4dSfinal}). 
For weighted hypersurface singularities this reads
\bea\label{Sweighted}
S \,&=&\, \frac{4\pi^2}{\kappa_4^2}\frac{d\left(-d+\sum_{i=1}^3w_i\right)^2}{4w_1w_2w_3}~.
\eea
As summarized in \cite{Martelli:2015kuk}, we may construct supersymmetric quotients 
$M_3\cong S^3/\Lambda$ where $\Lambda=\Lambda_{\mathrm{ADE}}\subset SU(2)$. These may equivalently be realized 
as links of ADE hypersurface singularities, and one can check that indeed
\bea\label{ADEsize}
\frac{4w_1w_2w_3}{d\left(-d+\sum_{i=1}^3w_i\right)^2} \,&=&\, |\Lambda_{\mathrm{ADE}}|~.
\eea
For example, the $E_8$ singularity has weights $(w_1,w_2,w_3)=(6,10,15)$ and degree $d=30$, for which 
the left hand side of (\ref{ADEsize}) gives $|\Lambda_{E_8}|=120$, which is the order of the binary icosahedral group. 

Our formula for the action (\ref{4dSfinal}) reproduces all known large $N$ field theory results, 
summarized in section \ref{sec:FTCasimir}. In particular, we may realize squashed three-spheres, 
with rational Reeb vector $\xi=b_1\partial_{\varphi_1}+b_2\partial_{\varphi_2}$, where $b_1/b_2=p/q\in\mathbb{Q}$,
as links of hypersurface 
singularities with weights $(w_1,w_2,w_3)=(p,q,1)$ and degree $d=1$, for which (\ref{Sweighted}) 
reproduces the field theory result
(\ref{FS3}). Similarly, we may realize Lens spaces $L(p,1)=S^3/\Z_p=S^3/\Lambda_{A_{p-1}}$ 
as links of $A_{p-1}$ singularities, with weights $(w_1,w_2,w_3)=(2,p,p)$ and degree $d=2p$. 
Here $|\Lambda_{A_{p-1}}|=p$, and we reproduce the field theory result of 
\cite{Alday:2012au} that the large $N$ free energy is simply 
$\frac{1}{p}$ times the free energy on $S^3$. The formula (\ref{4dSfinal}) was derived by
assuming supergravity solutions with appropriate general properties exist. For more general 
$M_3$, and in particular for $M_3$ with \emph{negative} $c_1(\Sigma_2)$, more work 
needs to be done to investigate such solutions. We leave this interesting question for future work.

\section{Five-dimensional supergravity}\label{sec:5dsugra}

In the remaining part of the paper we turn to five-dimensional supergravity. We start by constructing a very general AlAdS$_5$ supersymmetric solution of minimal gauged supergravity, in a perturbative expansion near the conformal boundary. 
Then we perform holographic renormalization, extract the holographic energy-momentum tensor and R-current and compare with the field theory results reviewed in section~\ref{sec:fieldtheory}. We will show that standard holographic renormalization violates the field theory supersymmetric Ward identities. However, we will prove that the latter can be restored by introducing new, unconventional boundary terms. For solutions satisfying suitable global assumptions, we also evaluate the on-shell action and conserved charges.

\subsection{The perturbative solution}\label{5d_solution}

Differently from what we did in four-dimensional supergravity, we will initially work in Lorentzian signature $(-,+,+,+,+)$ and discuss an analytic continuation later. In this way we take advantage of the known technology for constructing the solution and postpone the complexification of the supergravity fields.

The bosonic action of minimal gauged supergravity in five dimensions reads~\cite{Gunaydin:1983bi}\footnote{This section is independent of section~\ref{sec:4dgravity}. We will thus adopt the same notation for the five-dimensional supergravity fields as for the four-dimensional ones with no risk of confusion.}
\be
S_{\rm bulk} \, = \, \frac{1}{2\kappa_5^2}\int \left[ \diff^5x \sqrt{G}\left( R_G - \mathcal{F}_{\mu\nu}\mathcal{F}^{\mu\nu} +12 \right)  - \frac{8}{3\sqrt 3} \mathcal{A} \wedge \mathcal{F}\wedge \mathcal{F}  \right]\,.
\label{sugraction}
\ee
Here $R_G$ denotes the Ricci scalar of the five-dimensional metric $G_{\mu\nu}$, $G=|\det G_{\mu\nu}|$, $\mathcal{A}$ is the Abelian graviphoton and $\mathcal{F} = \diff \mathcal{A}$. Moreover, $\kappa_5^2$ is the five-dimensional gravitational coupling constant, and the cosmological constant has been normalized to $\Lambda = -6$.
The Einstein and Maxwell equations read
\bea
\label{EinsteinEq}
R_{\mu\nu} + 2 \mathcal{F}_{\mu\rho}\mathcal{F}^\rho{}_\nu + G_{\mu\nu} \left( 4 + \frac{1}{3}\mathcal{F}_{\rho\sigma}\mathcal{F}^{\rho\sigma} \right) \,& = &\, 0 \ ,\\
\label{MaxwellEq}
\diff *\mathcal{F} + \frac{2}{\sqrt 3} \mathcal{F} \wedge \mathcal{F} \,& = &\, 0\ .
\eea
All solutions of these equations uplift to solutions of type IIB supergravity~\cite{Buchel:2006gb,Gauntlett:2007ma}.\footnote{As for 
the four-dimensional supergravity solutions discussed in section \ref{sec:4dgravity}, this statement holds locally, see {\it e.g.}~\cite{Cassani:2015upa} for some global issues.}

A bosonic field configuration is supersymmetric if there exists a non-trivial Dirac spinor $\epsilon$ satisfying the generalized Killing spinor equation
\be\label{KillingSpEqDirac}
\left[ \nabla_\mu + \frac{\ii}{4\sqrt 3}\left( \Gamma_{\mu}{}^{\nu\lambda} - 4\delta^\nu_\mu\Gamma^\lambda\right) \mathcal{F}_{\nu\lambda} - \frac{1}{2}  \big( \Gamma_\mu - 2\sqrt 3\,\ii \mathcal{A}_\mu \big)\right]\epsilon  \,=\, 0\,,
\ee
where the $\Gamma_\mu$ generate Cliff$(1, 4)$, with $\{\Gamma_\mu,\Gamma_\nu\} = 2 G_{\mu\nu}$. The conditions for a bosonic supersymmetric solution were worked out in~\cite{Gauntlett:2003fk} and discussed further in~\cite{Cassani:2015upa}. The solutions relevant to us are those in the timelike class of~\cite{Gauntlett:2003fk} and are largely determined by a certain four-dimensional K\"ahler structure. In appendix~\ref{app:constr_5d_sol} we review such conditions and solve them in a perturbative expansion. A suitable ansatz for the K\"ahler structure eventually yields a metric and a gauge field on the conformal boundary of the five-dimensional solution which, after a Wick rotation, match the field theory Euclidean background fields~\eqref{bckgnd_metric}, \eqref{A4d}. Here we present the final result after having cast it in Fefferman-Graham form, which is most convenient for extracting the holographic data.

The Fefferman-Graham form of the five-dimensional metric is
\be\label{FG_5d_metric}
\diff s^2_5 \ = \ \frac{\diff \tr^2}{\tr^2} +  h_{ij}(\tx,\tr) \diff \tx^i\diff \tx^j  \,,    
\ee 
 with the induced metric on the hypersurfaces at constant $\rho$ admitting the expansion
\be\label{FGexpansion}
h(\tx,\tr) \ = \ \frac{1}{\tr^2} \left[h^{(0)} + h^{(2)} \tr^2 + \left(h^{(4)} + \tilde h^{(4)}\log \tr^2\right)\tr^4 + \mathcal{O}(\tr^5)\right]\,.
\ee
The gauge field is of the form
\be\label{FGexp_A}
\mathcal{A}(\tx,\tr) \ = \ A^{(0)} + \big(A^{(2)} + \tilde A^{(2)}\log \tr^2\big)\tr^2 + \mathcal{O}(\tr^3) \ ,
\ee
with $\mathcal{A}_{\tr} = 0$.

The hypersurfaces at constant $\rho$ will be described by coordinates $x^i=\{t,z,\bar z,\psi\}$.
As discussed in detail in appendix~\ref{app:constr_5d_sol}, we find that the solution depends on six arbitrary functions $u(z,\bar z)$, $w(z,\bar z)$, $k_1(z,\bar z)$, $k_2(z,\bar z)$, $k_3(z,\bar z)$, $k_4(z,\bar z)$. The functions $u$ and $w$ control the boundary geometry and will be referred to as the {\it boundary data}; these are the same functions appearing in the field theory background~\eqref{bckgnd_metric}, \eqref{A4d}. The functions $k_1$, $k_2$, $k_3$, $k_4$ first show up in the $h^{(4)}$ and $A^{(2)}$ subleading terms of the Fefferman-Graham expansion and will be denoted as the {\it non-boundary data} of the solution.

The first two terms in the expansion of the induced metric read
\begin{align}\label{termsFGmetric}
h^{(0)}  \,&=\, -\diff t^2 + (\diff \psi + a)^2 + 4\me^{\wzero}\diff z \diff\bar z\ ,\nn\\[2mm]
h^{(2)} \,&=\, \frac{8\square \wzero + \wone^2}{96}\diff t^2 - \frac{8\square \wzero +7\wone^2}{96} (\diff\psi+a)^2 +\frac{16 \square \wzero+5\wone^2}{24} \me^{\wzero}\diff z\diff\bar z\nn\\[1mm]
& \quad - \frac{1}{4}(*_2\diff \wone) (\diff\psi+a)\,  ,
\end{align}
where $a$ satisfies \eqref{eq:3dBoundaryDa} as in the field theory background.
Moreover,
 $\square  = \ex^{-\wzero}\partial^2_{z\bar{z}} $ is the Laplacian of the two-dimensional part of the boundary metric $h^{(0)}$, which coincides with \eqref{2dMetric},
and we are using the notation
\be
*_2\! \diff \ = \ 
\ii(\diff \bar{z}\,\partial_{\bar z} - \diff z\,\partial_{z})\ .
\ee
One can check that $h^{(2)}$ is determined by $h^{(0)}$ according to the general relation~\cite{deHaro:2000vlm,Taylor:2000xw}
\be\label{g(2)general}
h_{ij}^{(2)} \ = \ \frac{1}{12} \left(R \, h_{ij} - 6R_{ij} \right)^{(0)}\,.
\ee Here and in the formulae below, a superscript $\,^{(0)}$ outside the parenthesis means that all quantities within the parenthesis are computed using the boundary metric $h^{(0)}$ (and, as far as the formulae below are concerned, the boundary gauge field $A^{(0)}$).

In order to determine the on-shell action and the holographic charges we will also need the $\tilde h^{(4)}$ and $h^{(4)}$ terms in the Fefferman-Graham expansion~\eqref{FGexpansion}. We have verified that $\tilde h^{(4)}$ is determined by the boundary data as 
\be\label{tildeg4fromWeylAn}
\tilde h^{(4)}_{ij} \ =\ -\frac{1}{8}\left(B_{ij}+8 F_{ik}F_j{}^k -2 h_{ij}F_{kl}F^{kl}\right)^{(0)}  ,
\ee
where $B_{ij}$ is the Bach tensor, see appendix \ref{app_curvature} for its definition. 
 Recalling that the variation of the integrated Euler density vanishes identically in four dimensions, we can write
\be
\tilde h^{(4)}_{ij} \ =\ \frac{1}{16\sqrt{h^{(0)}}} \frac{\delta}{\delta h^{(0)ij}}\int \diff^4 x \sqrt{h^{(0)}} \left(- E^{(0)} + C^{(0)}_{klmn}C^{(0)klmn} - 8 F^{(0)}_{kl}F^{(0)kl} \right) \ ,
\ee
where $E^{(0)}$ and $C^{(0)}_{ijkl}$ are the Euler scalar and the Weyl tensor of the boundary metric $h^{0}$ (again see appendix~\ref{app_curvature}).
This means that $\tilde h^{(4)}_{ij}$
is proportional to the metric variation of the integrated holographic Weyl anomaly, 
 a fact that for vanishing gauge field  was first observed in~\cite{deHaro:2000vlm}. 

As for $h^{(4)}_{ij}$, this contains the four non-boundary functions $k_1$, $k_2$, $k_3$, $k_4$, as well as the boundary functions $\wone,\wzero$ (hit by up to six derivatives); we will not give its explicit expression here as it is extremely cumbersome and can only be dealt with using a computer algebra system as Mathematica. 
As a sample we provide two simple relations between some of the components:
\begin{align}
h^{(4)}_{tt} - h^{(4)}_{\psi\psi} \,&=\, -k_3 + \frac{1}{6} k_2^2 + \frac{1}{24}\square k_2 + \frac{1}{24} (2\square\wzero + \wone^2)k_2 + \frac{17}{6144}\wone^4 - \frac{3}{256}\square \wone^2  \nn \\[1mm]
\,&\quad  + \frac{1}{96}\me^{-\wzero}\partial_z \wone \partial_{\bar z}\wone + \frac{1}{192}\left(\wone^2 \square \wzero - \frac{5}{2} \square^2 \wzero - (\square \wzero)^2\right)\ ,
\end{align}
\be
 h^{(4)}_{tt}+ h^{(4)}_{\psi\psi} - 2 h^{(4)}_{t\psi} \,=\, -\frac{1}{2}\wone k_1 -\frac{1}{6}\wone^2 k_2 + \frac{1}{128} \wone^4 + \frac{1}{48}\wone^2 \square \wzero\ .
\ee
We also checked that the trace is determined by boundary data as
\be
 h^{(0)\,ij}h^{(4)}_{ij} \,=\,  \frac{1}{48} \left( 4R_{ij}R^{ij} - R^2 \right)^{(0)} \ .
\ee

As a consequence of supersymmetry, the gauge field is entirely determined by the metric and does not contain new functions (apart for the gauge choice to be discussed momentarily). In particular, $A^{(0)}$ and $\tilde A^{(2)}$ just depend on the boundary metric functions, while $A^{(2)}$ also depends on $k_1$, $k_2$, $k_3$. The explicit expressions are
\begin{align}
\label{gaugefield_FG0} A^{(0)} \, &= \, -\frac{1}{\sqrt 3}\left[-\frac{1}{8}\wone\,\diff t + \frac{1}{4}\wone (\diff \psi + a) + \frac{1}{4}*_2 \!\diff \wzero + \diff \lambda + \cpsi\,\diff \psi  + \ct \diff t \right] \, ,\\[1mm]
\tilde A^{(2)}\,&=\, \frac{1}{32\sqrt 3}\left[ - \square \wone\, \diff t + \Big( 2\square \wone - \wone \square \wzero - \frac{1}{2}\wone^3\Big)(\diff \psi + a) + *_2 \diff \left( 2\square \wzero + \wone^2 \right)\right] ,\\[1mm]
A^{(2)}\,&=\, \frac{1}{64\sqrt 3}\left[ \left(96 k_1 + 32\wone k_2 - 4\wone\square \wzero - \frac{3}{2}\wone^3 \right) \diff t - *_2 \diff \left(32 k_2 + \wone^2  \right)\right. \nn\\[1mm]
\,&\,\quad  + \left. \frac{1}{\wone}\left( 128k_3 -32 \wone k_1 -\frac{64}{3}k_2^2 + 16\square k_2 - \frac{32}{3}k_2\square \wzero -16 \wone^2 k_2 + 3 \square (\square \wzero + \wone^2) \right.\right.\nn\\[1mm]
\,&\,\quad - \left.\left.  2(\square \wzero)^2 - \frac{5}{3}\wone^2\square \wzero - 3\me^{-\wzero} \partial_z \wone\partial_{\bar z}\wone - \frac{5}{12} \wone^4 \right)\left(\diff t+ \diff \psi +a \right)\right]\,.
\end{align}

Clearly, upon performing the Wick rotation $t = -\ii \tau$ we can identify $h^{(0)} = g$, $A^{(0)}= -\frac{1}{\sqrt 3} A$, where $g$ and $A$ were given in~\eqref{bckgnd_metric}, \eqref{A4d} and define the four-dimensional SCFT background. 
We recall that the last three terms in \eqref{gaugefield_FG0} are gauge choices: $\cpsi,\ct$ are two constants while $\lambda$ is a function of $z,\bar z$; these will play an important role in the following.

One can check that
\be\label{expr_tildeA2}
\tilde A^{(2)}_i = -\frac{1}{4}(\nabla^j F_{ji})^{(0)}\ .
\ee 
In analogy with $\tilde h^{(4)}$, we see that $\tilde A^{(2)}$ is obtained by varying the integrated holographic Weyl anomaly, this time with respect to the boundary gauge field $A^{(0)}$. 

Generically, the boundary is not conformally flat and the solution is asymptotically {\it locally} AdS$_5$. In the particular case where the boundary is conformally flat and the boundary gauge field strength vanishes --- {\it i.e.}\ when the solution is AAdS rather than AlAdS --- both $\tilde h^{(4)}$ and $\tilde A^{(2)}$ vanish.  This is in agreement with the general fact that the logarithmic terms in the Fefferman-Graham expansion vanish for a   conformally flat boundary.

The solutions described above preserve at least (and generically no more than) two real supercharges.
We have also verified that the five-dimensional metric and gauge field discussed above satisfy the Einstein and Maxwell equations at order $\mathcal{O}(\tr^3)$, which is the highest we have access to given the order at which we worked out the solution.

\subsection{Standard holographic renormalization}\label{Stand_Holo_Reno}
 
Following the standard procedure of holographic renormalization,\footnote{See~\cite{Taylor:2000xw,Bianchi:2001kw} for the modifications due to the inclusion of a Maxwell field.} a finite on-shell action $S$ is obtained by considering a regularized five-dimensional space $M_\epsilon$ where the radial coordinate $\tr$ does not extend until the conformal boundary at $\tr = 0$ but is cut off at $\rho=\epsilon$, so that $\partial M = \lim_{\epsilon\to 0}\partial M_\epsilon$. Then one evaluates the limit
\be
S \ =\  \lim_{\epsilon\to 0}  \left(S_{\rm bulk} + S_{\rm GH} + S_{\rm ct}+ S_{\rm ct,finite}\right)\ .
\ee
Here, $S_{\rm bulk}$ is the bulk action \eqref{sugraction}, where the integral is carried out over $M_\epsilon$.
$S_{\rm GH}$ is the Gibbons-Hawking-York boundary term,
\be
S_{\rm GH}\ = \  \frac{1}{\kappa_5^2}\!\int_{\partial M_\epsilon}\!\! \diff^4x \sqrt{h}\, K\,,
\ee
where $K = h^{ij}K_{ij}$ is the trace of the extrinsic curvature $K_{ij} \,=\, -\frac{\rho}{2} \frac{\partial h_{ij}}{\partial \tr}$ of $\partial M_\epsilon$. 
The counterterm action $S_{\rm ct}$ is a boundary term cancelling all divergences that appear in $S_{\rm bulk}+S_{\rm GH}$ as $\epsilon\to 0$; it reads
\be\label{Counterterm}
S_{\rm ct}\ = \ -\frac{1}{\kappa_5^2}\int_{\partial M_\epsilon} \!\!\diff^4x\sqrt{h} \left[3 + \frac{1}{4} R  +\frac{1}{16} \left(E - C_{ijkl}C^{ijkl} + 8 \mathcal{F}_{ij}\mathcal{F}^{ij} \right)\log  \epsilon \right]\,.
\ee
The first two terms cancel power-law divergences while the logarithmically divergent term removes the holographic Weyl anomaly. Here, $E$ is the Euler scalar and $C_{ijkl}$ is the Weyl tensor of the induced metric $h_{ij}$. Note that since $\sqrt{h}(E - C_{ijkl}C^{ijkl} + 8 \mathcal{F}_{ij}\mathcal{F}^{ij})$ remains finite as $\epsilon\to 0$, it can equivalently be computed using the boundary metric~$h^{(0)}_{ij}$ and boundary gauge field $A^{(0)}_i$. 

Finally, $S_{\rm ct,finite}$ comprises local counterterms that remain finite while sending $\epsilon\to 0$. In general, these may describe ambiguities in the renormalization scheme or be necessary in order to restore some desired symmetry that is broken by the rest of the action. In our case, requiring diffeomorphism and gauge invariance the linearly independent such terms may be parameterized as
\be\label{ambij}
S_{\rm ct,finite} \ =\ \frac{1}{\kappa_5^2} \int_{\partial M_\epsilon}\diff^4x \sqrt h \left( \varsigma\, R^2 - \varsigma' \mathcal{F}_{ij}\mathcal{F}^{ij} + \varsigma''\, C_{ijkl}C^{ijkl} \right),
\ee
where $\varsigma$, $\varsigma'$, $\varsigma''$ are {\it a priori} arbitrary numerical constants.\footnote{We could also include in the linear combination the terms $\int \diff^4 x\sqrt h E$, $\int \diff^4 x\sqrt h\mathcal{P}$ and $\int \diff^4 x\sqrt h\epsilon^{ijkl}\mathcal{F}_{ij} \mathcal{F}_{kl}$, where $\mathcal{P}$ is the Pontryagin density on $\partial M_\epsilon$, however these are topological quantities that have a trivial variation; moreover, as we will see below they vanish identically in the geometries of interest for this paper.}

The holographic energy-momentum tensor is defined as the variation of the on-shell action with respect to the boundary metric
\be\label{HoloEnMomTensor0}
 T_{ij} \ = \  -\frac{2}{\sqrt{g}}\frac{\delta S}{\delta g^{ij}}\ ,
\ee
and can be computed by means of the general formula
\begin{align}\label{HoloEnMomTensor}
 T_{ij} \, & = \, \frac{1}{\kappa_5^2}\lim_{\epsilon \to 0}\, \frac{1}{\epsilon^2}  \,\bigg[ -K_{ij} + K h_{ij} - 3 h_{ij}  + \frac{1}{2}\left( R_{ij}-\frac 12 R\, h_{ij}\right) \nn \\[2mm]
\ &\ \hspace{29mm} +\,\frac{1}{4}\! \left( B_{ij} + 8 \mathcal{F}_{ik}\mathcal{F}_j{}^k - 2 h_{ij} \mathcal{F}_{kl}\mathcal{F}^{kl} \right)\log  \epsilon  \nn\\[2mm]
\ &\ \hspace{29mm} +\,   \left( 2\varsigma H_{ij} + 4 \varsigma''  B_{ij} + \varsigma' \left( 4\mathcal{F}_{ik}\mathcal{F}_j{}^k - h_{ij} \mathcal{F}_{kl}\mathcal{F}^{kl} \right) \right)\bigg]\ ,
\end{align}
where all quantities in the square bracket are evaluated on $\partial M_\epsilon$, and we refer to appendix~\ref{app_curvature} for the definition of the tensor
$H_{ij}$.

The holographic $U(1)_R$ current is defined as
\be\label{HoloRcurrent}
 j^i \ = \  \frac{1}{\sqrt{g}} \frac{\delta S}{\delta A_i}\,,
\ee
Note that we defined the variation in terms of  the rescaled boundary gauge field $A = -\sqrt 3 A^{(0)}$. In this way the holographic R-current is normalized in the same way as the field theory R-current.
This yields the expression:
\be\label{ExpressionRcurrent}
 j^i  = -\frac{2}{\sqrt 3\kappa_5^2} \lim_{\epsilon \to 0}\, \frac{1}{\epsilon^4}\left\{ *_4\! \left[ \diff x^i \wedge \left(*_5 \mathcal{F} + \frac{4}{3\sqrt 3} \mathcal{A} \wedge \mathcal{F}\right) \right]  + \nabla_j \mathcal{F}^{ji}\log\epsilon + 2 \varsigma' \nabla_j \mathcal{F}^{ji} \right\} ,
\ee
where the first term comes from varying the bulk action $S_{\rm bulk}$, the second from $S_{\rm ct}$ and the third from $S_{\rm ct,finite}$.

Given the definitions~\eqref{HoloEnMomTensor0} and \eqref{HoloRcurrent}, the variation of the renormalized on-shell action under a generic deformation of the boundary data can be expressed via the chain rule as
\be\label{deltaS_general}
\delta S \ = \ \int_{\partial M} \diff^4 x \sqrt{g} \left( -\frac{1}{2} T_{ij} \delta g^{ij} +  j^i \delta A_i  \right)\ .
\ee
Starting from this formula, one can check several Ward identities holding in the holographic renormalization scheme defined above. Invariance of the action under a boundary diffeomorphism generated by an arbitrary  vector on $\partial M$ yields the expected conservation equation for the holographic energy-momentum tensor,
\be\label{divT}
\nabla^i  T_{ij} \ =\ F_{ji} j^i  - A_j\nabla_i  j^i  \ ,
\ee
where $\nabla_i$ is the Levi-Civita connection of $g_{ij}$.
Studying the variation of the on-shell action under a boundary Weyl transformation such that $\delta g_{ij}=2g_{ij}\delta \sigma$, $\delta A_i =0$, one finds for the trace of the holographic energy-momentum tensor \cite{Henningson:1998gx}:
\be\label{traceT_general}
 T{}_{i}{}^i  \ = \ \frac{1}{16\kappa_5^2}\left(- E + C_{ijkl}C^{ijkl} - \frac{8}{3} F_{ij}F^{ij} \right) -\frac{12\varsigma}{\kappa_5^2}\, \nabla^2 R \ ,
\ee
which reproduces the known expression for the Weyl anomaly of a superconformal field theory \cite{Anselmi:1997am,Cassani:2013dba}, with the standard identifications $\a=\c=\pi^2/\kappa_5^2$. The variation under a gauge transformation at the boundary leads to~\cite{Witten:1998qj,Cassani:2013dba}:
\be\label{divRcurrent}
\nabla_i  j^i  \, = \, 
\frac{1}{27\kappa_5^2} \,\epsilon^{ijkl}F_{ij}F_{kl}\ ,
\ee
which again is consistent with the chiral anomaly of the superconformal R-symmetry.

\subsection{The new boundary terms}\label{sec:new_dry_terms}

We now specialize to the family of asymptotic supersymmetric solutions constructed in section~\ref{5d_solution} and test whether the supersymmetric Ward identities reviewed in section~\ref{sec:fieldtheory} are satisfied holographically. We will consider variations of the boundary functions that preserve the complex structure(s) on $M_4 \cong \partial M_5$, and compute the corresponding variation of the on-shell action via~\eqref{deltaS_general}. As discussed in section~\ref{sec:fieldtheory}, the input from field theory is that this variation should vanish if supersymmetry is preserved.
{\it A priori} one might expect that there is at least a choice of the $\varsigma$-coefficients in the standard finite counterterms~\eqref{ambij} such that the supersymmetric Ward identity is satisfied. However, we will show that this is {\it not} the case and that {\it new}, non-standard finite counterterms are required.

Before going into this, it will be useful to notice that the boundary metric and gauge field in \eqref{termsFGmetric}, \eqref{gaugefield_FG0} satisfy 
\be
E \ = \  \mathcal{P} \ = \ \epsilon^{ijkl}F_{ij}F_{kl} \ = \ 0\ ,
\ee
where $\mathcal{P}$ is the Pontryagin density on $\partial M$.
Moreover, supersymmetry implies~\cite{Cassani:2013dba}
\begin{align}\label{relWeylsqFsq}
C_{ijkl}C^{ijkl} - \frac{8}{3} F_{ij}F^{ij} \ &= \ 0 \ .
\end{align}
It follows that~\eqref{divT}--\eqref{divRcurrent} simplify to
\be\label{conseqs_susy}
\nabla_i  j^i   \ = \ 0 \ ,\qquad \
\nabla^i  T_{ij} \ =\ F_{ji} j^i \ , \qquad \ T{}_i{}^i   \ =\ -\frac{12\varsigma}{\kappa_5^2}\, \nabla^2 R   \ .
\ee

Relation \eqref{relWeylsqFsq} also implies that by redefining the coefficients $\varsigma'$, $\varsigma''$ we can set $\varsigma''=0$ in the finite counterterm action~\eqref{ambij} as well as in all its variations that preserve supersymmetry at the boundary. Below we will assume this has been done.

As explained in section~\ref{sec:FTCasimir}, a variation of the boundary data that preserves the complex structures $I_\pm$ on the boundary corresponds to deformations $\wone\to \wone+\delta\wone$, $\wzero\to \wzero+\delta\wzero$ such that $\delta\wone=\delta\wone(z,\bar z)$ and $\delta\wzero=\delta\wzero(z,\bar z)$ are {\it globally well-defined} functions. In the following we study the consequences of such variations.
We will also assume that $\partial M$ is compact and that the non-boundary functions $k_1$, $k_2$, $k_3$, $k_4$ are globally well-defined functions of their arguments $z,\bar z$. This will allow us to apply Stokes' theorem on the boundary and discard several total derivative terms.

We first vary $\wzero$ keeping the one-form $\phizero$ fixed.  From~\eqref{eq:3dBoundaryDa}, we see that this is possible provided the variation preserves $\me^{\wzero}\wone$, hence we also need to take $\delta \wone = - \wone\, \delta \wzero$. Plugging the explicit expression of $T_{ij}$ and $j^i$ into~\eqref{deltaS_general} and dropping several total derivative terms involving the boundary functions and $k_2(z,\bar z)$, we find that the variation of the on-shell action is:
\begin{align}
\delta_{\wzero} S &=  \frac{1}{2^6 3\kappa_5^2}\int_{\partial M} \! \diff^4 x \sqrt{g}\, \delta \wzero\,\bigg[ \left(-1+96\varsigma - 16\varsigma' \right) \wone^2 R_{2d} -  \frac{1}{2}\left(1-96\varsigma +28\varsigma'\right) \square \wone^2 \nn \\[2mm] 
\ &\qquad\qquad \qquad +  \frac{1}{32}\left(19-288\varsigma+192\varsigma'\right) \wone^4  - \frac{8}{9}(\cpsi +2\ct) \left( 2\wone R_{2d} + 2\square \wone - \wone^3\right)  \nn \\[2mm]
\ &\qquad\qquad\qquad - 12\varsigma'  \wone\square \wone  + 8(-24\varsigma+\varsigma') (R_{2d}^2 + 2\square R_{2d})
 \bigg]\ ,\label{w_variation}
\end{align}
where we recall that $R_{2d}= -\square \wzero$ is the Ricci scalar of the two-dimensional metric~\eqref{2dMetric}.
If instead we vary $\wone$ while keeping $\wzero$ fixed we obtain
\begin{align}\label{u_variation}
\delta_{\wone} S \,&=\, \frac{1}{2^{9} 3^2\kappa_5^2}\int_{\partial M} \!\diff^4 x \sqrt{g}\,
 \delta \wone \bigg[ 24\left(1-96\varsigma + 16\varsigma'\right)\wone R_{2d}+ 288\varsigma' \square \wone\nn\\[1mm]
&\qquad\qquad\quad - \left(19-288\varsigma+192\varsigma'\right) \wone^3  -\frac{32}{3} (\cpsi+2\ct) (3 \wone^2 - 4R_{2d}) \bigg]\,,\quad
\end{align}
where again we dropped many total derivative terms, some of which containing the non-boundary data $k_2$, $k_3$. In order to do this, we used that $\delta \phizero$ is globally defined; this follows from the assumption that the complex structures are not modified.

Inspection of~\eqref{w_variation}, \eqref{u_variation} shows that there exists no choice of the coefficients $\varsigma,\varsigma'$ such that $\delta_\wzero S=\delta_\wone S= 0$. Therefore we conclude:

\begin{quote} \emph{Standard holographic renormalization does not satisfy the field theory supersymmetric Ward identities}.\end{quote}

\noindent Remarkably, we find that this can be cured by introducing {\it new} finite terms. Both variations $\delta_\wzero S$ and $\delta_\wone S$ vanish if we take $\varsigma=\varsigma'=0$ (that is, if we set $S_{\rm ct,finite}=0$) {\it and} add to the on-shell action the new terms
\be\label{newCttrms}
\Delta S_{\rm new} \, = \, \frac{1}{2^{11} 3^2 \kappa_5^2}\int_{\partial M} \diff^4 x \sqrt{g} \left[  19 \wone^4 - 48 \wone^2R_{2d} + \frac{128}{3}(2\ct + \cpsi)(\wone^3 - 4 \wone R_{2d}) \right]\,.
\ee
In other words, {\it the new renormalized action}
\be\label{susy_ren_action}
S_{\rm susy} \ =\ \lim_{\epsilon\to 0} \left(S_{\rm bulk} + S_{\rm GH} + S_{\rm ct}\right) + \Delta S_{\rm new}
\ee
{\it does satisfy the supersymmetric Ward identities}. We claim that this is the correct supersymmetric on-shell action that should be compared with the supersymmetric field theory partition function.

It should be clear that the terms $\Delta S_{\rm new}$ cannot be written as a local action that is: $i)$ invariant under four-dimensional diffeomorphisms, $ii)$ invariant under gauge transformations of $A$, and $iii)$ constructed using the boundary metric, the boundary gauge field and their derivatives only. If this was the case, $\Delta S_{\rm new}$ would fall in the family of standard finite counterterms \eqref{ambij}, which we have just proven not to be possible.
We will comment on this issue in the conclusions. Here we make a first step towards clarifying it by observing
 that the gauge-dependent part of $\Delta S_{\rm new}$ --- {\it i.e.}\ the term containing the gauge parameters $\cpsi,\ct$ --- has to come from a term linear in the boundary gauge potential $A = -\sqrt 3 A^{(0)}$. So we may write
\be\label{new_cttrm_split1}
\Delta S_{\rm new}\ = \ \frac{1}{\kappa_5^2}\int_{\partial M} (A \wedge \Phi + \Psi)\ ,
\ee
where $\Psi$ is gauge-invariant. Matching this with~\eqref{newCttrms}, we obtain
\begin{align}\label{new_cttrm_split2}
\Phi \ &=\ \frac{1}{2^3 3^3} \left(\wone^3-4 \wone R_{2d} \right)\ii\,\me^{\wzero}\diff z \wedge \diff \bar z\wedge (2\diff \psi - \diff t )\ ,\nn\\[1mm]
\Psi \ &= \ \frac{1}{2^{11}3^2}\left( 19 \wone^4-48 \wone^2 R_{2d} \right) \diff^4 x \sqrt{g} \ .
\end{align}
Notice that $\diff\Phi=0$, so $\Delta S_{\rm new}$ is invariant under small gauge transformations. However, it depends on the choice of flat connection for $A$ when $\partial M$ has one-cycles. Also notice that~\eqref{new_cttrm_split1} implies that $\Delta S_{\rm new}$ yields a new contribution to the holographic R-current~\eqref{HoloRcurrent}. Below we will show that this modifies the R-charge precisely as demanded by the superalgebra. 

\subsection{Evaluation of the on-shell action}\label{sec:5d_onshell_action}

In this section we evaluate the renormalized supergravity action \eqref{susy_ren_action} on the class of five-dimensional solutions constructed above. Since this involves performing a bulk integral, {\it a priori} one would need to know the full solution in the interior, while we just have it in a perturbative expansion near the boundary. However, we show that under certain  global assumptions the on-shell action reduces to a boundary term that can be evaluated exactly as a function of boundary data only.

The assumptions consist in requiring that the solution caps off regularly and with no boundary in the interior, and that the graviphoton $\mathcal{A}$ is a global 
one-form.\footnote{For example this excludes supersymmetric black hole solutions \cite{Gutowski:2004ez,Chong:2005hr}.} As shown in~\cite{Cassani:2014zwa},
this allows to express the bulk action of supersymmetric solutions in the timelike class as the boundary term
\be\label{SbulkAsBoundaryIntegral}
S_{\rm bulk}\  =\   \frac{1}{3\kappa_5^2} \int_{\partial M_\epsilon} \left(\diff y \wedge P \wedge J - 2 \mathcal{A}\wedge *_5\mathcal{F} \right)\,,
\ee
where the coordinate $y$, the Ricci one-form potential $P$ and the K\"ahler form $J$ are those of the ``canonical structure'' dictated by supersymmetry~\cite{Gauntlett:2003fk} and are defined in appendix~\ref{5dsusy_eqs}. We remark that while demanding that $\mathcal{A}$ is a global one-form we are also taking $P$ as a global one-form, see eq.~\eqref{susygaugefield}. Notice this implies that the canonical bundle 
of the $4d$ K\"ahler metric is trivial, {\it cf.}\ an analogous global assumption in  section \ref{sec:4dgravity}. The integral  on the hypersurface $\partial M_\epsilon$ at constant $\rho$  can be explicitly evaluated for our solution after passing to Fefferman-Graham coordinates as discussed in appendix~\ref{sec:ansatz}.

Even if the on-shell action has now reduced to a boundary term, generically it still depends on the arbitrary non-boundary functions appearing in the solution. However, generalizing an argument given in~\cite{Cassani:2014zwa} we can show that the assumption of global regularity also entails a relation between these non-boundary functions and the boundary ones that is precisely sufficient for determining the on-shell action. 

Let $\mathcal{C}$ be a Cauchy surface (namely, a hypersurface at constant $t$), with boundary $M_3 = \mathcal{C} \cap \partial M_5$, and consider the Page charge
\be\label{defGlobConstr}
\Theta \ = \ \int_{M_3} \Big(*_5\! \mathcal{F} + \frac{2}{\sqrt 3}\mathcal{A}\wedge \mathcal{F} \Big)\ .
\ee
Since $\mathcal{A}$ is globally defined and $\partial M_5$ is by assumption the only boundary of the space, we can apply Stokes' theorem and then use the Maxwell equation to infer that $\Theta$ must vanish:
\be
\Theta \, = \, \int_{M_3} \Big(*_5\! \mathcal{F} + \frac{2}{\sqrt 3}\mathcal{A}\wedge \mathcal{F} \Big) \ =\ \int_\mathcal{C} \Big(\diff *_5\! \mathcal{F} + \frac{2}{\sqrt 3}\mathcal{F}\wedge \mathcal{F} \Big) \ =\  0\ .
\ee
We now replace the Fefferman-Graham expansion of the graviphoton field strength
\be
\mathcal{F}  =  \diff A^{(0)} + \tr^2\big(\diff A^{(2)}+\diff \tilde A^{(2)}\log \tr^2 + \mathcal{O}(\tr)\big)  + 2\tr\diff\tr \wedge \big(A^{(2)}+ \tilde A^{(2)}+\tilde A^{(2)}\log \tr^2+\mathcal{O}(\tr)\big)
\ee
and its Hodge dual restricted to the hypersurfaces at constant $\tr$,
\be
(*_5 \mathcal{F})\big|_{\diff \tr = 0} \, = \, 2  *^{(0)} \left(A^{(2)}+ \tilde A^{(2)}+\tilde A^{(2)}\log \tr^2\right) + \mathcal{O}(\tr)\ ,
\ee
where $*^{(0)}$ is the Hodge star of the boundary metric $h^{(0)}$.\footnote{Note that the logarithmic divergence drops out of the quantities we are interested in. Indeed, from~\eqref{expr_tildeA2} we see that $*^{(0)}\tilde A^{(2)} \propto (\diff* F)^{(0)} $ is a total derivative, hence it drops from any boundary integral.}
It is easy to see that expression \eqref{defGlobConstr} then becomes
\be
\Theta \ =\  \int_{M_3} \Big(2\, {\rm vol}_3 \big(A^{(2)}_t + \tilde A^{(2)}_t \big) + \frac{2}{\sqrt 3}A^{(0)}\wedge \diff A^{(0)}\Big)\ ,
\ee
where we are using the notation ${\rm vol}_3 \equiv \diff^3x \sqrt{ g_3}$ for the Riemannian volume form on $M_3$.
The condition $\Theta=0$ is thus equivalent to the statement that the integrated time component of $A^{(2)}$, which {\it a priori} is controlled by non-boundary data and is thus not fixed by the equations of motion, is actually determined by boundary data.
Evaluating this on our perturbative solution, we find the following integral relation between the non-boundary functions $k_1,k_2,k_3$ and the boundary functions $\wone,\wzero$:
\begin{align}\label{GlobConstr}
 0 \,&=\, \Theta\,=\, \frac{1}{96\sqrt3}\int_{M_3} \!{\rm vol}_3\, \bigg[\frac{1}{\wone} \Big( 384\, k_3 -64k_2^2 + 48 \square  k_2  + 32 k_2 R_{2d} + 9\me^{-\wzero} \partial_z \wone \partial_{\bar z} \wone  \nn\\[1mm]
&\qquad\qquad\qquad\qquad\qquad\quad - 9 \square R_{2d} - 6 R_{2d}^2 \Big) + 48 \wone k_2 - \frac{15}{4} \wone^3  + 192 k_1 \nn\\[1mm]
&\qquad\qquad\qquad \qquad\qquad\quad   + 6\,\me^{\frac{1}{3}\wzero} \big[\nabla_z\big( \me^{-\frac{4}{3}\wzero}\partial_{\bar z} \wone \big)+ c.c\big] + (13 \wone - 16 \cpsi) R_{2d} \bigg]\nn\\[1mm]
&\qquad\quad  -\frac{1}{6\sqrt 3}\int_{M_3} \diff \psi \wedge \diff \big[\wone (\diff \lambda - \cpsi\, \phizero)\big]\ .
\end{align}

We can now give our result for the renormalized on-shell action.
Adding up all contributions to \eqref{susy_ren_action}, including the new counterterms \eqref{newCttrms}, and without making further assumptions, we obtain
\begin{align}\label{Sren_finite_withtotder}
S_{\rm susy} &= \frac{\int \diff t}{27\kappa_5^2} \bigg\{ \int_{M_3}  {\rm vol}_3\, \Big[(\ct-\cpsi) \cpsi R_{2d} + \frac{9}{8}\,\square \left( 4 k_2 - \cpsi \wone \right)\Big] \nn\\[1mm]
& \ \ \quad + \frac{1}{64} \int_{M_3}\!\!\diff \left[\diff \psi\wedge \left(96k_2 + 12R_{2d} - 3\wone^2 + 16 (\ct-\cpsi) \wone \right) (4\diff \lambda-4 \cpsi \phizero + *_2 \diff \wzero)\right]\nn\\[1mm]
& \ \ \quad + 6\sqrt 3 (\ct-\cpsi) \,\Theta \bigg\} \ .
\end{align}
The Laplacian term in the first line and the whole integrand in the second line are total derivatives of globally defined quantities and therefore vanish upon integration. The term $\Theta$ in the third line, given by \eqref{GlobConstr}, also vanishes as just seen.
So we obtain a very simple expression for the on-shell action, depending on boundary data only:
\be\label{ResultOnShActLorentzian}
S_{\rm susy}\ =\ \frac{(\ct-\cpsi) \cpsi }{27\kappa_5^2}\int \diff t \int_{M_3} {\rm vol}_3\, R_{2d} \ .
\ee
We next implement the analytic continuation $t =  -\ii \tau$, which renders the boundary metric Euclidean (while the bulk metric generally is complex), and assume that $\tau$ parameterizes a circle of length $\beta$. The expression for the on-shell action thus becomes\footnote{The overall sign change comes from the identification $\ii S_{{\rm Lorentzian,}\,t = -\ii\tau} = - S_{\rm Euclidean}$.}
\be\label{ResultOnShActEuclidean}
S_{\rm susy}\ =\ \frac{\beta(\cpsi- \ct) \cpsi }{27\kappa_5^2} \int_{M_3} {\rm vol}_3\, R_{2d} \ .
\ee
It is interesting to note that, as we show in appendix~\ref{app:Killingspinors}, the flat connection parameters $\cpsi$ and $\ct$ also correspond to the charge of the boundary Killing spinor $\zeta_+$ under $\partial_\psi$ and $\ii\partial_\tau$, respectively. Hence $\cpsi-\ct$ is twice the charge of $\zeta_+$ under the complex Killing vector $K$ introduced in section~\ref{sec:4dbackgrounds}.

Recall from section~\ref{sec:global} that the requirement that the boundary gauge field is globally defined fixes $\cpsi$ as
\be\label{choice_cpsi}
\cpsi \ =\ -\frac 14 \frac{\int_{M_3} {\rm vol}_3\, R_{2d}}{\int_{M_3}  \eta \wedge \diff\eta}\ .
\ee
Recalling~\eqref{def_eta}, \eqref{eq:3dBoundaryDa}, the contact volume of $M_3$ appearing in the denominator can also be expressed as $\int_{M_3}\! \eta\wedge \diff \eta = \frac 12\int_{M_3}\!  {\rm vol}_3\,\wone$. 

As far as the bosonic solution is concerned, expression~\eqref{ResultOnShActEuclidean} makes sense for any value of $\ct$. However, for $S_{\rm susy}$ to be the on-shell action of a proper supersymmetric solution we also need to impose that the Killing spinors are independent of $\tau$, so that they remain globally well-defined when this coordinate is made compact.
Since $\ct$ is the charge of the Killing spinors under $\ii\partial_\tau$, we must take $\ct=0$.

We conclude that for a regular, supersymmetric AlAdS$_5$ solution satisfying the global assumptions above, and such that the conformal boundary has a direct product form  $S^1\times M_3$, 
the supersymmetric on-shell action is given by 
\be\label{susy_onshell_action}
S_{\rm susy}\ =\ \frac{\beta \cpsi^2}{27\kappa_5^2} \int_{M_3} {\rm vol}_3\, R_{2d} \ ,
\ee
where $\cpsi$ is fixed as in \eqref{choice_cpsi}.
Note that because of the dependence on $\cpsi^2$, $S_{\rm susy}$ cannot itself be written as a local term in four dimensions. 

In section~\ref{sec:5dexamples} we will show that this result precisely matches the large $N$ limit of the SCFT partition function in all known examples (and beyond).

\subsection{Twisting the boundary}\label{sec:twisted_bckgnd}

We can easily discuss a slightly more general class of solutions, having different boundary geometry. 
This is obtained by making the local change of coordinates
\be\label{twisting}
\tau \ \to\ \cos\alpha\;\tau\ , \qquad  \psi \ \to \ \psi + \sin\alpha\; \tau\ ,
\ee
where $0< \alpha <\pi/2$ is a real parameter.\footnote{In Lorentzian signature, the change of coordinates reads $t\to \cosh \alpha_{\rm L} \, t$, $\ \psi \to \psi + \sinh\alpha_{\rm L}\, t$, with $\alpha_{\rm L}$ constant. This is related to \eqref{twisting} by $t=-\ii \tau$ and $\alpha_{\rm L}=\ii\alpha$.\label{foot:Lorentzian_twist}}
Then the old boundary metric and gauge field \eqref{bckgnd_metric}, \eqref{A4d} become
\begin{align}\label{bdrymetric_twisted}
\diff s^2_4 
\ &= \ \left(\diff \tau +\sin\alpha\, (\diff \psi + \phizero)\right)^2 +\cos^2 \alpha\,(\diff \psi + \phizero)^2 + 4\me^{\wzero}\diff z \diff\bar z\ ,\\[1mm]
A \, &= \, \left(\ii\cos\alpha+ 2\sin\alpha \right)\frac{\wone}{8} \diff \tau + \frac{\wone}{4} (\diff \psi + \phizero) + \frac{1}{4}*_2\! \diff \wzero \nn\\[1mm]
\,&\ \quad  + (\cpsi\sin\alpha-\ii\ct \cos\alpha)\diff \tau + \cpsi \,\diff \psi + \diff \lambda \label{bdryA_twisted}\ .
\end{align}
Although this configuration is locally equivalent to the original one, if we take for the new coordinates the same identifications as for the old ones (in particular $\tau \sim \tau + \beta$, $\psi\sim \psi$ as one goes around the $S^1$ parameterized by $\tau$ one full time), then the new boundary geometry with $\alpha\neq 0$ is globally distinct from the original one. From \eqref{bdrymetric_twisted} we see that the $S^1$ parameterized by $\tau$ is fibered over $M_3$, although in a topologically trivial way since $\diff \psi + a$ is globally defined; moreover, the term $(\diff \psi + a)^2$ in the $M_3$ part of the metric is rescaled by a factor $\cos^2\alpha$. We will denote as ``twisted'' the new four-dimensional background~\eqref{bdrymetric_twisted}, \eqref{bdryA_twisted}, as well as the corresponding five-dimensional solution obtained by implementing the transformation~\eqref{twisting} in the bulk.\footnote{An equivalent description would be to maintain the metric and gauge field \eqref{bckgnd_metric}, \eqref{A4d} and modify the identifications for the periodic coordinates, so that going around the circle parameterized by $\tau$ also advances the coordinate $\psi$ in $M_3$. This is what is commonly known as twisting, see {\it e.g.}~\cite{Closset:2013vra}.} In fact we can show that the complex structure of the twisted boundary is inequivalent to the complex structure with $\alpha=0$. Recall from section~\ref{sec:4dbackgrounds} that four-dimensional field theory backgrounds with two supercharges of opposite R-charge admit a globally defined, complex Killing vector $K$ of Hodge type $(0,1)$ with respect to two complex structures $I_\pm$. For our untwisted background, this was given in~\eqref{K4d_explicit}. For the twisted background, and in terms of a coordinate $\tilde\tau = \tau/\beta$ with canonical unit periodicity, it reads
\be\label{K_twisted}
K \ = \ \frac{1}{2\beta\cos\alpha}\left( \beta\me^{\ii\alpha}\,\partial_\psi - \ii\, \partial_{\tilde \tau}\right)\,.
\ee
We infer that $\beta \me^{\ii \alpha}$ is a complex structure parameter of the background (while the overall factor in $K$ does not affect the complex structure). Depending on the specifics of $M_3$, the background may admit additional complex structure moduli, however the one discussed here is a universal modulus of manifolds with $S^1 \times M_3$ topology and metric~\eqref{bdrymetric_twisted}.

The results of~\cite{Closset:2013vra} then imply that the supersymmetric partition function on the twisted background should be related to the one on the untwisted background by replacing $\beta \to \beta \me^{\ii\alpha}$. 
It would be interesting to check this expectation by an explicit localization computation. To date, only partial localization computations have been carried out for four-dimensional supersymmetric field theories on similarly twisted backgrounds~\cite{Closset:2013sxa}.\footnote{In \cite{Assel:2014paa} the two complex structure parameters of primary Hopf surfaces were assumed real, however in appendix D therein it was discussed how to generalize the background so that these take complex values. It would be interesting to evaluate the partition function of general supersymmetric gauge theories on such backgrounds.}

We can compare with the on-shell action of the twisted bulk solutions. This is evaluated in the same way as for $\alpha=0$, with just two differences: $i)$ the volume form on $M_3$ is  rescaled by a factor $\cos\alpha$, and $ii)$ the boundary Killing spinors are independent of the new time coordinate for a different value of $\ct$: as discussed in appendix~\ref{app:Killingspinors}, now we must take
\be\label{choice_ct}
\ct \ =\ -\ii \,\cpsi\tan\alpha\ .
\ee 
Starting from \eqref{ResultOnShActEuclidean} it is thus easy to see that the net result of  the twist by $\alpha$ is to multiply the on-shell action of the untwisted solution by a phase:
\be\label{onshact_twisted}
S_{\rm susy,\,\alpha} \ =\ \me^{\ii\alpha}\, S_{\rm susy,\, \alpha=0}\ ,
\ee
where $S_{\rm susy,\, \alpha=0}$ is given by~\eqref{susy_onshell_action}.
Here the imaginary part is a consequence of the choice of $\ct$, that is of the way the terms depending on large gauge transformations $A \to A + \, {\rm const}\, \diff \tau$ are fixed in the on-shell action.
Effectively, the phase $\me^{\ii\alpha}$ can be seen as a complexification of $\beta$. So we find that the twisting has the same consequence for the on-shell action as expected for the field theory partition function: the parameter $\beta$ is replaced by~$\beta \me^{\ii\alpha}$.

Besides being interesting {\it per se}, this complexification of the on-shell action will serve as a tool for computing the charges below.

\subsection{Conserved charges}\label{sec:holocharges}

We now compute the holographic conserved charges taking into account the contribution of the new counterterms $\Delta S_{\rm new}$ and verify that they satisfy the expected BPS condition.

Let us first consider the currents defined by standard holographic renormalization. Recall from~\eqref{conseqs_susy} that the R-current $j^i$ is conserved and thus provides a conserved R-charge. In addition, given any boundary vector $v$ preserving the boundary fields, {\it i.e.} such that $\mathcal{L}_v g =\mathcal{L}_v A =0$, we can introduce the current
\be\label{conscurrent}
Y^i \, = \, v^j (T_j{}^i + A_j j^i)\ .
\ee
Using the modified conservation equation of the energy-momentum tensor in~\eqref{conseqs_susy}, it is easy to see that $Y^i$ is conserved and thus defines a good  charge for the symmetry associated with $v$.

Although we do not know how exactly the new counterterms affect the energy-momentum tensor (because we do not know the variation of $\Delta S_{\rm new}$ with respect to the metric), we will show how the relevant charges can be computed anyway by varying the on-shell action with respect to appropriate parameters.
We will just need to assume that $\Delta S_{\rm new}$ {\it can} be expressed as a quantity invariant under diffeomorphisms and small gauge transformations, constructed from the boundary metric and the boundary gauge field (and necessarily other boundary fields), so that the chain rule~\eqref{deltaS_general} and the conservation equations make sense also after $S$ is replaced by $S_{\rm susy}$, and $T_{ij}$, $j^i$ are replaced by their supersymmetric counterparts defined by varying $S_{\rm susy}$.

We will discuss the charges for the untwisted background with $\alpha=0$, although it would be straightforward to extend this to general $\alpha$. The background with $\alpha \neq 0$ will however play a role in the computation of the angular momentum.

\paragraph{R-charge}
The supersymmetric holographic R-charge is defined as
\be
Q_{\rm susy} \ = \  -\int_{M_3} {\rm vol}_3\, j^t_{\rm susy} \ = \ -\ii \int_{M_3} {\rm vol}_3\, j^\tau_{\rm susy}\, \ ,
\ee
where
\be
j^i_{\rm susy} \ = \ j^i + \Delta j^i
\ee
is the sum of the current \eqref{ExpressionRcurrent}, evaluated in a minimal holographic renormalization scheme, and
\be
\Delta j^i \ = \  \frac{1}{\sqrt{g}} \frac{\delta}{\delta A_i} \Delta S_{\rm new}\ .
\ee
Using \eqref{ExpressionRcurrent}, the former contribution is found to be
\begin{align}
 \int_{M_3} {\rm vol}_3\,j^t  \ &= \  \frac{2}{\sqrt 3\kappa_5^2} \,\Theta + \frac{1}{108\kappa_5^2}\int_{M_3}\diff \psi \wedge \diff \left[\wone(4\diff \lambda-4 \cpsi \phizero + *_2 \diff \wzero)\right]\nn\\[1mm]
& \hspace{7mm} +  \frac{1}{216\kappa_5^2} \int_{M_3} {\rm vol}_3 \left( 8 \cpsi R_{2d} + 4\wone R_{2d} - \wone^3 \right) \ ,
\end{align}
where $\Theta$ is again given by expression~\eqref{GlobConstr}.
Both $\Theta$ and the other integral in the first line vanish due to the global assumptions we made in section~\ref{sec:5d_onshell_action}, so the R-charge in a minimal holographic renormalization scheme is given by the second line only. The shift in the current due to the new counterterms can be read from \eqref{new_cttrm_split1}, \eqref{new_cttrm_split2} and leads to 
\be\label{Deltajt}
 \int_{M_3}{\rm vol}_3\,\Delta j^t  \ = \ \frac{1}{216\kappa_5^2}\int_{M_3}\,{\rm vol}_3\left( -4\wone R_{2d} + \wone^3 \right)\ .
\ee
Adding the two contributions up, the expression for the supersymmetric holographic R-charge simplifies to
\be\label{charge_improved}
Q_{\rm susy} \ = \ - \frac{\cpsi }{27\kappa_5^2} \int_{M_3}  {\rm vol}_3\, R_{2d}\,=\, -\frac{1}{\beta\cpsi} S_{\rm susy}\,.
\ee
We notice that a faster way to arrive at the same result is to take the derivative $\frac{1}{\beta}\frac{\partial}{\partial\ct}$ of the action~\eqref{ResultOnShActEuclidean}. Indeed, a variation 
of the parameter $\ct$ amounts to shift by a constant the time component of the gauge field, which computes the electric charge. 

\paragraph{Energy}

We define the energy $H$ of the supergravity solution as the charge associated with the Killing vector $\partial_t$ (or $\partial_\tau$ in Euclidean signature). This is given by
\be\label{Ward1}
H \, = \, \int_{M_3} {\rm vol}_3 \left( T_{tt}  +  A_t j_t  \right) \,=\, \int_{M_3} {\rm vol}_3 \left( T_{\tau\tau}  +  A_\tau j_\tau  \right)\ .
\ee
Since we wish to compute the supersymmetric energy, we need to use the supersymmetric versions of the energy-momentum tensor and R-current, which receive contributions from the new boundary terms $\Delta S_{\rm new}$. Although we do not know the contribution to the holographic energy-momentum tensor, we notice that the chain rule~\eqref{deltaS_general} implies that $H$ is obtained by simply varying the on-shell action with respect to $\beta$. This is easily seen by rescaling $\tau$ so that it has fixed unit periodicity while $\beta$ appears in the expressions for the metric and gauge field. 
Hence we obtain
\be
H_{\rm susy} \ = \ \frac{\partial}{\partial\beta}S_{\rm susy} \ = \ \frac{ 1}{\beta} S_{\rm susy} \ .
\ee

\paragraph{Angular momentum}
 We denote as angular momentum the charge associated with $-\partial_\psi$. This is given by
\be\label{defangmom}
J \, = \,  -\int_{M_3} {\rm vol}_3 \left( T_{t\psi} +   A_\psi j_t \right) \,=\,  \ii \int_{M_3} {\rm vol}_3 \left( T_{\tau \psi} +  A_\psi j_\tau \right) ~.
\ee
Again we can circumvent the problem that we do not know how $\Delta S_{\rm new}$ affects the energy-momentum tensor by varying the supersymmetric on-shell action with respect to a parameter. In this case the relevant parameter is  $\alpha$ introduced via the twisting transformation of section~\ref{sec:twisted_bckgnd}. Using the chain rule \eqref{deltaS_general} and recalling \eqref{bdrymetric_twisted}, \eqref{bdryA_twisted}, we find that the variation of the on-shell action with respect to $\alpha$ (keeping $\ct$ fixed) gives:
\begin{align}\label{delS_delalpha}
 \frac{\partial }{\partial \alpha} S_{\rm susy}\bigg|_{\alpha=0}  \,&=\, \ \int \diff^4 x \sqrt{g} \left( T_{\tau\psi} + A_\psi j_\tau \right)_{\alpha=0} \, =\, \ -\ii\beta J_{\rm susy}\ ,
\end{align}
where as indicated all quantities are evaluated at $\alpha =0$, namely in the original, untwisted, background. On the other hand, we can vary the explicit expression for $S_{\rm susy}$. Since $\ct$ is kept fixed, we just need to vary the overall factor $\cos\alpha$. This gives  $\frac{\partial }{\partial \alpha} S_{\rm susy}|_{\alpha=0}=0$ and thus we conclude that
\be
J_{\rm susy} \, = \, 0\ ,
\ee
that is all untwisted solutions have vanishing angular momentum.

\paragraph{BPS relation}
In summary, we obtained the following expressions for the holographic charges associated with our supersymmetric, untwisted solutions:
\be\label{holo_charges_summary1}
H_{\rm susy} \ =\  - \cpsi \, Q_{\rm susy} \ = \ \frac{1}{\beta}S_{\rm susy}\ ,\qquad\qquad J_{\rm susy} \ =\ 0\ .
\ee
Via the AdS/CFT correspondence, these should be identified with the vacuum expectation values of the dual SCFT operators. The SCFT superalgebra implies that the latter satisfy the BPS relation
\be
\langle H \rangle +  \langle J \rangle + \cpsi  \langle Q\rangle  \ = \ 0\ ,
\ee
see appendix~\ref{app:Killingspinors} for its derivation. Of course, here it is assumed that the vacuum expectation values are computed in a supersymmetric scheme. We see that the holographic charges \eqref{holo_charges_summary1} do indeed satisfy the condition. This can be regarded as a further check that the proposed boundary terms $\Delta S_{\rm new}$ restore supersymmetry.

\section{Examples in five dimensions}\label{sec:5dexamples}

We now discuss some examples of increasing complexity. This will offer the opportunity to illustrate further the role of the new boundary terms and make contact with the existing literature.

\subsection{AdS$_5$}

It is instructive to start by discussing the simplest case, that is AdS$_5$ space. 

Euclidean AdS$_5$ is just five-dimensional hyperbolic space. In global coordinates, the unit metric can be written as
\be\label{metric_AdS5}
\diff s^2_{5} \ = \ \frac{\diff{\rho}^2}{\rho^2} +  \left(\frac{1}{\rho}+ \frac{\rho}{4r_3^2}\right)^2\diff \tau^{\,2} + \left(\frac{1}{\rho}- \frac{\rho}{4r_3^2}\right)^2 \diff s^2_{S^3} \ ,
\ee
where
\be\label{metric_S3r3}
\diff s^2_{S^3}\ =\ \frac{r_3^2}{4}\left[ \big(\diff \tilde\psi + \cos\theta \diff\varphi\big)^2+ \diff \theta^2 + \sin^2\theta\diff \varphi^2 \right]
\ee
is the round metric on a three-sphere of radius $r_3$, with canonical angular coordinates $\theta\in [0,\pi]$, $\varphi\in[0, 2\pi]$, $\tilde\psi\in [0, 4\pi]$. Here $\rho$ is a Fefferman-Graham radial coordinate, extending from the conformal boundary at $\rho=0$ until $\rho=2r_3$, where the three-sphere shrinks to zero size. The conformal boundary is $\mathbb{R}\times S^3$, equipped with the conformally-flat metric
\be\label{bdrymetric_AdS5}
\diff s^2_4 \ = \ \diff \tau^{\,2} +  \diff s^2_{S^3} \ .
\ee
We compactify the Euclidean time so that $\tau \sim \tau+\beta$ and the boundary becomes $S^1_\beta\times S^3_{r_3}$.
For the relevant Killing spinors to be independent of time, we need to switch on a flat gauge field on $S^1$,
\be\label{graviphoton_AdS5}
-\sqrt 3 \mathcal{A} \ = \ A \ = \ - \frac{\ii}{2 r_3}\, \diff \tau\ .
\ee

It is natural to assume that AdS$_5$ is dual to the vacuum state of a SCFT living on the conformal boundary $S^1_\beta\times S^3_{r_3}$.\footnote{The possibility that a different asymptotically AdS supergravity solution may be dual to the SCFT vacuum on $S^1_\beta\times S^3_{r_3}$ was considered in~\cite{Cassani:2015upa}. The analysis of that paper, though not exhaustive, indicates that this is not the case, and strongly suggests that AdS is the natural candidate.} In the following we illustrate how {\it the on-shell action and the holographic charges of AdS$_5$ match the SCFT supersymmetric vacuum expectation values only after holographic renormalization is supplemented with our new boundary terms}.

In the standard scheme of section~\ref{Stand_Holo_Reno}, the renormalized on-shell action and holographic energy are found to be
\be
S \ = \ \beta H \ = \ \frac{3(1-96\varsigma)\beta}{4r_3}\frac{\pi^2}{\kappa_5^2}\ ,
\ee
while both the angular momentum $J$ and the holographic R-charge $Q$ vanish. $Q=0$ follows from formula~\eqref{ExpressionRcurrent} using $\mathcal F=0$. 
Thus, by dialing $\varsigma$ the holographic energy $H$ may be set either to agree with $Q=0$, so that the BPS condition stating the proportionality between energy and charge is satisfied, or with the field theory result in~\eqref{Esusy_round}, but not with both. Hence even in the simple example of AdS we see that standard holographic renormalization disagrees with the supersymmetric field theory results.

Let us describe how this discrepancy is solved by the new terms introduced in section~\ref{sec:new_dry_terms}.
Starting from the general boundary geometry~\eqref{bckgnd_metric}, \eqref{A4d} we take $\wone = {\rm const} = -\frac{4}{r_3}$, $\me^{\frac{\wzero}{2}}= \frac{r_3}{2} \frac{1}{1+|z|^2}$, and make the change of coordinate $z = \cot \frac{\theta}{2}\, \me^{-\ii \varphi}$, $\psi =\frac{r_3}{2}\tilde\psi$.  
Then the two-dimensional metric, its curvature and the volume form are 
\be\label{2d_metric_AdS}
\diff s^2_{2}= \frac{r_3^2}{4}(\diff \theta^2 + \sin^2\theta \diff\varphi^2)\ , \qquad R_{2d} = \frac{8}{r_3^2}\ , \qquad {\rm vol}_{2} = \frac{r_3^2}{4}\sin\theta\, \diff \theta\wedge \diff \varphi\ ,
\ee
and eq.~\eqref{eq:3dBoundaryDa} for the connection one-form $\phizero$ is solved by $\phizero = \frac{r_3}{2}\cos\theta\diff\varphi$. Moreover to recover the correct gauge field we need to take
\be
\cpsi \, = \, \frac{1}{r_3}\  ,\qquad\qquad \ct \,=\, 0\ , \qquad\qquad \lambda=-\frac{\varphi}{2}\ ,
\ee
the value of $\cpsi$ being in agreement with~\eqref{choice_cpsi}.
In this way our general boundary metric and gauge field reduce to \eqref{bdrymetric_AdS5}, \eqref{graviphoton_AdS5}.

The new boundary terms~\eqref{new_cttrm_split1} then evaluate to (after Wick rotation):
\be
\Delta S_{\rm new} \, = \, -\frac{17\beta}{108r_3}\frac{\pi^2}{\kappa_5^2}\ ,
\ee
so that we obtain for the supersymmetric on-shell action of AdS$_5$: 
\be\label{SusyOnShActAdS}
S_{\rm susy} \ = \ S_{\varsigma = 0} + \Delta S_{\rm new} \ = \   \frac{16 \,\beta}{27 r_3}\frac{\pi^2}{\kappa_5^2}\ .
\ee
This result also follows directly from~\eqref{susy_onshell_action} since AdS$_5$ satisfies all global assumptions that were made in section~\ref{sec:5d_onshell_action} to derive it.\footnote{For generic asymptotically AdS solutions, conformal flatness of the boundary metric~\eqref{bckgnd_metric} on $S^1_\beta\times M_3$ amounts to $\wone = {\rm const}$ and $R_{2d}=\frac{\wone^2}{2}$;  it also implies $\diff A=0$. Then from~\eqref{choice_cpsi} we find
$\cpsi = -\frac{\wone}{4}$. If the solution satisfies the global assumptions made in section~\ref{sec:5d_onshell_action}, our formula~\eqref{susy_onshell_action} applies and the supersymmetric on-shell action reads
\begin{equation*}
S_{\rm susy}\ =\ \frac{\beta \wone^4}{2^5 3^3 \kappa_5^2} \int_{M_3}{\rm vol}_3\ .
\end{equation*}
For a round sphere $M_3 \cong S^3_{r_3}$, we set $\wone = -\frac{4}{r_3}$, $\int_{S^3} {\rm vol}_3 = 2\pi^2 r_3^3$ and the result~\eqref{SusyOnShActAdS} follows.} Then the energy is just $H=\frac{1}{\beta}S_{\rm susy}$ and the angular momentum vanishes, $J=0$.

Using eq.~\eqref{Deltajt}, we see that the new terms also shift the value of the holographic R-charge from zero to
\be
Q_{\rm susy} \ = \ -\frac{16}{27}\frac{\pi^2}{\kappa_5^2}\ .
\ee

Therefore we have found for the supersymmetric energy, charge and angular momentum:
\be\label{chargesAdS5}
H_{\rm susy} \ = \ -\frac{1}{r_3}Q_{\rm susy} \ = \ \frac{16}{27 r_3}\frac{\pi^2 }{\kappa_5^2}\ , \qquad\qquad J_{\rm susy}=0\ .
\ee
Besides respecting the BPS condition, these values precisely match the supersymmetric field theory vacuum expectation values of~\cite{Assel:2014paa,Assel:2015nca}, {\it cf.}\ eq.~\eqref{Esusy_round} for the energy.

It is worth pointing out that the choice~\eqref{graviphoton_AdS5} for the flat gauge field does not affect the conserved charges of AdS$_5$ computed via standard holographic renormalization, while it plays a crucial role in our new boundary terms. Indeed in the formulae of section~\ref{Stand_Holo_Reno} the only term potentially affected by a flat gauge connection is the bulk Chern-Simons term $\int \mathcal{A}\wedge \mathcal{F}\wedge \mathcal{F}$, which however vanishes in AdS$_5$ as $\mathcal{F}=0$. On the other hand, $\Delta S_{\rm new}$ in~\eqref{new_cttrm_split1} depends on a flat connection on $S^1$ since the three-form $\Phi$ does not vanish on the $S^3$ at the boundary of AdS$_5$, and this affects the holographic charges. In particular, it gives the full answer for the holographic R-charge associated with AdS$_5$.

\subsection{Twisted AdS$_{5}$}

We can take advantage of the very explicit example of AdS$_5$ to further illustrate the twisting of section~\ref{sec:twisted_bckgnd}.

Starting from the AdS$_5$ metric~\eqref{metric_AdS5}, \eqref{metric_S3r3} we make the change of coordinates
\be\label{twist_AdS5}
\tau \ \to\ \cos\alpha\;\tau\ , \qquad  \tilde \psi \ \to \  \tilde \psi + \frac{2}{r_3}\sin \alpha\,\tau\ ,
\ee
with $0< \alpha <\pi/2$. Then the new bulk metric reads
\begin{align}
\diff s^2_{5}  \, &=\, \frac{\diff{\rho}^2}{\rho^2} +  \left(\frac{1}{\rho}+ \frac{\rho}{4r_3^2}\right)^2\cos^2\alpha\,\diff \tau^{\,2} \nn\\[1mm]
& \,\quad+ \left(\frac{1}{\rho}- \frac{\rho}{4r_3^2}\right)^2 \frac{r_3^2}{4}\left[\Big(\diff\tilde\psi + \frac{2}{r_3}\sin\alpha\,\diff\tau+ \cos\theta\diff\varphi \Big)^2+ \diff \theta^2+\sin^2\theta\diff\varphi^2 \right] .
\end{align}
The new boundary metric may be written as
\be\label{metric_newcoords}
\diff s^2_4 
=  \left[\diff \tau + \frac{r_3}{2}\sin \alpha\big(\diff \tilde\psi +  \cos\theta \diff\varphi\big) \right]^{2} + \frac{r_3^2}{4}\left[ \cos^2 \alpha\big(\diff \tilde\psi + \cos\theta \diff\varphi\big)^2 + \diff \theta^2+\sin^2\theta\diff\varphi^2  \right] .
\ee
Since we do not transform the range of the coordinates, {\it i.e.}\ we take $\tau \in [0,\beta]$, $\tilde\psi \in [0,4\pi]$ also after the transformation, the new geometry is globally distinct from the original one.
However, both the boundary and the bulk metric remain regular.\footnote{Regularity of the boundary metric follows from the fact that $\diff \tilde\psi +  \cos\theta \diff\varphi$ is globally defined. Regularity of the bulk metric $G_{\mu\nu}$ as $\rho \to 2r_3$ can be seen by noting that the $G_{\tau\tau}$ component  remains finite, that the components $G_{\rho\rho}, G_{\theta\theta}, G_{\varphi\varphi}, G_{\tilde\psi\tilde\psi}$ and $G_{\tilde\psi\varphi}$ asymptote to the metric on the cone on a round $S^3$ ({\it i.e.}\ the flat metric on $\mathbb{R}^4$), and finally that the $G_{\tau\varphi}, G_{\tau\theta}$ components go to zero. It follows that as $\rho \to 2r_3$ the space looks like $S^1 \times \mathbb{R}^4$.}
The choice of boundary gauge field $A$ ensuring that the Killing spinors are independent of the new time coordinate on $S^1$ was explained in section \ref{sec:twisted_bckgnd}, {\it cf.}\ eqs.~\eqref{bdryA_twisted}, \eqref{choice_ct}.  For AdS$_5$ this also corresponds to the bulk gauge field: 
\be\label{A_twisted_AdS5}
-\sqrt 3\mathcal{A} \ = \ A \ = \ \frac{\ii}{2r_3} \left(-\cos \alpha + 2\ii \sin \alpha\right) \diff \tau\ .
\ee
Note that this has both a real and an imaginary part.

The on-shell action in the standard holographic scheme is found to be
\be
S \ = \  \cos\alpha\,\frac{3(1-96\varsigma)\beta}{4r_3}\frac{\pi^2}{\kappa_5^2}\ ,
\ee
 as the only consequence of the twist in the computation is to rescale the volume by $\cos\alpha$.
The new boundary terms~\eqref{new_cttrm_split1} are evaluated as for untwisted AdS$_5$, except that one must implement the transformation~\eqref{twist_AdS5} and use the gauge field~\eqref{A_twisted_AdS5}. This gives
\be
\Delta S_{\rm new}\ = \ \left(-\frac{17}{108}\cos \alpha + \frac{16}{27}\,\ii \,\sin \alpha \right)\frac{\beta}{r_3}\frac{\pi^2}{\kappa_5^2}\ .
\ee
Then the supersymmetric on-shell action evaluates to
\be
S_{\rm susy} \ = \ S_{\varsigma = 0} + \Delta S_{\rm new}\ = \  \frac{16\beta\,\me^{\ii\alpha}}{27r_3}\frac{\pi^2}{\kappa_5^2} \ .
\ee
This illustrates in a concrete example the general result of section~\ref{sec:twisted_bckgnd} that the on-shell action in the twisted background is related to the one in the untwisted background by the replacement $\beta\to \me^{\ii \alpha}\beta$.

\subsection{A simple squashing of AdS$_5$}

A different one-parameter supersymmetric deformation of AdS$_5$ was presented in~\cite{Cassani:2014zwa}. In this solution, the boundary geometry is non conformally flat as  $S^3\subset\,\partial$AdS$_5$ is squashed. The squashing is such that the Hopf fibre of $S^1\hookrightarrow S^3\rightarrow S^2$ is rescaled with respect to the $S^2$ base by a parameter $v$, which defines a Berger sphere $S^3_v$ with $SU(2)$-invariant metric. The boundary metric then reads
\be\label{boundary_Berger}
\diff s^2_4 \ = \ \diff \tau^2 + \frac{r_3^2}{4}\left[  v^2\big(\diff \tilde\psi + \cos\theta\diff\varphi\big)^2 +  \diff \theta^2 + \sin^2\theta \diff \varphi^2  \right]\,,
\ee
which for $v=1$ reduces to~\eqref{metric_S3r3}, \eqref{bdrymetric_AdS5}.
The boundary geometry is controlled by the three parameters $\beta,r_3,v$, however the complex structure on the boundary is determined just by the ratio $\frac{\beta}{v r_3}$ specifying the relative size of $S^1_\beta$ to the Hopf fibre, hence the supersymmetric field theory partition function depends on $\beta,r_3,v$ only through this combination~\cite{Closset:2013vra,Assel:2014paa}.

Similarly to the solutions in section~\ref{5d_solution}, the supergravity solution of~\cite{Cassani:2014zwa} was constructed in Lorentzian signature and then analytically continued so that the boundary is Riemannian, while the bulk metric becomes complex. It is known analytically at first order in the squashing and numerically for finite $v$. While we refer to~\cite{Cassani:2014zwa} for more details, here it will be sufficient to mention that the solution is regular and such that the global assumptions made in section~\ref{sec:5d_onshell_action} to derive the on-shell action formula~\eqref{susy_onshell_action} are satisfied. 
In fact, as already mentioned, the strategy followed in section~\ref{sec:5d_onshell_action} is a generalization of the one in~\cite{Cassani:2014zwa}.
Since its near-boundary behaviour falls in the larger family of perturbative solutions constructed in the present paper, the solution of~\cite{Cassani:2014zwa} also provides a concrete example that the latter can admit a smooth completion in the interior also when the boundary is not conformally flat.

While the field theory results predict that the on-shell action only depends on the ratio $\frac{\beta}{vr_3}$, it was found in~\cite{Cassani:2014zwa} that after performing standard holographic renormalization this depends both on $\frac{\beta}{vr_3}$ and $v$. Indeed, in a minimal scheme where the finite counterterms~\eqref{ambij} are set to zero one obtains\footnote{{\it cf.} eq.\ (4.15) of~\cite{Cassani:2014zwa}. The present variables are obtained setting $\Delta_t^{\rm there} = \frac{v}{r_3}\beta$ and $\frac{8\pi G}{\ell^2} = \kappa_5^2$.}
\be\label{S_squashedpaper}
S_{\rm min} \ = \ \frac{8v\beta}{r_3} \left( \frac{2}{27v^2} + \frac{2}{27} - \frac{13}{108}v^2 + \frac{19}{288}v^4 \right) \frac{\pi^2}{\kappa_5^2} \ ,
\ee
so only the first term in parenthesis yields the correct dependence on $\frac{\beta}{vr_3}$.
In addition, it was shown in~\cite[sect.$\:$5.3]{Cassani:2014zwa} that there is no combination of the ordinary finite counterterms~\eqref{ambij} that cancels all but the first term in~\eqref{S_squashedpaper}. 
 It was then proposed that a new counterterm should be added, and it was found that a certain term involving the Ricci form, combined with the standard finite counterterms, does the job ({\it cf.}\ eq.\ (5.51) therein). However, in the light of our more general analysis that specific prescription turns out incorrect, as the proposed term does not evaluate to $\Delta S_{\rm new}$ in~\eqref{newCttrms} for the more general boundary metric and gauge field considered in the present paper. This also follows from the fact that the term proposed in~\cite{Cassani:2014zwa} is gauge invariant, while in order to adjust the holographic R-charge so that the BPS condition is satisfied a dependence on large gauge transformations is needed.
  Therefore while the idea of correcting the holographic renormalization scheme by new boundary terms survives and is much strengthened by the general analysis performed in the present paper, a covariant form for these terms remains to be found.

Let us show how $\Delta S_{\rm new}$ removes the terms in~\eqref{S_squashedpaper} not depending solely on $\frac{\beta}{vr_3}$.
The metric~\eqref{boundary_Berger} on  $S^1 \times S^3_v$ is obtained from our general boundary metric~\eqref{bckgnd_metric} by modifying slightly the transformations made for the example of AdS$_5$. Again we take $\me^{\frac{\wzero}{2}}= \frac{r_3}{2} \frac{1}{1+|z|^2}$ and $z = \cot \frac{\theta}{2}\, \me^{-\ii \varphi}$, so that the two-dimensional formulae~\eqref{2d_metric_AdS} hold the same. Choosing $\wone = -\frac{4v}{r_3}$, the connection one-form $\phizero$ can be taken $\phizero = \frac{vr_3}{2}\cos\theta\diff\varphi$, while the coordinate on the Hopf fibre with canonical period $4\pi$ is $\tilde\psi = \frac{2}{v r_3}\psi$. In this way~\eqref{bckgnd_metric} reduces to~\eqref{boundary_Berger}. Also choosing
\be
\cpsi \ =\ \frac{1}{vr_3}\ ,\qquad\qquad \ct \,=\, 0\ , \qquad\qquad \lambda=-\frac{\varphi}{2}\ ,
\ee
where again the value of $\cpsi$ is in agreement with~\eqref{choice_cpsi}, the boundary gauge field~\eqref{A4d} reduces to the $SU(2)$-invariant expression\footnote{These boundary fields agree with those of~\cite{Cassani:2014zwa} upon identifying $\psi^{\rm there} = \tilde\psi$, $t^{\rm there} = \frac{iv}{r_3}\tau$ and $a_0^{\rm there} = \frac{r_3}{2}$.}
\be
-\sqrt 3 A^{(0)} \ = \ A \ = \   -\frac{\ii\, v}{2r_3}\diff \tau + \frac{1}{2}(1-v^2) (\diff \tilde\psi + \cos\theta \diff \varphi) \ .
\ee

Then our formula~\eqref{susy_onshell_action} for the supersymmetric on-shell action evaluates to
\be\label{Ssusy_BergerSph}
S_{\rm susy} \ = \  \frac{16\beta}{27v r_3} \frac{\pi^2}{\kappa_5^2} \ ,
\ee
that only depends on $\frac{\beta}{vr_3}$ as predicted by the field theory arguments. In fact
our new counterterms evaluate to 
\be
\Delta S_{\rm new} \, = \, -\frac{8v \beta}{r_3} \left( \frac{2}{27} - \frac{13}{108}v^2 + \frac{19}{288}v^4 \right)\frac{\pi^2}{\kappa_5^2}\ ,
\ee
which precisely accounts for the difference between \eqref{S_squashedpaper} and \eqref{Ssusy_BergerSph}. One could also consider twisting this five-dimensional solution by the parameter $\alpha$ as discussed in section~\ref{sec:twisted_bckgnd} and further illustrated in the example of AdS$_5$, thus introducing an overall phase $\me^{\ii\alpha}$ in the on-shell action.

Eq.~\eqref{holo_charges_summary1} gives for the holographic charges:
\be\label{susycharges_BergerSph}
H_{\rm susy} \, = \, -\frac{1}{vr_3} Q_{\rm susy} \, = \, \frac{16}{27 vr_3}\frac{\pi^2}{\kappa_5^2} \ ,
\qquad\qquad J_{\rm susy}\, =\, 0 \ .
\ee
The electric charge given in~\cite[sect.$\:$4]{Cassani:2014zwa} reads in the present normalization
\be
Q^{\rm there} \ = \  -\frac{16\pi^2}{27 \kappa_5^2}(v^2-1)^2\ ,
\ee 
while the shift \eqref{Deltajt} due to our new boundary terms evaluates to
\be
\Delta Q \ = \ -\int{\rm vol}_3\,\Delta j^t  \ = \  \frac{16\pi^2}{27 \kappa_5^2}(v^4-2v^2)\ ,
\ee
therefore $Q^{\rm there} + \Delta Q$ matches the supersymmetric charge in \eqref{susycharges_BergerSph}.
When comparing~\eqref{susycharges_BergerSph} with the energy and angular momentum computed in~\cite{Cassani:2014zwa} one needs to take into account both the contribution of the new boundary terms and the fact that in~\cite{Cassani:2014zwa} these quantities were defined in terms of the energy-momentum tensor alone (which for the present solution still yields conserved quantities), while here we presented the charges \eqref{Ward1}, \eqref{defangmom} computed from the current~\eqref{conscurrent} that is always conserved in the presence of a general background gauge field.

\subsection{Hopf surfaces at the boundary}\label{Hopfexample}

We can also evaluate our on-shell action formula \eqref{susy_onshell_action} for the more general boundary geometry with $S^1 \times S^3$ topology considered in~\cite{Assel:2014paa}. Contrarily to the previous examples in this section, in this case we do not have a general proof of existence of regular bulk fillings satisfying all the global properties we required in section~\ref{sec:5d_onshell_action} to evaluate the on-shell action. However, we are going to show that if we assume that such supergravity solutions exist, then eq.~\eqref{susy_onshell_action} gives the correct holographic dual of the supersymmetric Casimir energy of~\cite{Assel:2014paa,Assel:2015nca}.

In~\cite{Assel:2014paa} the three-sphere is described as a torus foliation: the torus coordinates are $\varphi_1 \in [0,2\pi]$, $\varphi_2 \in [0,2\pi]$, while the remaining coordinate is $\hat\rho \in [0,1]$.\footnote{The coordinate $\hat\rho$ is defined on the four-dimensional boundary and should not be confused with the radial coordinate $\rho$ used elsewhere in this paper.} The four-dimensional complex manifolds with topology $S^1\times S^3$ are Hopf surfaces, and in~\cite{Assel:2014paa} the complex structure moduli are two real, positive parameters $\beta b_1,\beta b_2$ (as above, $\beta$ denotes the circumpherence of the $S^1$ parameterized by $\tau$). These characterize the choice of complex Killing vector~\eqref{K4d} as
\be
K \, =\, \frac{1}{2} \left( \partial_\psi  - \ii\, \partial_\tau \right) \,=\, \frac{1}{2} \left( b_1 \partial_{\varphi_1} + b_2 \partial_{\varphi_2}  - \ii\, \partial_\tau \right)\ .
\ee
The four-dimensional metric is taken as
\begin{align}\label{4dmetrictoric}
\diff  s^2_4 \, &= \,  \Omega^2 \left[\diff\tau^2 + (\diff \psi + a_\chi \diff \chi)^2 + \Omega^{-2}f^2 \diff \hat\rho^2 + c^2 \diff \chi^2 \right] \nn \\[1mm]
\, & = \, \Omega^2 \diff\tau^2 + f^2 \diff\hat\rho^2 +  m_{IJ}   \diff\varphi_I \diff\varphi_J \ ,
\end{align}
where $I,J=1,2$. The first line is the canonical form dictated by supersymmetry (with $\diff s^2_{2}=\Omega^{-2}f^2 \diff \hat\rho^2 + c^2 \diff \chi^2$), while the expression in the second line is convenient for discussing global properties, since it uses periodic coordinates. When passing from the first to the second expressions one identifies the coordinates as
\be\label{Relphi1phi2Withpsichi}
\psi = \frac{1}{2}\left(\frac{\varphi_1}{b_1}+\frac{\varphi_2}{b_2}\right)\ , \qquad \chi = \frac{1}{2}\left(\frac{\varphi_1}{b_1}-\frac{\varphi_2}{b_2}\right)
\ee
and the functions as
\be
 a_\chi \,=\, \frac{1}{\Omega^2}\left( b_1^2 m_{11} - b_2^2 m_{22} \right)\ ,\qquad
c \,=\, \frac{2b_1b_2}{\Omega^2} \sqrt{\det m_{IJ}} \ .
\ee
Moreover supersymmetry imposes the relation
\be\label{OmegaToric}
\Omega^2 \ = \ b^I m_{IJ} b^J  \ ,
\ee
which ensures Hermiticity of the metric.
Here, $f$ and $m_{IJ}$ are functions of $\hat\rho$ satisfying suitable boundary conditions at $\hat\rho = 0$ and $\hat\rho =1$ so that the metric is regular and describes a smooth $S^3$ topology. 
As $\hat\rho \to 0$, one requires that
\be\label{conditionsrhoto0}
f \to f_2 \, , \quad m_{11} \to m_{11}(0) \,,\quad m_{22}= (f_2\hat\rho)^2 + \mathcal O( \hat\rho^3)\,,\quad m_{12} = \mathcal O(\hat\rho^2)~,
\ee
where $f_2>0$ and $m_{11}(0)>0$ are constants, and similarly for $\hat\rho\to1$ (see~\cite{Assel:2014paa}).  

In principle our on-shell action formula \eqref{susy_onshell_action} is derived for a boundary metric of the type~\eqref{bckgnd_metric}, thus with trivial conformal factor $\Omega=1$, however we now show that the same formula gives the correct result even for general $\Omega$
if it is evaluated using the metric in the square bracket of~\eqref{4dmetrictoric}.\footnote{Otherwise one can choose $m_{IJ}$ so that~\eqref{OmegaToric} is satisfied with $\Omega=1$, which is not a serious loss of generality since it still allows for general $b_1,b_2$.}

Using the expressions above, we can compute
\begin{align}
\int_{M_3}\!\! {\rm vol}_3 R_{2d} \, &= \,   -\int \partial_{\hat\rho} \left( \frac{c\,\Omega}{f}\, \partial_{\hat\rho}\log c^2  \right)\diff \hat\rho \wedge \diff\chi\wedge \diff \psi\,=\, -\frac{4\pi^2}{b_1b_2} \left[ \frac{\Omega}{f}\, \partial_{\hat\rho} c   \right]_{\hat\rho=0}^{\hat\rho=1} \nn\\[1mm]
\, &= \, 8\pi^2\frac{b_1+b_2}{b_1b_2}\ ,
\end{align}
where in the last equality we used the behaviour of the functions at the extrema of the $\hat\rho$ interval. Similarly,
\be
\int_{M_3} \eta\wedge \diff \eta \ = \  \int \partial_{\hat\rho} a_\chi\, \diff \hat\rho \wedge \diff \chi \wedge\diff \psi \ = \ \frac{2\pi^2}{b_1b_2} a_\chi\big |_{\hat\rho=0}^{\hat\rho=1} \ = \ -\frac{4\pi^2}{b_1b_2} \ .
\ee

Then formula~\eqref{choice_cpsi} for $\cpsi$ gives
\be
\cpsi \,=\, \frac{1}{2}(b_1+ b_2)
\ee 
and the on-shell action~\eqref{susy_onshell_action} evaluates to
\be
S_{\rm susy} \ = \ \frac{2\beta}{27} \,\frac{(b_1+b_2)^3}{b_1b_2}\frac{\pi^2}{\kappa_5^2}\ ,
\ee
which perfectly matches the field theory prediction~\eqref{Esusy_Hopf_gravity}.\footnote{This agrees with eq.~(5.18) of~\cite{Assel:2014paa}, upon identifying $|b_I|^{\rm there}=\frac{\beta}{2\pi}b_I^{\rm here}$ and $8\pi G^{\rm there}=\kappa_5^2$.} This result was the main point emphasized in our short communication~\cite{Genolini:2016sxe}.

\subsection{General $M_3$}

In section \ref{sec:5d_onshell_action} we derived the general formula
(\ref{susy_onshell_action}) for the supersymmetric on-shell action (evaluated with our new counterterms). 
Here the conformal boundary has topology $S^1\times M_3$, and the derivation 
of the formula requires certain global assumptions about the topology of the 
five-dimensional bulk supergravity solution that fills this boundary. In particular, 
we required the graviphoton field $\mathcal{A}$ to be a global one-form.
Particular explicit examples have been studied in the subsections above. 
In this subsection we present a more  general but abstract analysis, and 
show that our supergravity result (\ref{susy_onshell_action}) always reproduces the supersymmetric Casimir energy, 
as computed in field theory in \cite{Martelli:2015kuk}.\footnote{There are caveats to this statement, that we will clarify 
below.}

We begin by rewriting the supersymmetric on-shell supergravity action (\ref{susy_onshell_action})  
in terms of Seifert invariants of $M_3$. In particular, using equations (\ref{R2dc1}) and (\ref{eq:4dGammaFromGauge}) 
we may write
\bea\label{SsusySeifert}
S_{\rm susy} &=& \frac{2\pi^2 b\beta}{27\kappa_5^2}\frac{\left(\int_{\Sigma_2}c_1(\Sigma_2)\right)^3}{\left(\int_{\Sigma_2}c_1(\mathcal{L})\right)^2}~.
\eea
Recall here that $\psi$ has period $2\pi/b$, so that the Reeb vector $\xi=\partial_\psi=b\zetacharge$, where $\zetacharge$ is the 
normalized vector field which exponentiates to the corresponding $U(1)$ action on $M_3$.

Under the same global assumptions on $M_4\cong S^1_\beta\times M_3$, the supersymmetric Casimir energy $\Esusy$
was computed in field theory in \cite{Martelli:2015kuk}.  More precisely, in the path integral sector with 
trivial flat gauge connection on $M_3$, $\Esusy$ may be computed from an index-character that counts holomorphic 
functions on $X_0\cong \R_{>0}\times M_3$. The formula for weighted homogeneous hypersurface singularities was given in equation 
(\ref{Esusylink}), with large $N$ limit (\ref{EsusylargeNlink}). Substituting for $\int_{\Sigma_2}c_1(\Sigma_2)$ 
and $\int_{\Sigma_2}c_1(\mathcal{L})$ for  hypersurface singularities using 
formulas (\ref{c1whhs}), the supergravity result (\ref{SsusySeifert}) precisely agrees 
with the large $N$ field theory computation of $\beta\Esusy$, with $\Esusy$ given by (\ref{EsusylargeNlink})!

This agreement between exact field theory and supergravity calculations is already remarkable.
However, we can go further and present a very general derivation of this agreement, based on 
a formula for the index-character appearing in \cite{Martelli:2006yb}. 
Recall first that the $U(1)$ Seifert action on $M_3$ extends to a holomorphic 
$\C^*$ action on $X_0=\R_{>0}\times M_3$, and hence on $X=C(M_3)$. 
Following \cite{Martelli:2015kuk,Martelli:2006yb}, we denote the 
index-character that counts holomorphic functions on 
$X$ (or equivalently $X_0$) according to their weights under $q\in\C^*$
by $C(\bar\partial,q,X)$. If the $U(1)\subset \C^*$ action is \emph{free}, meaning 
that $\Sigma_2=M_3/U(1)$ is a smooth Riemann surface, then we may write
\bea
C(\bar\partial,q,X) \, &=& \,  \sum_{k\geq 0} q^k \int_{\Sigma_2}\ex^{-k c_1(\mathcal{L})}\cdot \mathrm{Todd}(\Sigma_2) \\
\,&=&\, \sum_{k\geq 0} q^k \int_{\Sigma_2} \left[-k\,c_1(\mathcal{L}) + \frac{1}{2}c_1(\Sigma_2)\right]~.
\eea
The first equality is the Riemann-Roch theorem, and the second equality uses
$\mathrm{Todd}=1+\frac{1}{2}c_1+\cdots$, where the higher order terms 
do not contribute in this dimension. We may then sum the series 
for $|q|<1$ to obtain the formula
\bea\label{charreg}
C(\bar\partial,q,X) \,&=&\, \frac{\int_{\Sigma_2}c_1(\Sigma_2) - q\left(\int_{\Sigma_2}2c_1(\mathcal{L})+c_1(\Sigma_2)\right)}{2(1-q)^2}~.
\eea
We emphasize that this formula is valid for regular Reeb vector fields, so that $\Sigma_2$ is a smooth Riemann surface, and is 
\emph{not} valid in the quasi-regular case, where $\Sigma_2$ has orbifold singularities. However, as we shall explain below, 
one may effectively still use this formula to compute the large $N$ supersymmetric Casimir energy even in the general quasi-regular case. 

The full character that computes the supersymmetric Casimir energy is given by~\cite{Martelli:2015kuk}
\bea\label{fullchar}
C(q,\lambdaNEW,X) \,&=&\, q^{-\int_{\Sigma_2}c_1(\Sigma_2)/2\int_{\Sigma_2}c_1(\mathcal{L})}\cdot\lambdaNEW\cdot C(\bar\partial, q,X)~.
\eea
Here the power of $q$ in the first factor is precisely $\gamma/b$, which arises as $\frac{1}{2}$ the charge of the holomorphic $(2,0)$-form under 
the canonically normalized vector field $\zetacharge$. The supersymmetric Casimir energy is then 
obtained by setting $q=\ex^{t b}$, $\lambdaNEW=\ex^{-tu}$, where $u=(r-1)\gamma$ for a matter multiplet 
of R-charge $r$,
 and extracting the coefficient of $-t$ in 
a Laurent series about $t=0$. For field theories with a large $N$ gravity dual 
in type IIB supergravity one has $\mathtt{a}=\mathtt{c}=\pi^2/\kappa_5^2$, where 
the trace anomaly coefficients may in turn be expressed in terms of certain cubic functions of the R-charges
$(r-1)$ of fermions. Using this prescription applied to (\ref{fullchar}), (\ref{charreg}), we find that the 
large $N$ field theory result gives
\bea\label{EsusySeifert}
\Esusy \,&=&\, \frac{2\pi^2 b}{27\kappa_5^2}\frac{\left(\int_{\Sigma_2}c_1(\Sigma_2)\right)^3}{\left(\int_{\Sigma_2}c_1(\mathcal{L})\right)^2}~,
\eea
so that the supergravity action $S_{\rm susy}$ in (\ref{SsusySeifert}) agrees with $\beta\Esusy$ computed in field theory.

Although (\ref{charreg}) only holds in the regular case, in fact this formula is sufficient to compute the correct
large $N$ supersymmetric Casimir energy in (\ref{EsusySeifert}) in the general quasi-regular case. 
The point is that when $\Sigma_2$ has orbifold singularities there are additional contributions to Riemann-Roch formula (\ref{charreg}). 
However, also as in \cite{Martelli:2006yb}, the general form of these contributions is such that they do not contribute 
to the relevant limit that gives (\ref{EsusySeifert}). Thus the latter formula holds in general (we have already shown 
independently that it holds for homogeneous hypersurface singularities, which are generically not regular).

Finally, although the agreement of the two computations is remarkable, without more work 
it is also somewhat formal. In particular, in the field theory computation we have assumed 
that the sector with trivial flat gauge connection dominates at large $N$, while  the general supergravity computation 
assumes the existence of an appropriate solution with the required global properties. Known 
examples suggest that these are not unreasonable assumptions, but 
there is clearly a need for further work to clarify how general a result this is. We leave these interesting questions for future work. 

\section{Outlook}\label{sec:outlook}

Since the early days of the AdS/CFT correspondence, it has been clear that in order to define observables holographically, infinities have to be subtracted 
\cite{Witten:1998qj,Henningson:1998gx,Balasubramanian:1999re}. These initial findings developed into the systematic framework  of holographic renormalization, which has taken  various incarnations 
\cite{deHaro:2000vlm,Bianchi:2001de,Bianchi:2001kw,deBoer:1999tgo,Martelli:2002sp,Skenderis:2002wp,Papadimitriou:2004ap,Olea:2006vd}. 
Despite the fact that this has proved to be very robust as a method for subtracting infinities in the context of AlAdS solutions,
 the problem of matching \emph{finite} boundary terms in holographic computations to choices of renormalization schemes in quantum field theory has remained a subtle question requiring further study. Recent exact results in supersymmetric quantum field theories, in part obtained through the technique of localization, have sharpened this question within a large class of holographic constructions. In this paper, we have presented a systematic study of the interplay of holographic renormalization and supersymmetry,
in the context of minimal $\mathcal{N}=2$ gauged supergravity theories in four and five dimensions. These theories are consistent truncations of eleven-dimensional and type IIB supergravity on very general classes of internal manifolds with known field theory duals. They thus give access to a vast set of examples of supersymmetric gauge/gravity dual pairs, where both sides are well understood   
 \cite{Romans:1991nq,Martelli:2011fu,Martelli:2011fw,Martelli:2012sz,Martelli:2013aqa,Huang:2014gca,Cassani:2014zwa,Farquet:2014kma,Cassani:2015upa,Genolini:2016sxe}.  

In this paper we have made certain simplifying assumptions; in particular our studies apply to AlAdS solutions of the given supergravities, where the boundary geometry admits at least 
a pair of Killing spinors.   Under these assumptions, our main results may be summarized as follows. In four-dimensional minimal $\mathcal{N}=2$ gauged supergravity, the on-shell action, renormalized using standard counterterms, is supersymmetric. In particular, as expected, we did not find any ambiguities related to finite counterterms.\footnote{This situation is radically different in supergravity models coupled to matter. The interplay of holographic renormalization and supersymmetry in the presence of scalar fields has been discussed for conformally flat boundaries in \cite{Bianchi:2001de,Freedman:2013ryh,Bobev:2013cja,Bobev:2016nua,Freedman:2016yue,Kol:2016ucd}.} In five-dimensional minimal gauged supergravity, we showed that there is \emph{no choice} of standard finite counterterms (\emph{i.e.}\ four-dimensional diffeomorphism and gauge invariants constructed with the boundary metric and graviphoton) that renders the holographically renormalized on-shell action supersymmetric. Thus, surprisingly, standard holographic renormalization breaks supersymmetry in five dimensions. We then found a specific set of new boundary terms that restores supersymmetry of the on-shell action, as well as the validity of certain supersymmetric Ward identities inferred from field theory~\cite{Closset:2013vra,Closset:2014uda}. We provided some independent tests of these new terms, illustrating their application in smooth AlAdS$_5$ solutions with topology  $\R\times \R^4$.

Although our analysis provides a very strong evidence that in order to formulate holographic renormalization in a supersymmetric fashion  a new set of boundary terms is needed, a more fundamental understanding of the origin of these terms is clearly desirable.  We emphasize that in the present work we assumed the validity of the gauge/gravity duality, and used this to obtain constraints on the gravity side from exact results originally derived on the field theory side.
It will be very interesting to perform a first principles analysis of supersymmetry of supergravities in asymptotically locally AdS space-times. Let us mention some possible avenues that could be pursued to achieve this goal. 
A direct approach to retrieve the correct boundary terms is to work on a space with a boundary at a finite distance and to impose 
that the combination of bulk plus boundary supergravity action is invariant under supersymmetry (of course the bulk action is invariant under supersymmetry \emph{up to boundary terms}). Notice that, in different situations, this approach has been recently advocated in \cite{Andrianopoli:2014aqa,Freedman:2016yue}. One could also attempt to derive the boundary terms by enforcing the holographic Ward identities stemming from supersymmetry, using the Hamilton-Jacobi approach \cite{Martelli:2002sp,Papadimitriou:2010as}.
It may also be fruitful to extend to higher dimensions the approach of  \cite{Belyaev:2007bg,Grumiller:2009dx}, where the standard holographic counterterms  in three-dimensional\footnote{An off-shell formulation of four dimensional supergravity in the presence of a boundary  has been considered in  \cite{Belyaev:2008ex}, however as 
far as we are aware the application to the study of holographic renormalization is lacking in the literature.} $\mathcal{N}=1$ supergravity were argued to preserve supersymmetry, by working in an off-shell formulation. It will be very interesting to see whether any of these methods, or possibly others, may be used to shed light on the origin of the boundary terms proposed in the present work.

We conclude by alluding to a few possible generalizations of our results. Perhaps the most straightforward extension will be to lift the simplifying assumption that the metric on the four-dimensional conformal boundary is locally of a direct product type $S^1 \times M_3$.  
We expect that the new boundary terms arising from this analysis will be more general than those found presently, and this could help achieving a better understanding of them.  One could also study the consequences on such terms following from a Weyl transformation of the boundary metric. In minimal gauged supergravity, to complete the program we initiated it will be necessary to address the supersymmetric solutions in the null class \cite{Gauntlett:2003fk}, which are known to comprise AlAdS$_5$ solutions.
Another obvious generalization would be to investigate similar gauged supergravities in three, six, and seven space-time dimensions. In particular, it is expected that defining 
 two- and six-dimensional SCFTs in curved backgrounds leads to suitable versions of the supersymmetric Casimir energy \cite{Bobev:2015kza}, and reproducing these in dual holographic computations remains an open problem. The fact that in odd bulk dimension one has  anomalies and ambiguities in holographic renormalization suggests that at least in these dimensions a supersymmetric formulation of holographic renormalization will lead to a set of new boundary terms, analogous to those we uncovered in five-dimensional supergravity. 

Finally, we emphasize that in the derivation of the boundary terms, we made no assumptions on the properties of the supersymmetric solutions in the bulk. In particular, our boundary terms should be included in holographic studies of supersymmetric solutions with topologies different from $\R\times \R^4$. For example, it will be nice to investigate how the analysis of the properties of supersymmetric AlAdS$_5$ black holes \cite{Gutowski:2004ez,Chong:2005hr}
(or topological solitons \cite{Cvetic:2005zi,Cassani:2015upa}) will be affected by our findings. 

\section*{Acknowledgments}

P.~B.~G. is supported by EPSRC and a Scatcherd European Scholarship.  P.~B.~G. gratefully acknowledges support from the Simons Center for Geometry and Physics, Stony Brook University during the ``Simons Summer Workshop 2016.''
The work of D.~C. has been done within the Labex ILP (reference ANR-10-LABX-63) part of the Idex SUPER, and received financial state aid managed by the Agence Nationale de la Recherche, as part of the programme Investissements d'avenir under the reference ANR-11-IDEX-0004-02.  The work of D.~M. is  supported by the ERC Starting Grant N. 304806,  
``The gauge/gravity duality and geometry in string theory''. 

\appendix

\section{Curvature tensors}\label{app_curvature}

Our sign convention on the Riemann tensor is fixed by  
\be
R^i{}_{jkl} = \partial_k \Gamma^i_{jl} - \partial_l \Gamma^i_{jk} + \Gamma^i_{km}\Gamma^m_{jl} - \Gamma^i_{lm}\Gamma^m_{jk} \ ,
\ee
and the Ricci tensor is $R_{ij}= R^k{}_{ikj}$. Hence a round sphere has positive scalar curvature.

We next give some formulae by specializing to four dimensions; these are used in section~\ref{sec:5dsugra}. The Weyl tensor of a metric $g_{ij}$ is given by
\be\label{defWeyl}
C_{ijkl} \ = \ R_{ijkl} -  g_{i[k} R_{l]j} + g_{j[k} R_{l]i}  + \frac 13 R\, g_{i[k} g_{l]j}\ .
\ee
Its square can be expressed as  
\be
C_{ijkl}C^{ijkl}\ =\ R_{ijkl}R^{ijkl} - 2 R_{ij}R^{ij} + \frac 13 R^2\ .
\ee
The Euler scalar can be written as
\be 
E \ =\ R_{ijkl}R^{ijkl} - 4 R_{ij}R^{ij} + R^2\ ,
\ee
while the Pontryagin scalar is given by
\be
\mathcal{P} \ =\ \frac{1}{2}\epsilon^{ijkl} R_{ijmn}R_{kl}{}^{mn}\ .
\ee

From the metric and the Levi-Civita symbol we can construct four linearly independent functionals: $\int \diff^4x\sqrt{g} \,E$ (proportional to the Euler characteristic), $\int \diff^4x\sqrt{g} \, \mathcal{P}$ (proportional to the signature invariant), $\int \diff^4x\sqrt{g} \,C_{ijkl}C^{ijkl}$ (the conformal gravity action) and $\int \diff^4x\sqrt{g}\, R^2$ (which is neither topological nor conformal). While the metric variation of the first and the second vanishes identically in four dimensions, varying the third defines the Bach tensor
\begin{align}\label{BachTensor}
B_{ij} &= -\frac{1}{2\sqrt{g}} \frac{\delta}{\delta g^{ij}}\int \diff^4 x \sqrt{g} \, C_{klmn}C^{klmn} \nn \\[2mm] 
&= \frac 13\nabla_i\nabla_j R - \nabla^2 R_{ij} + \frac 16 g_{ij}\, \nabla^2 R - 2 R_{ ikjl}R^{kl} + \frac 23 R R_{ij} + \frac 12 g_{ij}\left( R_{kl}R^{kl} - \frac 13 R^2 \right).
\end{align}
This is covariantly conserved and traceless.
Varying the fourth functional yields the tensor
\be
H_{ij} = -\frac{1}{\sqrt{g}} \frac{\delta}{\delta g^{ij}}\int \diff^4 x \sqrt{g} \, R^2 \ =\ 2 \nabla_i \nabla_j R - 2 g_{ij} \nabla^2 R + \frac 12 g_{ij} R^2 - 2 R \, R_{ij}\ .
\ee 
which is covariantly conserved and satisfies $H_i{}^i = -6\nabla^2 R$.

\section{Construction of the five-dimensional solution}\label{app:constr_5d_sol}

In this appendix we provide details on how our five-dimensional supersymmetric solution is constructed.

\subsection{The general equations}\label{5dsusy_eqs}

We start by summarizing the conditions for bosonic solutions of minimal gauged supergravity in five dimensions to be supersymmetric, first obtained in~\cite{Gauntlett:2003fk} and recently revisited in~\cite{Cassani:2015upa}.
The analysis of~\cite{Gauntlett:2003fk} shows that the supersymmetry equation \eqref{KillingSpEqDirac} implies the existence of a Killing vector field $V$ that is either timelike or null. In this paper we just consider the timelike case.
Choosing coordinates such that $V = \partial/\partial y$, the five-dimensional metric takes the form
\be\label{5dMetricTimelike}
\diff s^2_5 \,=\, - f^2 \left( \diff y + \omega \right)^2 + f^{-1}\, \diff s^2_B\ ,
\ee
where $\diff s^2_B$ is a K\"ahler metric on a four-dimensional base $B$ transverse to $V$, while $f$ and $\omega$ are a positive function and a one-form on $B$, respectively. 
We will work with a K\"ahler form $J$ that is anti-self-dual on $B$, namely, $*_B J = -J$, so that the orientation on $B$ is fixed as ${\rm vol}_B = -\frac 12 J \wedge J$.
 We will also need the Ricci form $\mathcal R$ and its potential $P$, satisfying $\mathcal{R}=\diff P$. The Ricci form is defined as $\mathcal R_{mn}= \frac{1}{2}R_{mnpq}J^{pq}$, where $R_{mnpq}$ is the Riemann tensor of the K\"ahler metric and $m,n=1,\ldots,4$ are curved indices on $B$. The Ricci potential also appears in the relation $\nabla_m\Omega_{np} + \ii  P_m \Omega_{np} = 0$, where $\nabla_m$ is the Levi-Civita connection of the K\"ahler metric and $\Omega$ is a complex $(2,0)$-form normalized as $\Omega \wedge \overline \Omega =2J\wedge J$.
 
The geometry of the K\"ahler base determines the whole solution. The function $f$ in \eqref{5dMetricTimelike} is given by
\be\label{solf}
f \,=\, - \frac{24}{R}\ ,
\ee
where $R$ is the Ricci scalar of the K\"ahler metric, and is required to be non-zero everywhere. 
The equations for the one-form $\omega$ are
\be\label{eqforomega1}
\diff \omega + *_B \diff \omega \ =\ \frac{R}{24}\Big( \mathcal R - \frac{1}{4}R J  \Big)\ ,
\ee
and
\be\label{eqforomega2}
(\dd\omega)_{mn} J^{mn}\ = \ -\frac{1}{12}\left( \frac 12 \nabla^2 R + \frac 23 R_{mn}R^{mn} - \frac 13 R^2 \right)\ .
\ee
It was shown in~\cite{Cassani:2015upa} that for these conditions to admit a solution the K\"ahler metric on $B$ must necessarily satisfy the highly non-trivial sixth-order equation\footnote{The specialization of this equation for a particular K\"ahler metric appeared earlier in~\cite{Figueras:2006xx}.}

\be\label{SixthOrderEq}
\nabla^2\left( \frac 12 \nabla^2 R + \frac 23 R_{mn}R^{mn} - \frac 13 R^2 \right) + \nabla^m(R_{mn}\partial^n R) \ =\ 0\ .
\ee
Finally, the expression for the Maxwell field strength is
\be\label{susygaugefield}
\mathcal{F} \, =\, -\sqrt 3\, \diff \Big[ f(\diff y + \omega) +\frac{1}{3}P \Big] \ .
\ee

The solutions obtained from \eqref{5dMetricTimelike}--\eqref{susygaugefield} preserve at least  (and generically no more than) two real supercharges.

\subsection{The perturbative solution}\label{sec:ansatz}

We will make the assumption that the four-dimensional base $B$ admits an isometry. This is motivated by the fact that (after Wick rotation) we want the boundary metric to reproduce the field theory background metric~\eqref{bckgnd_metric}, and has the obvious advantage of simplifying the supersymmetry equations. With no further loss of generality, for the metric on $B$ we can choose 
\be\label{Kahleransatz}
\diff s^2_B \ =\ U(r,z,\bar{z})^2\left[\frac{\diff r^2}{r^2} + 4r^2W(r,z,\bar{z})^2 \diff z\diff\bar{z}\right]+ \frac{r^4}{U(r,z,\bar{z})^2}(\diff \hat{\psi} + \phi)^2\ ,
\ee
where $z$ is a complex coordinate, $\hat{\psi}$ is the Killing coordinate (to be redefined later) and $r$ will play the role of the radial coordinate. Moreover, $U(r,z,\bar z)$, $W(r,z,\bar z)$ are functions while $\phi$ is a $\hat{\psi}$-independent one-form transverse to $\partial/\partial\hat{\psi}$. This type of metric ansatz has been studied by \cite{Lebrun:1991,Tod:1995} where it is shown to be the \emph{generic} form satisfying our assumptions. The explicit powers of $r$ in \eqref{Kahleransatz} have been introduced for convenience: they are chosen so that the asymptotic expansions of $U$ and $W$ start at order one -- see below. 
We fix the orientation choosing the volume form on $B$ as
\be
{\rm vol}_B \ =\ 2\ii r^3 U^2W^2 \diff z \wedge \diff\bar z\wedge\diff \hat{\psi} \wedge \diff r \ .
\ee
The ansatz for the K\"ahler form is
\be\label{Kahlform}
J \ =\  2\ii r^2U^2W^2\,  \diff z\wedge \diff \bar{z} + r\, \diff r \wedge (\diff \hat{\psi}+\phi)\ ,
\ee
which defines an almost complex structure, {\it i.e.}\ $J_m{}^pJ_p{}^n=-\delta_m{}^n$. The metric is K\"ahler if $\diff J = 0$ and the almost complex structure $J_m{}^n$ is integrable. Together, these two conditions are equivalent to imposing
\be\label{eqforphi}
\diff \phi \ =\ \frac{1}{r}\partial_r\left(r^2U^2W^2 \right)2\ii\, \diff z\wedge \diff\bar{z} + \ii(\diff \bar{z}\,\partial_{\bar z} - \diff z\,\partial_{z})U^2 \wedge\frac{\diff r}{r^3} ~,
\ee
which determines the connection one-form $\phi$ in terms of other metric data. 
Acting on this equation with the exterior derivative, we find the integrability condition
\bea\label{integrabilityagain}
\partial_{z}\partial_{\bar{z}}U^2 + r^3\partial_r\left[r^{-1}\partial_r(r^2U^2W^2)\right] &=& 0~,
\eea
which constrains the functions $U,W$.
Using \eqref{eqforphi}, the Ricci scalar of the K\"ahler metric can be written as
\be\label{Rscalar}
R \ =\ -\frac{2}{r^2U^2 W^2}\left[\partial_{z}\partial_{\bar{z}}\log W + \partial_r\left(rW\partial_r(r^3W)\right) + W\partial_r(r^3W)\right]~,
\ee
and the Ricci connection as
\be\label{P}
P \ =\ -\frac{1}{U^2W}\partial_r(r^3W)(\diff\hat{\psi}+\phi) - \ii (\diff \bar{z}\,\partial_{\bar z} - \diff z\,\partial_{z}) \log W~,
\ee
with the Ricci form following from $\mathcal{R}= \diff P$. 

We will solve the supersymmetry equations in an asymptotic expansion around $r= \infty$. To do so, we express all functions entering in the ansatz in a suitable expansion involving powers of $1/r$ and $\log r$. The requirement that the solution be AlAdS$_5$ fixes the leading order terms in the expansions, as explained in detail in~\cite{Cassani:2012ri}.

For the function $U(r,z,\bar z)$ we take:
\begin{align}
U \ &= \ \sum_{m\geq 0}\;\sum_{0 \leq n\leq m}U_{2m,n}\frac{(\log r)^n}{r^{2m}}\nn\\[1mm]
 \ &=\ U_{0,0} + \frac{1}{r^2}(U_{2,0}+U_{2,1}\log r) + \frac{1}{r^4}(U_{4,0}+U_{4,1}\log r+U_{4,2}(\log r)^2) + \ldots\ ,
\end{align}
with $U_{2m,n}=U_{2m,n}(z,\bar z)$. Similarly, for $W$ we take
\be
W \ = \ W_{0,0} + \frac{1}{r^2}(W_{2,0}+W_{2,1}\log r) + \frac{1}{r^4}(W_{4,0}+W_{4,1}\log r+W_{4,2}(\log r)^2) + \ldots\ ,
\ee
with all coefficients also being functions of $z,\bar z$.
As for the one-form $\phi$, note that by redefining the coordinate $\hat{\psi}$ in~\eqref{Kahleransatz} we can always take the radial component $\phi_r = 0$, namely we can take $\phi = \phi_z(r,z,\bar{z})\diff z + \overline{\phi_z(r,z,\bar{z})}\diff \bar z$. The expansion of $\phi_z$ is analogous to those of $U$ and $W$ (albeit with complex coefficients), in particular it starts at order $\mathcal{O}(1)$.

We also need to expand the one-form $\omega$ appearing in the five-dimensional metric~\eqref{5dMetricTimelike}. By a redefinition of the coordinate $y$ we can always choose $\omega_r = 0$. Then $\omega$ can be parameterized as
\be\label{ansatz_omega}
\omega \ =\ c(r,z,\bar{z})(\diff\hat{\psi}+\phi) + C_z(r,z,\bar z) \diff z + \overline{C_z(r,z,\bar z)} \diff \bar z~.
\ee
The expansion of the real function $c$ starts at order $\mathcal{O}(r^2)$,
\be
c \ = \ c_{-2,0}\,r^2 + (c_{0,0} + c_{0,1} \log r) + \frac{1}{r^2}\left(c_{2,0}+c_{2,1}\log r+c_{2,2}(\log r)^2\right)  + \ldots\ ,
\ee
and a similar expansion is taken for $C_z$.

We next solve order by order the conditions on the four-dimensional metric on $B$. The explicit expressions are too cumbersome to be presented here and can only be dealt with using a computer algebra system like Mathematica; we will nevertheless describe in detail the procedure we followed. The constraints on the four-dimensional base metric amount to the equation~\eqref{eqforphi} for $\phi$, its integrability condition~\eqref{integrabilityagain}, and the sixth-order equation~\eqref{SixthOrderEq}. 
We start from~\eqref{integrabilityagain}, that we solve for $U_{2,1}$, $U_{4,0}$, $U_{4,1}$, $U_{4,2}$, $U_{6,0}$, $U_{6,1}$, $U_{6,2}$, $U_{6,3}$ in terms of $U_{0,0}$, $U_{2,0}$ and the coefficients of $W$. Then we solve the sixth-order equation \eqref{SixthOrderEq} at the first two non-trivial orders, which are $\mathcal{O}(1/r)$ and $\mathcal{O}(1/r^3)$ (together with the associated logarithmic terms). This fixes $W_{4,2}$, $W_{6,1}$, $W_{6,2}$, $W_{6,3}$ in terms of $U_{0,0}$, $U_{2,0}$, $W_{0,0}$, $W_{2,0}$, $W_{2,1}$, $W_{4,0}$, $W_{4,1}$, $W_{6,0}$, which thus remain undetermined at this stage.
Finally we solve~\eqref{eqforphi} for $\phi$; the latter is explicitly determined, up to the leading $\mathcal O(1)$ term $\phi_{0,0}$, which has to obey the equation 
\be\label{eq_phi00}
\diff \phi_{0,0} \ = \ 4\ii\, (U_{0,0}W_{0,0})^2\diff z\wedge \diff\bar z\ .
\ee 

Having fulfilled the constraints on the four-dimensional base $B$ with metric \eqref{Kahleransatz}, we can solve the equations  \eqref{eqforomega1}, \eqref{eqforomega2} for the connection $\omega$. Using the ansatz \eqref{ansatz_omega}, these become equations for $c$ and $C_z$, that again we can solve order by order. We find that both $c$ and $C_z$ are fully determined (in particular, the divergent $\mathcal{O}(r^2)$ term in the expansion of $C_z\diff z + \overline{C_z}\diff \bar z$ vanishes), 
 except for the $\mathcal{O}(1)$ term $C_{0,0}$ in the expansion of $C_z\diff z + \overline{C_z}\diff \bar z$, which is left free. In addition, from the $\mathcal{O}(\log r/r^2)$ term in the expansion of~\eqref{eqforomega1} we obtain a differential equation involving $U_{0,0}, W_{0,0}, W_{2,0}, W_{2,1}, W_{4,1}$ and $C_{0,0}$, that can most easily be solved for $W_{4,1}$ as the latter appears linearly and with no derivatives.\footnote{This is a new constraint on the K\"ahler base metric, that may be unexpected since we have already solved  all the conditions reviewed above for obtaining a supersymmetric solution from such metric. There is no contradiction here: {\it a priori} we could avoid to further constrain the K\"ahler metric by interpreting the equation under examination as a differential equation for the boundary function $C_{0,0}$. However, shortly we will impose a boundary condition setting $C_{0,0}=0$; consistency with the present equation then fixes $W_{4,1}$.}

We can next obtain the function $f$ from~\eqref{solf}. This concludes the construction of the metric~\eqref{5dMetricTimelike} and the gauge field~\eqref{susygaugefield} near to $r\to \infty$. 
At leading order, we find that the five-dimensional metric is
\be
\diff s^2_5 \ =\ \frac{\diff r^2}{r^2} + r^2 \diff s^2_4\ ,
\ee
where the metric $\diff s^2_4$ on the conformal boundary is
\begin{align}\label{general_bdry_metric}
\diff s^2_4 \ &=\ \frac{1}{4U_{0,0}^4W_{0,0}^2}\left[2W_{0,0}W_{2,1} - 2\ii U_{0,0}^2 (\diff C_{0,0})_{z\bar z} - \partial_z\partial_{\bar z}\log W_{0,0} \right]\big( \diff \hat{\psi} + \phi_{0,0} \big)^2 \nn\\[1mm]
\ & \ \quad - 2 \left(\diff y + C_{0,0} \right)\big( \diff \hat{\psi} + \phi_{0,0} \big) +4W_{0,0}^2\diff z\diff\bar{z} \ .
\end{align}
This is in agreement with the general form of a supersymmetric Lorentzian boundary metric, as  can be seen by comparison with~\cite[eq.~(4.12)]{Cassani:2012ri}. In fact, it is even too general for our purposes, as it does not admit a simple Wick rotation to Euclidean signature. In order to be able to perform a simple Wick rotation and match \eqref{bckgnd_metric}, we will fix part of the free functions in~\eqref{general_bdry_metric} as
\be\label{bdrycond}
C_{0,0} \ =\ 0\,,\qquad W_{2,1} \ =\ 2 U_{0,0}^4 W_{0,0} + \frac{1}{2W_{0,0}}\partial_{z}\partial_{\bar z}\log W_{0,0}\ .
\ee
In this way, the perturbative solution takes a simpler form, and only depends on the free functions
$U_{0,0}, \ U_{2,0}, \ W_{0,0}, \ W_{2,0},\ W_{4,0},\ W_{6,0}$,
where $U_{0,0}$ and $W_{0,0}$ are boundary data, while the remaining four functions only appear at subleading order in the five-dimensional metric. For convenience we will rename the boundary data as
\be
U_{0,0} \ = \ \frac{1}{2}\wone^{1/2}\ , \qquad W_{0,0}\ =\ {\me}^{\wzero/2}\ , \qquad \phi_{0,0}\  =\ \phizero \ = \ \phizero_z \diff z + \overline{\phizero_z}\diff \bar z\ ,
\ee
and the subleading functions as
\be 
U_{2,0} \ =\ {\me}^{\wzero/2}k_1\ , \quad W_{2,0} \ =\ {\me}^{\wzero/2}k_2\ ,\quad W_{4,0}={\me}^{\wzero/2}k_3\ ,\quad W_{6,0}={\me}^{\wzero/2}k_4\ ,
\ee
where we recall that all functions depend on $z,\bar z$.
Also redefining the Killing coordinates $\{y,\hat \psi\}$ into new coordinates $\{t,\psi\}$ as
\be
y \,=\, t \ ,\qquad \hat{\psi} \,=\, \psi + t\ ,
\ee
the boundary metric becomes
\be
\diff s^2_4 = -\diff t^2 + (\diff \psi+\phizero)^2 + 4 \me^{\wzero}\diff z \diff \bar z \ ,
\ee
with eq.~\eqref{eq_phi00} now being
\be
\diff \phizero \ = \ \ii\, \wone\, \me^{\wzero}\diff z\wedge \diff\bar z\ .
\ee

At leading order, the gauge field strength reads
\be
\diff A^{(0)} \ = \ -\frac{1}{\sqrt 3}\,\diff \left[-\frac{\wone}{8} \diff t + \frac{\wone}{4} (\diff \psi + \phizero) + \frac{1}{4}*_2 \!\diff \wzero \right]\ ,
\ee
where we denote
$*_2 \diff  =  \ii(\diff \bar{z}\,\partial_{\bar z} - \diff z\,\partial_{z})$.
The corresponding gauge potential is determined up to a gauge choice that will play an important role. We see that after taking $t=-\ii\tau$, these agree with the field theory background fields~\eqref{bckgnd_metric}, \eqref{A4d}.

At subleading order the canonical form \eqref{5dMetricTimelike} of our five-dimensional metric is not of the Fefferman-Graham type~\eqref{FG_5d_metric}, \eqref{FGexpansion}. Besides being more standard, the latter is desirable as it makes it simpler to extract the holographic data from the solution. We find that Fefferman-Graham coordinates are reached after implementing a suitable asymptotic transformation, sending $\{t,z^{\rm old},\psi^{\rm old},r\}$ into $\{t,z^{\rm new},\psi^{\rm new},\tr\}$ and having the form:
\begin{align}\label{change_to_FGcoords}
r  &=  \frac{1}{\tr}\big[ 1 + \tr^2(m_{r,2,0} + m_{r,2,1}\log \tr) + \tr^4(m_{r,4,0} +m_{r,4,1}\log \tr + m_{r,4,2}(\log \tr)^2) + \mathcal{O}(\tr^5) \big] , \nn\\[1mm]
z^{\rm old}  &=  z^{\rm new} + \tr^4 \left( m_{z,4,0} + m_{z,4,1}\log \tr\right) + \mathcal{O}(\tr^5)\ ,  \nn\\[1mm]
\psi^{\rm old} &= \psi^{\rm new} + \tr^4 \left( m_{\psi,4,0} + m_{\psi,4,1}\log \tr \right) + \mathcal{O}(\tr^5) \ ,
\end{align}
where all the $m$ coefficients are specific functions of $z,\bar z$.
It should be noted that the conformal boundary, originally located at $r = \infty$, is now found at $\tr =0$.
In section~\ref{5d_solution} we give further details on the subleading terms in the metric and in the gauge field in Fefferman-Graham coordinates. 
There we drop the label ``new'', being understood that we always work in the new, Fefferman-Graham coordinates. 
Notice that since the metric can be cast in Fefferman-Graham form it is AlAdS.

\section{Supersymmetry at the boundary}\label{app:Killingspinors}

\subsection{Killing spinors}

At the boundary of an AlAdS$_5$ solution, the supersymmetry condition~\eqref{KillingSpEqDirac} gives rise to the charged conformal Killing spinor equation
\be\label{susycond_CSG}
\nabla_i^A \zeta_\pm \,=\, -\frac{1}{4} \sigma_{\pm\,i} \sigma_\mp^{\,j}\nabla_j^A \zeta_\pm  \ ,
\ee
where we are using the two-component spinor notation introduced in section~\ref{sec:4dbackgrounds} and $\nabla_i^A \zeta_\pm = \left(\nabla_i \mp \ii A_i\right) \zeta_\pm $ is the spinor covariant derivative, with $\nabla_i$ the Levi-Civita connection constructed with the boundary vierbein and $A = - \sqrt 3 A^{\rm (0)}$ the canonically normalized gauge connection.  This holds both in Euclidean and Lorentzian signature, for details see~\cite{Klare:2012gn} and~\cite{Cassani:2012ri}, respectively.
Here we are identifying the $\Gamma^1,\Gamma^2,\Gamma^3,\Gamma^4$ matrices of Cliff$(5)$ with those of Cliff$(4)$, and the $\Gamma^5$ of Cliff$(5)$ with the chirality matrix of Cliff$(4)$; then we pass to two-component notation.
The same equation ensures that some supersymmetry is preserved when a four-dimensional SCFT is coupled to background conformal supergravity, and (for spinors with no zeros) can be mapped into the equation arising when one couples the theory to new minimal supergravity~\cite{Klare:2012gn,Dumitrescu:2012ha,Cassani:2012ri}.

One can see that the four-dimensional metric \eqref{bckgnd_metric} and gauge field \eqref{A4d} 
allow for solutions to~\eqref{susycond_CSG} and thus define a supersymmetric field theory background as well as supersymmetric boundary conditions for the bulk supergravity fields. 
Our scope here is to illustrate the gauge choice that makes the spinors independent of the  coordinate $\tau$, so that they are globally well-defined when this is made compact. 

We choose the vierbein
\be\label{framechoice}
e^1+ \ii\,e^2 \,=\, 2\, \me^{\frac{\wzero}{2}}\diff z \ ,  \qquad e^3 \,=\, \diff \psi + \phizero\ , \qquad e^4 \,=\, \diff \tau\ .
\ee
By studying \eqref{susycond_CSG} we find that in the generic case where $\wone$ is non-constant, the solution reads
\be\label{KillingSpGeneric}
\zeta_+ \,=\, \frac{1}{\sqrt 2}\,\me^{ \ct \tau + \ii \cpsi \psi +\ii \lambda}
{0\choose1}\ , \qquad \zeta_- \,=\, \frac{1}{\sqrt 2}\,\me^{- \ct \tau - \ii \cpsi \psi -\ii \lambda}
{1\choose0}\ ,
\ee
where we have fixed an arbitrary overall constant.
 In the special case $\wone = {\rm const}$ there exist additional solutions, however this enhancement of supersymmetry is not relevant for the present paper and we will not discuss it further. 

Kosmann's spinorial Lie derivative along a vector $v$ is defined as
\be
\mathcal{L}_v \zeta_\pm \, = \, v^i \nabla_i \zeta_\pm + \frac{1}{2} \nabla_{i}v_{j} \sigma_\pm^{ij} \zeta_\pm\ . 
\ee
 For the Killing vectors in our background, we find:
\begin{align}\label{LieDerSpinor}
\mathcal{L}_{\partial_\psi}\zeta_\pm \ &= \ \partial_\psi\zeta_\pm \ = \ \pm\ii \cpsi\, \zeta_\pm\ ,\nn\\[1mm]
\mathcal{L}_{\partial_\tau}\zeta_\pm \ &= \ \partial_\tau\zeta_\pm \ = \  \pm \gamma'\,\zeta_\pm\ ,
\end{align}
hence $\pm\cpsi$ and $\pm\ct$ are the charge of the spinors $\zeta_\pm$ under $\partial_\psi$ and $\ii\partial_\tau$, respectively. It follows that the condition for $\zeta_\pm$ to be independent of $\tau$ is
\be\label{choicegamma'}
\ct \ = \ 0\ .
\ee

\subsection{Superalgebra}

The algebra of field theory supersymmetry transformations generated by a pair of spinors $\zeta_+,\zeta_-$ solving~\eqref{susycond_CSG} reads \cite{Klare:2012gn,Dumitrescu:2012ha,Cassani:2012ri} (see also~\cite[sect.$\:$5.1]{Cassani:2014zwa} for some more details):
\be\label{superalgebranm}
 [\delta_{\zeta_+} ,\delta_{\zeta_-}]\Phi \, = \, 2\ii \left( \mathcal L_{K}  -  \ii\, q\, K\lrcorner A^{\rm nm} \right) \Phi \ , \qquad \ \delta_{\zeta_\pm}^2  =\, 0\ ,
\ee
where $\mathcal{L}_{K}$ denotes the Lie derivative 
 along the complex Killing vector $K$ defined in \eqref{K4d} and $q$ is the R-charge of a generic field $\Phi$ in the field theory. The gauge field $A^{\rm nm}$ is defined as $A^{\rm nm} = A + \frac{3}{2} V^{\rm nm}$, where $V^{\rm nm}$ is a well-defined one-form satisfying 
\be\label{eqsforVnm}
\nabla^i V^{\rm nm}_i \,=\, 0\ ,\qquad  2\ii\, \sigma_\mp^i V^{\rm nm}_i \zeta_\pm \ = \ \pm\sigma_\mp^i \nabla_i^A \zeta_\pm\ .
\ee
This actually only fixes $K^i V_i^{\rm nm}$. In this way, $A^{\rm nm}$ and $V^{\rm nm}$ can be interpreted as the auxiliary fields of background new minimal supergravity (hence the label ``nm'').

Let us now evaluate these quantities in our background \eqref{bckgnd_metric}, \eqref{A4d}.
With the choice~\eqref{KillingSpGeneric}, the vector $K$ takes precisely the form \eqref{K4d_explicit}, $K = \frac{1}{2}(\partial_\psi - \ii \partial_\tau)
$, while its dual one-form is
\be
K^\flat \ = \ \frac{1}{2} \left( \diff \psi + \phizero -\ii\, \diff \tau \right)\,.
\ee
 As long as $\wone\neq 0$ this has non-vanishing twist,
\be
K^\flat \wedge \diff K^\flat \ = \ \frac{\ii}{4}\wone \,\me^{\wzero} \left(\diff \psi -\ii\, \diff \tau\right)\wedge \diff z \wedge \diff \bar z\ .
\ee
As discussed in~\cite{Cassani:2012ri}, after Wick rotating to Lorentzian signature by $\tau = \ii t$ this implies that the five-dimensional bulk solution falls in the timelike class of~\cite{Gauntlett:2003fk}.

Eqs.~\eqref{eqsforVnm} for $V^{\rm nm}$ are solved by
\be
V^{\rm nm} \ = \ -\frac{\wone}{4} (\diff \psi + \phizero ) + \kappa\, K^\flat\ ,
\ee
where $\kappa$ is an undetermined complex function satisfying $K^i \partial_i \kappa=0$.
 Then $A^{\rm nm}$ reads:
\be
A^{\rm nm} \, = \, A + \frac{3}{2} V^{\rm nm} \,=\,  \frac{1}{2}\left(3\kappa - \wone \right)K^\flat + \frac{\ii}{4}(\diff \bar{z}\,\partial_{\bar z}\wzero - \diff z\,\partial_{z}\wzero)  - \ii\ct \diff \tau + \cpsi\, \diff \psi + \diff \lambda \ .
\ee
Contracting with $K$ gives
\be
K \hook A^{\rm nm} \ = \ \frac{1}{2}\left(\cpsi-\ct\right)\ .
\ee
Note from~\eqref{LieDerSpinor} that this is also the charge of the Killing spinor under $K$, $\mathcal{L}_{K} \zeta_+ = \frac{\ii}{2}(\cpsi - \ct)\zeta_+$.

We conclude that in the background of interest, and with the choice \eqref{choicegamma'}, the superalgebra reads
\be\label{superalgebranm}
 [\delta_{\zeta_+} ,\delta_{\zeta_-}]\Phi \ = \ \ii \Big( -\ii \mathcal L_{ \partial_\tau} + \mathcal L_{ \partial_\psi}  - \ii\cpsi\, q \Big) \Phi \ .
\ee
 Passing to the corresponding generators gives 
\be
\{ \mathcal{Q}_+,\mathcal{Q}_- \} \,=\, H + J + \cpsi \,Q \ ,
\ee
where $H$ and $J$ are the charges associated with $\partial_\tau$ and $-\partial_\psi$, respectively, while $Q$ is the R-charge.
Taking the expectation value in a supersymmetric vacuum leads to the BPS condition 
\be
\langle H \rangle +  \langle J \rangle + \cpsi  \langle Q\rangle  \ = \ 0\ .
\ee

\subsection{Twisted background}

For the twisted background \eqref{bdrymetric_twisted}, \eqref{bdryA_twisted}, requiring that the Killing spinors $\zeta_\pm$ are independent of the new time coordinate and recalling relations~\eqref{LieDerSpinor}, valid in the old coordinates, immediately leads to
\be
\ct = -\ii\,\cpsi\tan\alpha\ .
\ee
It is also straightforward to implement the change of coordinates and obtain the new $K$ (given in~\eqref{K_twisted}) and the new form of the superalgebra.

\bibliographystyle{JHEP}
\bibliography{bibHoloCasimir}

\providecommand{\href}[2]{#2}\begingroup\raggedright\begin{thebibliography}{10}

\bibitem{Witten:1998qj}
E.~Witten, \emph{{Anti-de Sitter space and holography}}, {\emph{Adv. Theor.
  Math. Phys.} {\bf 2} (1998) 253--291},
  [\href{http://arxiv.org/abs/hep-th/9802150}{{\tt hep-th/9802150}}].

\bibitem{Gubser:1998bc}
S.~S. Gubser, I.~R. Klebanov and A.~M. Polyakov, \emph{{Gauge theory
  correlators from noncritical string theory}},
  \href{http://dx.doi.org/10.1016/S0370-2693(98)00377-3}{\emph{Phys. Lett.}
  {\bf B428} (1998) 105--114}, [\href{http://arxiv.org/abs/hep-th/9802109}{{\tt
  hep-th/9802109}}].

\bibitem{Henningson:1998gx}
M.~Henningson and K.~Skenderis, \emph{{The Holographic Weyl anomaly}},
  \href{http://dx.doi.org/10.1088/1126-6708/1998/07/023}{\emph{JHEP} {\bf 9807}
  (1998) 023}, [\href{http://arxiv.org/abs/hep-th/9806087}{{\tt
  hep-th/9806087}}].

\bibitem{Balasubramanian:1999re}
V.~Balasubramanian and P.~Kraus, \emph{{A Stress tensor for Anti-de Sitter
  gravity}},
  \href{http://dx.doi.org/10.1007/s002200050764}{\emph{Commun.Math.Phys.} {\bf
  208} (1999) 413--428}, [\href{http://arxiv.org/abs/hep-th/9902121}{{\tt
  hep-th/9902121}}].

\bibitem{deBoer:1999tgo}
J.~de~Boer, E.~P. Verlinde and H.~L. Verlinde, \emph{{On the holographic
  renormalization group}},
  \href{http://dx.doi.org/10.1088/1126-6708/2000/08/003}{\emph{JHEP} {\bf 08}
  (2000) 003}, [\href{http://arxiv.org/abs/hep-th/9912012}{{\tt
  hep-th/9912012}}].

\bibitem{deHaro:2000vlm}
S.~de~Haro, S.~N. Solodukhin and K.~Skenderis, \emph{{Holographic
  reconstruction of space-time and renormalization in the AdS / CFT
  correspondence}},
  \href{http://dx.doi.org/10.1007/s002200100381}{\emph{Commun. Math. Phys.}
  {\bf 217} (2001) 595--622}, [\href{http://arxiv.org/abs/hep-th/0002230}{{\tt
  hep-th/0002230}}].

\bibitem{Bianchi:2001de}
M.~Bianchi, D.~Z. Freedman and K.~Skenderis, \emph{{How to go with an RG
  flow}}, \href{http://dx.doi.org/10.1088/1126-6708/2001/08/041}{\emph{JHEP}
  {\bf 08} (2001) 041}, [\href{http://arxiv.org/abs/hep-th/0105276}{{\tt
  hep-th/0105276}}].

\bibitem{Bianchi:2001kw}
M.~Bianchi, D.~Z. Freedman and K.~Skenderis, \emph{{Holographic
  renormalization}},
  \href{http://dx.doi.org/10.1016/S0550-3213(02)00179-7}{\emph{Nucl. Phys.}
  {\bf B631} (2002) 159--194}, [\href{http://arxiv.org/abs/hep-th/0112119}{{\tt
  hep-th/0112119}}].

\bibitem{Martelli:2002sp}
D.~Martelli and W.~Mueck, \emph{{Holographic renormalization and Ward
  identities with the Hamilton-Jacobi method}},
  \href{http://dx.doi.org/10.1016/S0550-3213(03)00060-9}{\emph{Nucl. Phys.}
  {\bf B654} (2003) 248--276}, [\href{http://arxiv.org/abs/hep-th/0205061}{{\tt
  hep-th/0205061}}].

\bibitem{Skenderis:2002wp}
K.~Skenderis, \emph{{Lecture notes on holographic renormalization}},
  \href{http://dx.doi.org/10.1088/0264-9381/19/22/306}{\emph{Class. Quant.
  Grav.} {\bf 19} (2002) 5849--5876},
  [\href{http://arxiv.org/abs/hep-th/0209067}{{\tt hep-th/0209067}}].

\bibitem{Witten:1988ze}
E.~Witten, \emph{{Topological Quantum Field Theory}},
  \href{http://dx.doi.org/10.1007/BF01223371}{\emph{Commun.Math.Phys.} {\bf
  117} (1988) 353}.

\bibitem{Nekrasov:2002qd}
N.~A. Nekrasov, \emph{{Seiberg-Witten prepotential from instanton counting}},
  \href{http://dx.doi.org/10.4310/ATMP.2003.v7.n5.a4}{\emph{Adv.Theor.Math.Phys.}
  {\bf 7} (2004) 831--864}, [\href{http://arxiv.org/abs/hep-th/0206161}{{\tt
  hep-th/0206161}}].

\bibitem{Pestun:2007rz}
V.~Pestun, \emph{{Localization of gauge theory on a four-sphere and
  supersymmetric Wilson loops}},
  \href{http://dx.doi.org/10.1007/s00220-012-1485-0}{\emph{Commun.Math.Phys.}
  {\bf 313} (2012) 71--129}, [\href{http://arxiv.org/abs/0712.2824}{{\tt
  0712.2824}}].

\bibitem{Klare:2012gn}
C.~Klare, A.~Tomasiello and A.~Zaffaroni, \emph{{Supersymmetry on Curved Spaces
  and Holography}},
  \href{http://dx.doi.org/10.1007/JHEP08(2012)061}{\emph{JHEP} {\bf 1208}
  (2012) 061}, [\href{http://arxiv.org/abs/1205.1062}{{\tt 1205.1062}}].

\bibitem{Cassani:2012ri}
D.~Cassani, C.~Klare, D.~Martelli, A.~Tomasiello and A.~Zaffaroni,
  \emph{{Supersymmetry in Lorentzian Curved Spaces and Holography}},
  \href{http://dx.doi.org/10.1007/s00220-014-1983-3}{\emph{Commun.Math.Phys.}
  {\bf 327} (2014) 577--602}, [\href{http://arxiv.org/abs/1207.2181}{{\tt
  1207.2181}}].

\bibitem{Closset:2013vra}
C.~Closset, T.~T. Dumitrescu, G.~Festuccia and Z.~Komargodski, \emph{{The
  Geometry of Supersymmetric Partition Functions}},
  \href{http://dx.doi.org/10.1007/JHEP01(2014)124}{\emph{JHEP} {\bf 1401}
  (2014) 124}, [\href{http://arxiv.org/abs/1309.5876}{{\tt 1309.5876}}].

\bibitem{Closset:2014uda}
C.~Closset, T.~T. Dumitrescu, G.~Festuccia and Z.~Komargodski, \emph{{From
  Rigid Supersymmetry to Twisted Holomorphic Theories}},
  \href{http://dx.doi.org/10.1103/PhysRevD.90.085006}{\emph{Phys. Rev.} {\bf
  D90} (2014) 085006}, [\href{http://arxiv.org/abs/1407.2598}{{\tt
  1407.2598}}].

\bibitem{Alday:2013lba}
L.~F. Alday, D.~Martelli, P.~Richmond and J.~Sparks, \emph{{Localization on
  Three-Manifolds}}, \href{http://dx.doi.org/10.1007/JHEP10(2013)
  095}{\emph{JHEP} {\bf 1310} (2013) 095},
  [\href{http://arxiv.org/abs/1307.6848}{{\tt 1307.6848}}].

\bibitem{Assel:2014paa}
B.~Assel, D.~Cassani and D.~Martelli, \emph{{Localization on Hopf surfaces}},
  \href{http://dx.doi.org/10.1007/JHEP08(2014)123}{\emph{JHEP} {\bf 1408}
  (2014) 123}, [\href{http://arxiv.org/abs/1405.5144}{{\tt 1405.5144}}].

\bibitem{Assel:2015nca}
B.~Assel, D.~Cassani, L.~Di~Pietro, Z.~Komargodski, J.~Lorenzen and
  D.~Martelli, \emph{{The Casimir Energy in Curved Space and its Supersymmetric
  Counterpart}}, \href{http://dx.doi.org/10.1007/JHEP07(2015)043}{\emph{JHEP}
  {\bf 07} (2015) 043}, [\href{http://arxiv.org/abs/1503.05537}{{\tt
  1503.05537}}].

\bibitem{Cassani:2013dba}
D.~Cassani and D.~Martelli, \emph{{Supersymmetry on curved spaces and
  superconformal anomalies}},
  \href{http://dx.doi.org/10.1007/JHEP10(2013)025}{\emph{JHEP} {\bf 1310}
  (2013) 025}, [\href{http://arxiv.org/abs/1307.6567}{{\tt 1307.6567}}].

\bibitem{Gauntlett:2003fk}
J.~P. Gauntlett and J.~B. Gutowski, \emph{{All supersymmetric solutions of
  minimal gauged supergravity in five-dimensions}},
  \href{http://dx.doi.org/10.1103/PhysRevD.70.089901,
  10.1103/PhysRevD.68.105009}{\emph{Phys. Rev.} {\bf D68} (2003) 105009},
  [\href{http://arxiv.org/abs/hep-th/0304064}{{\tt hep-th/0304064}}].

\bibitem{Cassani:2014zwa}
D.~Cassani and D.~Martelli, \emph{{The gravity dual of supersymmetric gauge
  theories on a squashed $S^1\times S^3$}},
  \href{http://dx.doi.org/10.1007/JHEP08(2014)044}{\emph{JHEP} {\bf 1408}
  (2014) 044}, [\href{http://arxiv.org/abs/1402.2278}{{\tt 1402.2278}}].

\bibitem{Genolini:2016sxe}
P.~Benetti~Genolini, D.~Cassani, D.~Martelli and J.~Sparks, \emph{{The
  holographic supersymmetric Casimir energy}},
  \href{http://arxiv.org/abs/1606.02724}{{\tt 1606.02724}}.

\bibitem{Festuccia:2011ws}
G.~Festuccia and N.~Seiberg, \emph{{Rigid Supersymmetric Theories in Curved
  Superspace}}, \href{http://dx.doi.org/10.1007/JHEP06(2011)114}{\emph{JHEP}
  {\bf 1106} (2011) 114}, [\href{http://arxiv.org/abs/1105.0689}{{\tt
  1105.0689}}].

\bibitem{Dumitrescu:2012ha}
T.~T. Dumitrescu, G.~Festuccia and N.~Seiberg, \emph{{Exploring Curved
  Superspace}}, \href{http://dx.doi.org/10.1007/JHEP08(2012)141}{\emph{JHEP}
  {\bf 1208} (2012) 141}, [\href{http://arxiv.org/abs/1205.1115}{{\tt
  1205.1115}}].

\bibitem{Closset:2012ru}
C.~Closset, T.~T. Dumitrescu, G.~Festuccia and Z.~Komargodski,
  \emph{{Supersymmetric Field Theories on Three-Manifolds}},
  \href{http://dx.doi.org/10.1007/JHEP05(2013)017}{\emph{JHEP} {\bf 1305}
  (2013) 017}, [\href{http://arxiv.org/abs/1212.3388}{{\tt 1212.3388}}].

\bibitem{Farquet:2014kma}
D.~Farquet, J.~Lorenzen, D.~Martelli and J.~Sparks, \emph{{Gravity duals of
  supersymmetric gauge theories on three-manifolds}},
  \href{http://dx.doi.org/10.1007/JHEP08(2016)080}{\emph{JHEP} {\bf 08} (2016)
  080}, [\href{http://arxiv.org/abs/1404.0268}{{\tt 1404.0268}}].

\bibitem{Farquet:2014bda}
D.~Farquet and J.~Sparks, \emph{{Wilson loops on three-manifolds and their
  M2-brane duals}},
  \href{http://dx.doi.org/10.1007/JHEP12(2014)173}{\emph{JHEP} {\bf 12} (2014)
  173}, [\href{http://arxiv.org/abs/1406.2493}{{\tt 1406.2493}}].

\bibitem{Martelli:2015kuk}
D.~Martelli and J.~Sparks, \emph{{The character of the supersymmetric Casimir
  energy}}, \href{http://dx.doi.org/10.1007/JHEP08(2016)117}{\emph{JHEP} {\bf
  08} (2016) 117}, [\href{http://arxiv.org/abs/1512.02521}{{\tt 1512.02521}}].

\bibitem{Sohnius:1981tp}
M.~F. Sohnius and P.~C. West, \emph{{An Alternative Minimal Off-Shell Version
  of N=1 Supergravity}},
  \href{http://dx.doi.org/10.1016/0370-2693(81)90778-4}{\emph{Phys.Lett.} {\bf
  B105} (1981) 353}.

\bibitem{Benini:2015noa}
F.~Benini and A.~Zaffaroni, \emph{{A topologically twisted index for
  three-dimensional supersymmetric theories}},
  \href{http://dx.doi.org/10.1007/JHEP07(2015)127}{\emph{JHEP} {\bf 07} (2015)
  127}, [\href{http://arxiv.org/abs/1504.03698}{{\tt 1504.03698}}].

\bibitem{Benini:2016hjo}
F.~Benini and A.~Zaffaroni, \emph{{Supersymmetric partition functions on
  Riemann surfaces}},  \href{http://arxiv.org/abs/1605.06120}{{\tt
  1605.06120}}.

\bibitem{Closset:2016arn}
C.~Closset and H.~Kim, \emph{{Comments on twisted indices in 3d supersymmetric
  gauge theories}},
  \href{http://dx.doi.org/10.1007/JHEP08(2016)059}{\emph{JHEP} {\bf 08} (2016)
  059}, [\href{http://arxiv.org/abs/1605.06531}{{\tt 1605.06531}}].

\bibitem{Martelli:2011fu}
D.~Martelli, A.~Passias and J.~Sparks, \emph{{The gravity dual of
  supersymmetric gauge theories on a squashed three-sphere}},
  \href{http://dx.doi.org/10.1016/j.nuclphysb.2012.07.019}{\emph{Nucl. Phys.}
  {\bf B864} (2012) 840--868}, [\href{http://arxiv.org/abs/1110.6400}{{\tt
  1110.6400}}].

\bibitem{Drukker:2010nc}
N.~Drukker, M.~Marino and P.~Putrov, \emph{{From weak to strong coupling in
  ABJM theory}},
  \href{http://dx.doi.org/10.1007/s00220-011-1253-6}{\emph{Commun.Math.Phys.}
  {\bf 306} (2011) 511--563}, [\href{http://arxiv.org/abs/1007.3837}{{\tt
  1007.3837}}].

\bibitem{Benini:2011nc}
F.~Benini, T.~Nishioka and M.~Yamazaki, \emph{{4d Index to 3d Index and 2d
  TQFT}}, \href{http://dx.doi.org/10.1103/PhysRevD.86.065015}{\emph{Phys. Rev.}
  {\bf D86} (2012) 065015}, [\href{http://arxiv.org/abs/1109.0283}{{\tt
  1109.0283}}].

\bibitem{Alday:2012au}
L.~F. Alday, M.~Fluder and J.~Sparks, \emph{{The Large N limit of M2-branes on
  Lens spaces}}, \href{http://dx.doi.org/10.1007/JHEP10(2012)057}{\emph{JHEP}
  {\bf 10} (2012) 057}, [\href{http://arxiv.org/abs/1204.1280}{{\tt
  1204.1280}}].

\bibitem{Closset:2013sxa}
C.~Closset and I.~Shamir, \emph{{The $\mathcal{N}=1$ Chiral Multiplet on
  $T^2\times S^2$ and Supersymmetric Localization}},
  \href{http://dx.doi.org/10.1007/JHEP03(2014)040}{\emph{JHEP} {\bf 1403}
  (2014) 040}, [\href{http://arxiv.org/abs/1311.2430}{{\tt 1311.2430}}].

\bibitem{Romelsberger:2005eg}
C.~Romelsberger, \emph{{Counting chiral primaries in N = 1, d=4 superconformal
  field theories}},
  \href{http://dx.doi.org/10.1016/j.nuclphysb.2006.03.037}{\emph{Nucl.Phys.}
  {\bf B747} (2006) 329--353}, [\href{http://arxiv.org/abs/hep-th/0510060}{{\tt
  hep-th/0510060}}].

\bibitem{Kinney:2005ej}
J.~Kinney, J.~M. Maldacena, S.~Minwalla and S.~Raju, \emph{{An Index for 4
  dimensional super conformal theories}},
  \href{http://dx.doi.org/10.1007/s00220-007-0258-7}{\emph{Commun.Math.Phys.}
  {\bf 275} (2007) 209--254}, [\href{http://arxiv.org/abs/hep-th/0510251}{{\tt
  hep-th/0510251}}].

\bibitem{Assel:2014tba}
B.~Assel, D.~Cassani and D.~Martelli, \emph{{Supersymmetric counterterms from
  new minimal supergravity}},
  \href{http://dx.doi.org/10.1007/JHEP11(2014)135}{\emph{JHEP} {\bf 1411}
  (2014) 135}, [\href{http://arxiv.org/abs/1410.6487}{{\tt 1410.6487}}].

\bibitem{Lorenzen:2014pna}
J.~Lorenzen and D.~Martelli, \emph{{Comments on the Casimir energy in
  supersymmetric field theories}},
  \href{http://dx.doi.org/10.1007/JHEP07(2015)001}{\emph{JHEP} {\bf 07} (2015)
  001}, [\href{http://arxiv.org/abs/1412.7463}{{\tt 1412.7463}}].

\bibitem{Bobev:2015kza}
N.~Bobev, M.~Bullimore and H.-C. Kim, \emph{{Supersymmetric Casimir Energy and
  the Anomaly Polynomial}},
  \href{http://dx.doi.org/10.1007/JHEP09(2015)142}{\emph{JHEP} {\bf 09} (2015)
  142}, [\href{http://arxiv.org/abs/1507.08553}{{\tt 1507.08553}}].

\bibitem{Ardehali:2015hya}
A.~Arabi~Ardehali, J.~T. Liu and P.~Szepietowski, \emph{{High-Temperature
  Expansion of Supersymmetric Partition Functions}},
  \href{http://dx.doi.org/10.1007/JHEP07(2015)113}{\emph{JHEP} {\bf 07} (2015)
  113}, [\href{http://arxiv.org/abs/1502.07737}{{\tt 1502.07737}}].

\bibitem{Brunner:2016nyk}
F.~Brunner, D.~Regalado and V.~P. Spiridonov, \emph{{Supersymmetric Casimir
  Energy and $SL(3,\mathbb{Z})$ Transformations}},
  \href{http://arxiv.org/abs/1611.03831}{{\tt 1611.03831}}.

\bibitem{Nishioka:2014zpa}
T.~Nishioka and I.~Yaakov, \emph{{Generalized indices for $ \mathcal{N} $ = 1
  theories in four-dimensions}},
  \href{http://dx.doi.org/10.1007/JHEP12(2014)150}{\emph{JHEP} {\bf 1412}
  (2014) 150}, [\href{http://arxiv.org/abs/1407.8520}{{\tt 1407.8520}}].

\bibitem{Dunajski:2010uv}
M.~Dunajski, J.~B. Gutowski, W.~A. Sabra and P.~Tod, \emph{{Cosmological
  Einstein-Maxwell Instantons and Euclidean Supersymmetry: Beyond
  Self-Duality}}, \href{http://dx.doi.org/10.1007/JHEP03(2011)131}{\emph{JHEP}
  {\bf 03} (2011) 131}, [\href{http://arxiv.org/abs/1012.1326}{{\tt
  1012.1326}}].

\bibitem{Freedman:1976aw}
D.~Z. Freedman and A.~K. Das, \emph{{Gauge Internal Symmetry in Extended
  Supergravity}},
  \href{http://dx.doi.org/10.1016/0550-3213(77)90041-4}{\emph{Nucl. Phys.} {\bf
  B120} (1977) 221--230}.

\bibitem{Gauntlett:2007ma}
J.~P. Gauntlett and O.~Varela, \emph{{Consistent Kaluza-Klein reductions for
  general supersymmetric AdS solutions}},
  \href{http://dx.doi.org/10.1103/PhysRevD.76.126007}{\emph{Phys. Rev.} {\bf
  D76} (2007) 126007}, [\href{http://arxiv.org/abs/0707.2315}{{\tt
  0707.2315}}].

\bibitem{Martelli:2012sz}
D.~Martelli, A.~Passias and J.~Sparks, \emph{{The supersymmetric NUTs and bolts
  of holography}},
  \href{http://dx.doi.org/10.1016/j.nuclphysb.2013.04.026}{\emph{Nucl. Phys.}
  {\bf B876} (2013) 810--870}, [\href{http://arxiv.org/abs/1212.4618}{{\tt
  1212.4618}}].

\bibitem{Dunajski:2010zp}
M.~Dunajski, J.~Gutowski, W.~Sabra and P.~Tod, \emph{{Cosmological
  Einstein-Maxwell Instantons and Euclidean Supersymmetry: Anti-Self-Dual
  Solutions}},
  \href{http://dx.doi.org/10.1088/0264-9381/28/2/025007}{\emph{Class. Quant.
  Grav.} {\bf 28} (2011) 025007}, [\href{http://arxiv.org/abs/1006.5149}{{\tt
  1006.5149}}].

\bibitem{Futaki:2006cc}
A.~Futaki, H.~Ono and G.~Wang, \emph{{Transverse Kahler geometry of Sasaki
  manifolds and toric Sasaki-Einstein manifolds}}, {\emph{J. Diff. Geom.} {\bf
  83} (2009) 585--636}, [\href{http://arxiv.org/abs/math/0607586}{{\tt
  math/0607586}}].

\bibitem{Gunaydin:1983bi}
M.~Gunaydin, G.~Sierra and P.~K. Townsend, \emph{{The Geometry of N=2
  Maxwell-Einstein Supergravity and Jordan Algebras}},
  \href{http://dx.doi.org/10.1016/0550-3213(84)90142-1}{\emph{Nucl. Phys.} {\bf
  B242} (1984) 244--268}.

\bibitem{Buchel:2006gb}
A.~Buchel and J.~T. Liu, \emph{{Gauged supergravity from type IIB string theory
  on $Y^{p,q}$ manifolds}},
  \href{http://dx.doi.org/10.1016/j.nuclphysb.2007.03.001}{\emph{Nucl. Phys.}
  {\bf B771} (2007) 93--112}, [\href{http://arxiv.org/abs/hep-th/0608002}{{\tt
  hep-th/0608002}}].

\bibitem{Cassani:2015upa}
D.~Cassani, J.~Lorenzen and D.~Martelli, \emph{{Comments on supersymmetric
  solutions of minimal gauged supergravity in five dimensions}},
  \href{http://dx.doi.org/10.1088/0264-9381/33/11/115013}{\emph{Class. Quant.
  Grav.} {\bf 33} (2016) 115013}, [\href{http://arxiv.org/abs/1510.01380}{{\tt
  1510.01380}}].

\bibitem{Taylor:2000xw}
M.~Taylor, \emph{{More on counterterms in the gravitational action and
  anomalies}},  \href{http://arxiv.org/abs/hep-th/0002125}{{\tt
  hep-th/0002125}}.

\bibitem{Anselmi:1997am}
D.~Anselmi, D.~Freedman, M.~T. Grisaru and A.~Johansen, \emph{{Nonperturbative
  formulas for central functions of supersymmetric gauge theories}},
  \href{http://dx.doi.org/10.1016/S0550-3213(98)00278-8}{\emph{Nucl.Phys.} {\bf
  B526} (1998) 543--571}, [\href{http://arxiv.org/abs/hep-th/9708042}{{\tt
  hep-th/9708042}}].

\bibitem{Gutowski:2004ez}
J.~B. Gutowski and H.~S. Reall, \emph{{Supersymmetric AdS$_5$ black holes}},
  \href{http://dx.doi.org/10.1088/1126-6708/2004/02/006}{\emph{JHEP} {\bf 02}
  (2004) 006}, [\href{http://arxiv.org/abs/hep-th/0401042}{{\tt
  hep-th/0401042}}].

\bibitem{Chong:2005hr}
Z.~W. Chong, M.~Cvetic, H.~Lu and C.~N. Pope, \emph{{General non-extremal
  rotating black holes in minimal five-dimensional gauged supergravity}},
  \href{http://dx.doi.org/10.1103/PhysRevLett.95.161301}{\emph{Phys. Rev.
  Lett.} {\bf 95} (2005) 161301},
  [\href{http://arxiv.org/abs/hep-th/0506029}{{\tt hep-th/0506029}}].

\bibitem{Martelli:2006yb}
D.~Martelli, J.~Sparks and S.-T. Yau, \emph{{Sasaki-Einstein manifolds and
  volume minimisation}},
  \href{http://dx.doi.org/10.1007/s00220-008-0479-4}{\emph{Commun. Math. Phys.}
  {\bf 280} (2008) 611--673}, [\href{http://arxiv.org/abs/hep-th/0603021}{{\tt
  hep-th/0603021}}].

\bibitem{Papadimitriou:2004ap}
I.~Papadimitriou and K.~Skenderis, \emph{{AdS / CFT correspondence and
  geometry}}, \href{http://dx.doi.org/10.4171/013-1/4}{\emph{IRMA Lect. Math.
  Theor. Phys.} {\bf 8} (2005) 73--101},
  [\href{http://arxiv.org/abs/hep-th/0404176}{{\tt hep-th/0404176}}].

\bibitem{Olea:2006vd}
R.~Olea, \emph{{Regularization of odd-dimensional AdS gravity: Kounterterms}},
  \href{http://dx.doi.org/10.1088/1126-6708/2007/04/073}{\emph{JHEP} {\bf 04}
  (2007) 073}, [\href{http://arxiv.org/abs/hep-th/0610230}{{\tt
  hep-th/0610230}}].

\bibitem{Romans:1991nq}
L.~J. Romans, \emph{{Supersymmetric, cold and lukewarm black holes in
  cosmological Einstein-Maxwell theory}},
  \href{http://dx.doi.org/10.1016/0550-3213(92)90684-4}{\emph{Nucl. Phys.} {\bf
  B383} (1992) 395--415}, [\href{http://arxiv.org/abs/hep-th/9203018}{{\tt
  hep-th/9203018}}].

\bibitem{Martelli:2011fw}
D.~Martelli and J.~Sparks, \emph{{The gravity dual of supersymmetric gauge
  theories on a biaxially squashed three-sphere}},
  \href{http://dx.doi.org/10.1016/j.nuclphysb.2012.08.015}{\emph{Nucl. Phys.}
  {\bf B866} (2013) 72--85}, [\href{http://arxiv.org/abs/1111.6930}{{\tt
  1111.6930}}].

\bibitem{Martelli:2013aqa}
D.~Martelli and A.~Passias, \emph{{The gravity dual of supersymmetric gauge
  theories on a two-parameter deformed three-sphere}},
  \href{http://dx.doi.org/10.1016/j.nuclphysb.2013.09.012}{\emph{Nucl. Phys.}
  {\bf B877} (2013) 51--72}, [\href{http://arxiv.org/abs/1306.3893}{{\tt
  1306.3893}}].

\bibitem{Huang:2014gca}
X.~Huang, S.-J. Rey and Y.~Zhou, \emph{{Three-dimensional SCFT on conic space
  as hologram of charged topological black hole}},
  \href{http://dx.doi.org/10.1007/JHEP03(2014)127}{\emph{JHEP} {\bf 03} (2014)
  127}, [\href{http://arxiv.org/abs/1401.5421}{{\tt 1401.5421}}].

\bibitem{Freedman:2013ryh}
D.~Z. Freedman and S.~S. Pufu, \emph{{The holography of $F$-maximization}},
  \href{http://dx.doi.org/10.1007/JHEP03(2014)135}{\emph{JHEP} {\bf 03} (2014)
  135}, [\href{http://arxiv.org/abs/1302.7310}{{\tt 1302.7310}}].

\bibitem{Bobev:2013cja}
N.~Bobev, H.~Elvang, D.~Z. Freedman and S.~S. Pufu, \emph{{Holography for $N =
  2^*$ on $S^4$}}, \href{http://dx.doi.org/10.1007/JHEP07(2014)001}{\emph{JHEP}
  {\bf 07} (2014) 001}, [\href{http://arxiv.org/abs/1311.1508}{{\tt
  1311.1508}}].

\bibitem{Bobev:2016nua}
N.~Bobev, H.~Elvang, U.~Kol, T.~Olson and S.~S. Pufu, \emph{{Holography for $
  \mathcal{N} $ = 1$^{*}$ on S$^{4}$}},
  \href{http://dx.doi.org/10.1007/JHEP10(2016)095}{\emph{JHEP} {\bf 10} (2016)
  095}, [\href{http://arxiv.org/abs/1605.00656}{{\tt 1605.00656}}].

\bibitem{Freedman:2016yue}
D.~Z. Freedman, K.~Pilch, S.~S. Pufu and N.~P. Warner, \emph{{Boundary Terms
  and Three-Point Functions: An AdS/CFT Puzzle Resolved}},
  \href{http://arxiv.org/abs/1611.01888}{{\tt 1611.01888}}.

\bibitem{Kol:2016ucd}
U.~Kol, \emph{{Holography for $\mathcal{N}=1^*$ on $S^4$ and Supergravity}},
  \href{http://arxiv.org/abs/1611.09396}{{\tt 1611.09396}}.

\bibitem{Andrianopoli:2014aqa}
L.~Andrianopoli and R.~D'Auria, \emph{{N=1 and N=2 pure supergravities on a
  manifold with boundary}},
  \href{http://dx.doi.org/10.1007/JHEP08(2014)012}{\emph{JHEP} {\bf 08} (2014)
  012}, [\href{http://arxiv.org/abs/1405.2010}{{\tt 1405.2010}}].

\bibitem{Papadimitriou:2010as}
I.~Papadimitriou, \emph{{Holographic renormalization as a canonical
  transformation}},
  \href{http://dx.doi.org/10.1007/JHEP11(2010)014}{\emph{JHEP} {\bf 11} (2010)
  014}, [\href{http://arxiv.org/abs/1007.4592}{{\tt 1007.4592}}].

\bibitem{Belyaev:2007bg}
D.~V. Belyaev and P.~van Nieuwenhuizen, \emph{{Tensor calculus for supergravity
  on a manifold with boundary}},
  \href{http://dx.doi.org/10.1088/1126-6708/2008/02/047}{\emph{JHEP} {\bf 02}
  (2008) 047}, [\href{http://arxiv.org/abs/0711.2272}{{\tt 0711.2272}}].

\bibitem{Grumiller:2009dx}
D.~Grumiller and P.~van Nieuwenhuizen, \emph{{Holographic counterterms from
  local supersymmetry without boundary conditions}},
  \href{http://dx.doi.org/10.1016/j.physletb.2009.11.022}{\emph{Phys. Lett.}
  {\bf B682} (2010) 462--465}, [\href{http://arxiv.org/abs/0908.3486}{{\tt
  0908.3486}}].

\bibitem{Belyaev:2008ex}
D.~V. Belyaev and P.~van Nieuwenhuizen, \emph{{Simple d=4 supergravity with a
  boundary}},
  \href{http://dx.doi.org/10.1088/1126-6708/2008/09/069}{\emph{JHEP} {\bf 0809}
  (2008) 069}, [\href{http://arxiv.org/abs/0806.4723}{{\tt 0806.4723}}].

\bibitem{Cvetic:2005zi}
M.~Cvetic, G.~W. Gibbons, H.~Lu and C.~N. Pope, \emph{{Rotating black holes in
  gauged supergravities: Thermodynamics, supersymmetric limits, topological
  solitons and time machines}},
  \href{http://arxiv.org/abs/hep-th/0504080}{{\tt hep-th/0504080}}.

\bibitem{Figueras:2006xx}
P.~Figueras, C.~A.~R. Herdeiro and F.~Paccetti~Correia, \emph{{On a class of 4D
  Kahler bases and AdS$_5$ supersymmetric Black Holes}},
  \href{http://dx.doi.org/10.1088/1126-6708/2006/11/036}{\emph{JHEP} {\bf 11}
  (2006) 036}, [\href{http://arxiv.org/abs/hep-th/0608201}{{\tt
  hep-th/0608201}}].

\bibitem{Lebrun:1991}
C.~LeBrun, \emph{Explicit self-dual metrics on $\mathbb{CP}_2 \#
  \cdots\#\mathbb{CP}_2$}, {\emph{J. Differential Geometry} {\bf 34} (1991)
  223--253}.

\bibitem{Tod:1995}
K.~P. Tod, \emph{{Scalar-flat K\"ahler and hyper-K\"ahler metrics from
  Painlev\'e-III}}, {\emph{Classical and Quantum Gravity} {\bf 12} (1995)
  1535}.

\end{thebibliography}\endgroup

\end{document}